\documentclass[prx,twocolumn,showpacs,amsmath,amssymb,superscriptaddress,floatfix,reprint, nofootinbib]{revtex4-2}
\usepackage{amsmath}
\usepackage{amssymb}
\usepackage{amsthm}
\usepackage{amsfonts}
\usepackage{dsfont}
\usepackage[normalem]{ulem}
\usepackage{listings}
\usepackage{macros}
\usepackage{enumerate}
\usepackage{latexsym}
\usepackage{ulem}
\usepackage{hyperref}

\usepackage{bbold}

\usepackage{psfrag}
\usepackage{xcolor}
\usepackage{float}

\usepackage{comment}

\usepackage{graphicx}
\usepackage{subfigure}

\begin{document}

\tolerance 10000
\title{Hilbert Space Fragmentation and Commutant Algebras}
\author{Sanjay Moudgalya}
\email{sanjaym@caltech.edu}
\affiliation{Department of Physics and Institute for Quantum Information and Matter,
California Institute of Technology, Pasadena, California 91125, USA}
\affiliation{Walter Burke Institute for Theoretical Physics, California Institute of Technology, Pasadena, California 91125, USA}
\author{Olexei I. Motrunich}
\affiliation{Department of Physics and Institute for Quantum Information and Matter,
California Institute of Technology, Pasadena, California 91125, USA}
\begin{abstract}
We study the phenomenon of Hilbert space fragmentation in isolated Hamiltonian and Floquet quantum systems using the language of commutant algebras, the algebra of all operators that commute with \textit{each term} of the Hamiltonian or each gate of the circuit. 
We provide a precise definition of Hilbert space fragmentation in this formalism as the case where the dimension of the commutant algebra grows exponentially with the system size.
Fragmentation can hence be distinguished from systems with conventional symmetries such as $U(1)$ or $SU(2)$, where the dimension of the commutant algebra grows polynomially with the system size. 
Further, the commutant algebra language also helps distinguish between ``classical" and ``quantum" Hilbert space fragmentation, where the former refers to fragmentation in the product state basis.
We explicitly construct the commutant algebra in several systems exhibiting classical fragmentation, including the $t-J_z$ model and the spin-1 dipole-conserving model, and  we illustrate the connection to previously-studied ``Statistically Localized Integrals of Motion" (SLIOMs).
We also revisit the Temperley-Lieb spin chains, including the spin-1 biquadratic chain widely studied in the literature, and show that they exhibit quantum Hilbert space fragmentation. 
Finally, we study the contribution of the full commutant algebra to the Mazur bounds in various cases.
In fragmented systems, we use expressions for the commutant to analytically obtain new or improved Mazur bounds for autocorrelation functions of local operators that agree with previous numerical results.
In addition, we are able to rigorously show the localization of the on-site spin operator in the spin-1 dipole-conserving model.
\end{abstract}
\date{\today}
\maketitle
%

%%%%%%%%%%%%%%%%%%%%%%%%%%%%%%%%
%
\tableofcontents
\section{Introduction}
\label{sec:intro}
The study of ergodicity and its breaking in isolated quantum systems has been a subject of active research. 
Ergodicity in isolated quantum systems is defined by the Eigenstate Thermalization Hypothesis (ETH), a conjecture about the matrix elements of local operators in between the eigenstates of the Hamiltonian~\cite{srednicki1994chaos, deutsch1991quantum}. 
Restricting to diagonal matrix elements, which is sometimes referred to as diagonal ETH~\cite{shiraishimorireply2018}, it states that the expectation value of any local operator in eigenstates of the Hamiltonian is a smooth function of energy, determined by its thermal expectation value in an appropriate Gibbs density matrix~\cite{rigol2008thermalization, polkovnikov2011colloquium, d2016quantum}.
Its strong version, known as \textit{strong ETH}, states that \textit{all} eigenstates satisfy diagonal ETH, whereas its weak version, known as \textit{weak ETH}, states that \textit{almost all} eigenstates satisfy diagonal ETH, which allows for a small (measure-zero) set of eigenstates to violate it. 
While strong ETH is believed to hold in generic isolated quantum systems, its complete violations (i.e., violations of strong and weak ETH) are well-known in two cases.
First are integrable systems, where an extensive number of conserved quantities leads to quasiparticle descriptions and complete solvability of the spectrum. 
Second are Many-Body Localized systems, where strong disorder or quasiperidocity results in the existence of emergent integrals of motion that leads to localized eigenstates throughout the spectrum, although its stability for large system sizes in more than one dimension is a subject of much debate~\cite{suntajs2020quantum, abanin2021distinguishing}.
In addition to complete violations, several examples of partial violations of ETH have been recently found. 
One such family of examples is comprised of quantum scarred systems, which possess a small number of ``quantum scars," i.e., ETH-violating eigenstates amidst a sea of ETH-satisfying eigenstates. 
The quantum scars in all such systems form a measure zero set of eigenstates in the thermodynamic limit, and weak ETH is satisfied while strong ETH is violated. 
Nevertheless, quantum scars have a striking impact on the dynamics of particular initial states including in systems with experimental relevance, as seen in Rydberg atom experiments~\cite{bernien2017probing, turner2017quantum}. 
Several examples of exactly solvable quantum scars have been found in the literature~\cite{serbyn2020review}, including ones with equally spaced towers of states~\cite{moudgalya2018a, moudgalya2018b, choi2018emergent, schecter2019weak, iadecola2020quantum, shibata2020onsager, mark2020unified, moudgalya2020large, moudgalya2020eta, mark2020eta, pakrouski2020many, ren2020quasisymmetry, odea2020from} that lead to perfect revivals from particular initial states.
Many of these examples can be captured by unified formalisms~\cite{mori2017eth, mark2020unified, moudgalya2020eta, mark2020eta, pakrouski2020many, ren2020quasisymmetry, odea2020from} which typically involve starting from a highly symmetric Hamiltonian and adding perturbations that preserve some of its eigenstates. 
Another family of recently discovered violations of ETH is Hilbert space fragmentation found in constrained Hamiltonian systems as well as random unitary circuits, which will be the focus of this work. 
The term typically refers to the phenomenon where the system possesses exponentially many dynamically disconnected subspaces, referred to as Krylov subspaces. 
To be more precise, the Hilbert space $\mH$ of any isolated quantum system can be generically described as
\begin{equation}
    \mH = \bigoplus_{\alpha = 1}^K{\mK_\alpha},\;\;\;\mK_\alpha \defn \textrm{span}_t\{U^t \ket{\psi_\alpha}\},
\label{eq:Hilbertsplitting}
\end{equation}
where $U^t$ is the unitary governing the time evolution (i.e., $e^{-i H t}$ for systems with Hamiltonian $H$), $\textrm{span}_t\{U^t \ket{\psi_\alpha}\}$ denotes the subspace spanned by time-evolution of the state $\ket{\psi_\alpha}$, and $K$ is the number of Krylov subspaces of the system. 
While the decomposition of Eq.~(\ref{eq:Hilbertsplitting}) is trivial if $\ket{\psi_\alpha}$'s are eigenstates of $U$, typical examples of Hilbert space fragmentation focus on cases where $\ket{\psi_\alpha}$'s are ``simple" states (e.g., product states).
The existence of dynamically disconnected subspaces is also not surprising in the presence of symmetries, and  $\ket{\psi_\alpha}$'s in Eq.~(\ref{eq:Hilbertsplitting}) differ by some symmetry quantum numbers that are preserved under time-evolution. 
However, Hilbert space fragmentation differs from regular symmetries in two important ways. 
Firstly, in the case of conventional symmetries, the number of Krylov subspaces $K$ either stays constant or grows polynomially with increasing system size whereas it grows exponentially in fragmented systems, e.g., $K \sim \exp(c L)$ for one-dimensional systems with $L$ sites. 
Secondly, the Krylov subspaces in fragmentation systems $\mK_\alpha$ do not seem to be distinguished by quantum numbers corresponding to any obvious local symmetries of the Hamiltonian $H$. 
The phenomenon of Hilbert space fragmentation was explicitly pointed out in the context of dipole-moment conserving models~\cite{pai2018localization, sala2020fragmentation, khemani2020localization}, although similar phenomena have been known or implicitly assumed in earlier literature~\cite{ritort2003glassy, bergholtz2005half, bergholtz2006one, olmos2010thermalization, sikora2011extended,nakamura2012exactly, horssen2015dynamics, lan2018quantum, gopalakrishnan2018facilitated}.
The Krylov subspaces $\{\mK_\alpha\}$ can have any dimension, ranging from one-dimensional ``frozen" product states~\cite{sala2020fragmentation, khemani2020localization} where all terms of the Hamiltonian act trivially, to ones with exponentially large dimension that can be studied in terms of a restricted effective Hamiltonian~\cite{moudgalya2019thermalization, moudgalya2020quantum}.
From the perspective of the full Hilbert space $\mH$ (within a quantum number sector of a ``conventional" symmetry), fragmented systems can either violate strong ETH or violate weak (and hence also strong) ETH. 
Referring to the dimension of the largest Krylov subspace as $D_{\textrm{max}} \defn \textrm{max}_\alpha\{\textrm{dim}(\mK_\alpha)\}$, and the Hilbert space dimension as $D \defn \dim(\mH)$, Ref.~\cite{sala2020fragmentation} further classified fragmented systems into two classes: strongly fragmented and weakly fragmented, where $D_{\max}/D \rightarrow 0$ and $D_{\max}/D \rightarrow 1$ respectively in the thermodynamic limit.
Weakly fragmented systems have a dominant Krylov subspace in the thermodynamic limit, and hence while they violate strong ETH due to the small Krylov subspaces, they satisfy weak ETH. These systems share a lot of phenomenology with quantum many-body scars. 
Strongly fragmented systems, on the other hand, do not have a dominant Krylov subspace, and hence violate weak ETH as well.
Nevertheless, signatures of ETH within sufficiently large Krylov subspaces $\mK_\alpha$ (referred to as Krylov-restricted thermalization~\cite{moudgalya2019thermalization, serbyn2020review}) were found in models exhibiting both strong and weak Hilbert space fragmentation~\cite{moudgalya2019thermalization, yang2019hilbertspace, hahn2021information}, provided the Hamiltonian restricted to the studied Krylov subspace is non-integrable and not many-body localized~\cite{detomasi2019dynamics,herviou2021mbl}.  
Several examples of Hilbert space fragmentation that do not involve dipole-conserving models have been found in Refs.~\cite{sliom2020, yang2019hilbertspace, lee2021frustration, langlett2021hilbert, mukherjee2021constraint, mukherjee2021minimal, hahn2021information,  khudorozhkov2021hilbert,richter2021anomalous}. 
In spite of several examples of Hilbert space fragmentation, many aspects about fragmentation remain vague or unanswered.
Since the existence of Krylov subspaces of the form of Eq.~(\ref{eq:Hilbertsplitting}) by definition implies that the Hamiltonian is block diagonal in a certain basis, an obvious question that arises is whether the system has non-obvious non-local conserved quantities. 
While the existence and construction of such non-obvious conserved quantities are well-known in the study of quantum integrable systems, Hilbert space fragmentation differs from quantum integrability since it is completely determined by the local terms of the Hamiltonian, and does not require additional symmetries such as translation invariance that are key to integrability. 
That is, in all examples that defined the fragmentation concept, if a Hamiltonian of the form $H = \sum_j{\hh_{j}}$ shows Hilbert space fragmentation, where $\hh_{j}$ denotes a local few-site term, so does the entire family of Hamiltonians $H = \sum_j{J_j \hh_j}$, where $J_j$'s are arbitrary coefficients.
An important step towards understanding the nature of conserved quantities in fragmented systems was made in Ref.~\cite{sliom2020}. 
It turns out that highly non-local conserved operators, called ``Statistically Localized Integrals of Motion" or SLIOMs, uniquely label all the different Krylov subspaces in certain fragmented systems. 
However, the construction of SLIOMs there does not directly extend to all systems exhibiting fragmentation, and it is natural to wonder if SLIOMs are generic to systems exhibiting fragmentation. 
Furthermore, in all systems exhibiting fragmentation of the form of Eq.~(\ref{eq:Hilbertsplitting}), the projectors onto the Krylov subspaces $\{\mK_\alpha\}$ are conserved quantities, which are conserved by definition, and it is not clear if and how they are related to SLIOMs. 
In this work, we resolve these questions by studying fragmentation in the language of so-called commutant algebras, which is the algebra of conserved quantities [a.k.a. Integrals of Motion (IoMs)] for an entire family of Hamiltonians.
For the family $H = \sum_j J_j \hh_j$ discussed above, the commutant algebra is the algebra of all operators that commute with all $\hh_j$'s. 
This formalism allows us to compare and contrast fragmentation versus conventional symmetries, and also makes clear the relation between SLIOMs and the Krylov subspace projectors, both of which are conserved quantities and hence just different ``vectors'' in a uniquely defined commutant algebra associated with local terms in the Hamiltonian.
Another important aspect that is seldom discussed is the basis in which fragmentation occurs.
Most examples of Hilbert space fragmentation are in the product state basis, i.e., where $\ket{\psi_\alpha}$'s in Eq.~(\ref{eq:Hilbertsplitting}) are product states.
This means that the phenomenon is essentially classical in nature, and also exists in classical Markov processes with the same set of allowed local transitions, which has in fact been discussed in previous works~\cite{ritort2003glassy, morningstar2020kineticallyconstrained}. 
This leads us to wonder if truly quantum version of fragmentation can exist, which would be the case if fragmentation happens in a non-obvious entangled basis such that some of the $\ket{\psi_\alpha}$'s in Eq.~(\ref{eq:Hilbertsplitting}) necessarily need to have entanglement. 
Naively allowing for non-product state basis without a more precise definition of fragmentation is not helpful since for any finite-size system, the eigenstates are themselves one-dimensional dynamically disconnected Krylov subspaces, which leads to the meaningless conclusion that all finite-size systems are fragmented. 
Similar confusions also exist for the definition of quantum integrability in finite-size systems, as discussed in Refs.~\cite{yuzbashyan2013integrability, scaramazza2016integrable}. 
Hence, we first need a precise rigorous definition of fragmentation, which we provide in this work using the language of commutant algebras. 
This also allows us to distinguish between classical fragmentation in the product state basis and quantum fragmentation in entangled bases, and we find models that exist in the literature that are examples of the latter.
The rest of this paper is organized as follows. 
In Sec.~\ref{sec:commutants}, we introduce the concept of commutant algebras, which will be central to our discussion of Hilbert space fragmentation, and we discuss examples of conventional symmetries in this language. 
In Sec.~\ref{sec:tJz} we work out the Hilbert space fragmentation of the $t-J_z$ model in the language of commutant algebras, and also illustrate the connections between the commutant algebras and previously-constructed SLIOMs in the 1d $t-J_z$ model with open boundary conditions. 
In Sec.~\ref{sec:pairflip}, we similarly demonstrate the fragmentation in yet another model, which we refer to as Pair-Flip (PF) model, where the definition of SLIOMs is not a priori clear.
While these models are examples of ``classical fragmentation" in the product state basis, in Sec.~\ref{sec:TL} we study the Temperley-Lieb (TL) spin chains, which we show are examples of models exhibiting quantum fragmentation in an entangled basis; a distinguishing feature of the TL family is that its commutant algebra is non-abelian, while classical fragmentation examples have abelian commutant algebras.
In Sec.~\ref{sec:dipole}, we discuss the well-known strongly-fragmented spin-1 dipole-conserving model~\cite{sala2020fragmentation, sliom2020} in the commutant algebra language, and we analytically construct and count the full commutant.
Finally in Sec.~\ref{sec:Mazur}, we study the effect of the full commutant algebra on the Mazur bounds for autocorrelation functions of local operators, and we illustrate standard results for conventional symmetries.
Further, as a result of our analytical understanding of the full commutant, we are able to analytically compute the Mazur bounds in detail in many of the fragmented models.
In the $t-J_z$ model with open and periodic boundary conditions, we show that the commutant provides improved bounds that are not fully captured by SLIOMs, and we also analytically recover many of the numerical results of Ref.~\cite{sliom2020}.
In the spin-1 dipole-conserving model, we are able to analytically compute a large part of the full Mazur bound, which rigorously proves the localization of the on-site spin operator.
We also present numerical and analytical results of enhanced Mazur bounds corresponding to certain local operators in the PF and TL models.
We conclude with open questions in Sec.~\ref{sec:conclusion}. 
\section{Commutant Algebras}\label{sec:commutants}
\begin{figure}[t!]
    \centering
    \includegraphics[scale=0.4]{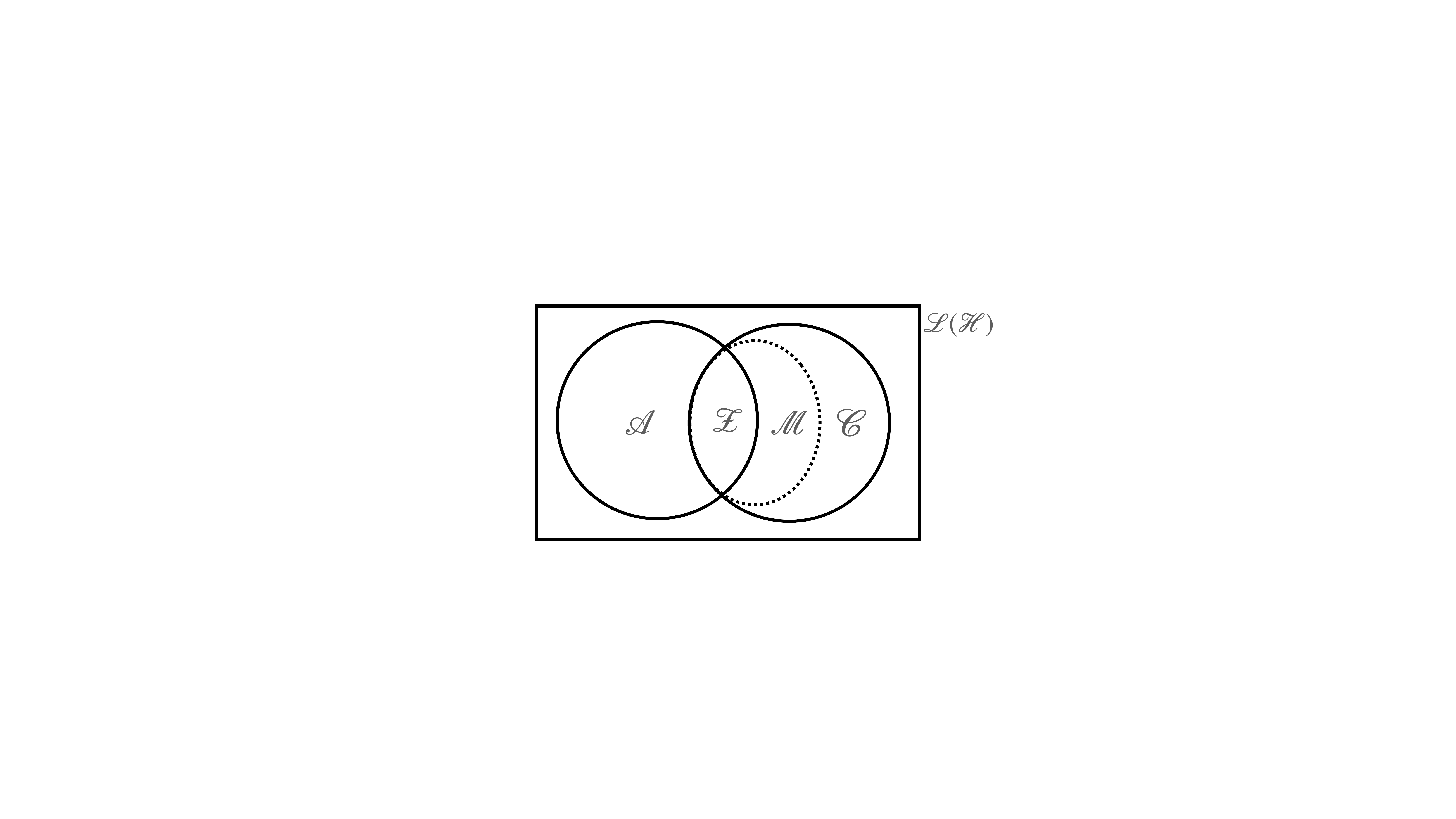}
    \caption{Depiction of the algebras studied in this work. $\mL(\mH)$ is the algebra of all linear operators on a finite-dimensional Hilbert space $\mH$.
    The left solid circle is the ``bond algebra" $\mA$, the algebra generated by the local terms $\{\hh_j\}$ of the family of Hamiltonians $H = \sum_j{J_j \hh_j}$.
    The right solid circle is the ``commutant algebra" $\mC$, the algebra of all elements that commute with every element of $\mA$.
    $\mA$ and $\mC$ are centralizers of each other in $\mL(\mH)$ as a consequence of the double commutant theorem~\cite{landsman1998lecture}.
    $\mZ = \mA \cap \mC$ is the center of both algebras.
    The dotted circle is a maximal Abelian subalgebra $\mM$ of $\mC$, which might not be unique. 
    Hilbert space fragmentation corresponds to the case when the dimension of $\mC$ grows exponentially with system size.
    When $\mC$ is Abelian (e.g., for classical fragmentation), we get $\mC = \mM = \mZ \subseteq \mA$.
    }
    \label{fig:commutant}
\end{figure}
\subsection{Definition and Properties}\label{subsec:defn}
As discussed in the previous section, Hilbert space fragmentation depends only on the local terms of the Hamiltonian, and the Hilbert space decomposition of Eq.~(\ref{eq:Hilbertsplitting}) is the same for a family of Hamiltonians $H$
\begin{equation}
    H = \sumal{j}{}{J_j \hh_{j}},  
\label{eq:genhamil}
\end{equation}
where $\hh_j$ is a strictly local (generically multi-site) operator in the vicinity of site $j$ such that $\hh_i$ and $\hh_j$ for $i \neq j$ need not commute and $J_j$ are arbitrary coefficients.  
Since we are interested in the block-diagonal structure of the family of Hamiltonians in Eq.~(\ref{eq:genhamil}) that does not depend on local couplings (similar in this aspect to conventional on-site symmetries), we are interested in operators $\hO$ that commute commutes with \textit{each term}, i.e.,
\begin{equation}
    [\hh_{j}, \hO] = 0\;\;\;\forall j.  
\label{eq:commutant}
\end{equation}
We refer to such operators $\hO$ either as conserved quantities or Integrals of Motion (IoMs) associated with the family of Hamiltonians of Eq.~(\ref{eq:genhamil}).
Denoting the set of operators $\hO$ that satisfy Eq.~(\ref{eq:commutant}) as $\mC$, we note that operators in $\mC$ form a closed associative algebra, i.e. 
\begin{align}
    &\hO_1 \in \mC,\;\; \hO_2 \in \mC\nn \\
    \implies&\;\; \left\{\begin{array}{ll}
    &\alpha_1 \hO_1 + \alpha_2 \hO_2 \in \mC\\
    &\hO_1 \hO_2, \hO_2 \hO_1  \in \mC\\
    \end{array}\right.\;\;\textrm{for any}\;\;\alpha_1, \alpha_2 \in \mathbb{C},
\label{eq:algebradefn}
\end{align}
and hence we refer to $\mC$ as the commutant algebra. 
If the family of systems contains several types of terms, e.g., Hamiltonians $H = \sum_j{J_j \hh_{j}} + \sum_j{K_j \hg_{j}}$ with arbitrary coefficients $J_j$ and $K_j$, we define the commutant as the algebra of operators $\hO$ that individually commute with all types of local terms, i.e., $[\hh_{j}, \hO] = [\hg_{j}, \hO] = 0$ for all $j$.   
This definition also implies that operators in the commutant commute with unitary circuits $U$ built by local gates using the same terms $\{\hh_j\}$ such as
\begin{equation}
    U = \prodal{j}{}{\exp\left(-i J_j \hh_j\right)},
\label{eq:genUnitary}
\end{equation}
where the ordering of terms in the product in $U$ does not matter for the commutant considerations (but of course matters for specific model instances).
Hence, while we only explicitly describe Hamiltonian systems in this work, all of our results also hold for unitary circuits of the form of Eq.~(\ref{eq:genUnitary}).
An alternate equivalent definition of $\mC$ involves the algebra $\mA$ generated by arbitrary linear combinations of arbitrary products of the terms of the Hamiltonian $\{\hh_j\}$, and we refer to this as ``bond algebra,"~\cite{nussinov2009bond, cobanera2010unified, cobenera2011bond} since the Hamiltonian terms are typically associated with bonds on the lattice, although we will sometimes include single/multi-site Hamiltonian terms. 
While the identity operator $\mathds{1}$ is not necessarily generated by these terms, we can always add it to the definition of $\mA$ since adding a constant to the family of Hamiltonians does not affect their symmetries.
Denoting the algebra of all linear operators on the (finite-dimensional) Hilbert space $\mH$ as $\mL(\mH)$, both $\mA$ and $\mC$ are subalgebras of $\mL(\mH)$.
As a consequence of Eq.~(\ref{eq:commutant}), the commutant algebra $\mC$ is the centralizer of the algebra $\mA$ in $\mL(\mH)$.
Note that $\mA$ and $\mC$ are both associative algebras that contain the $\mathds{1}$ operator.
Further, $\mA$ is generated starting from Hermitian $\hh_j$'s, $\mA$ and $\mC$ are also closed under Hermitian conjugation (i.e., $\hO^\dagger \in \mA/\mC$ if $\hO \in \mA/\mC$).
Due to these properties, $\mA$ and $\mC$ are examples of von Neumann algebras~\cite{landsman1998lecture,harlow2017}.
Von Neumann algebras are subject to the double commutant theorem~\cite{landsman1998lecture, harlow2017}, which states that $\mA$ and $\mC$ are centralizers of each other in $\mL(\mH)$.
Hence, the centers of the algebras $\mA$ and $\mC$ coincide, given by $\mZ \defn \mA \cap \mC$. 
All these algebras are depicted in Fig.~\ref{fig:commutant}.
Note that throughout this work, we will reserve the use of ``commutant" to denote $\mC$ and use ``centralizer" to denote the algebra that commutes with the given algebra. 
Further, we always restrict ourselves to systems with tensor product Hilbert spaces with total dimension $D$.
Hence we do not attempt to distinguish between the algebra and its $D$-dimensional representation, and we always mean the latter when we say the former.
The operators in the commutant $\mC$ then have a naturally defined Hilbert-Schmidt inner product, and we can construct an orthonormal basis of $\mC$.
The number of such basis elements is referred to as the dimension of the commutant, denoted by $\textrm{dim}(\mC)$.
In general, this is distinct from the number of generators of the algebra, which we denote by $\textrm{gen}(\mC)$, which is the minimal number of operators required to generate the entire algebra by means of arbitrary sums and products.
These algebras that are centralizers of each other can be used to construct a virtual bipartition~\cite{zanardi2001virtual, bartlett2007reference, lidar2014dfs} of the Hilbert space, i.e., the full Hilbert space $\mH$ can be decomposed into representations of $\mA \times \mC$ as follows~\cite{fulton2013representation,readsaleur2007}
\begin{equation}
    \mH = \bigoplus_{\lambda}{\left(\mH^{(\mA)}_\lambda \otimes \mH^{(\mC)}_\lambda\right)},
\label{eq:Hilbertdecomp}
\end{equation}
where $\mH^{(\mA)}_\lambda$ and $\mH^{(\mC)}_\lambda$ denote $D_\lambda$ and $d_\lambda$ dimensional irreducible representations of the algebras $\mA$ and $\mC$ respectively.
Specifically, for each $\lambda$, the $\mH^{(\mA)}_\lambda \otimes \mH^{(\mC)}_\lambda$ represents a subspace of dimension $D_\lambda d_\lambda$ that can be formally tensored such that operators of the bond algebra $\mA$ act only on the ``degrees of freedom'' in the first factor while operators in the commutant algebra $\mC$ act only in the second factor.
That is, for each $\lambda$ there exists a tensored basis $\{\ket{u_{\lambda,\alpha}} \otimes \ket{v_{\lambda,\beta}}\}$, $\alpha = 1,\dots,D_\lambda, \beta=1,\dots,d_\lambda$, in which operators $\hh_{\mA} \in \mA$  and $\hh_{\mC} \in \mC$ act as $\hh_{\mA} \ket{u_{\lambda, \alpha}} \otimes \ket{v_{\lambda, \beta}} = \sum_{\alpha'} M^\lambda_{\alpha,\alpha'}(\hh_{\mA}) \ket{u_{\lambda, \alpha'}} \otimes \ket{v_{\lambda, \beta}}$ and $\hh_{\mC} \ket{u_{\lambda, \alpha}} \otimes \ket{v_{\lambda, \beta}} = \sum_{\beta'} N^\lambda_{\beta,\beta'}(\hh_{\mC}) \ket{u_{\lambda, \alpha}} \otimes \ket{v_{\lambda, \beta'}}$, where $M^\lambda(\hh_{\mA})$ and $N^\lambda(\hh_{\mC})$ are some $D_\lambda \times D_\lambda$ and $d_\lambda \times d_\lambda$ matrices respectively.
Operators in $\hh_{\mA}$ and $\hh_{\mC}$ thus have the matrix representations
\begin{equation}
    \hh_{\mA} = \bigoplus_\lambda{\left(M^\lambda(\hh_{\mA}) \otimes \mathds{1}_{d_\lambda}\right)},\;\;\;\hh_{\mC} = \bigoplus_\lambda{\left(\mathds{1}_{D_\lambda} \otimes N^\lambda(\hh_{\mC})\right)}.
\label{eq:matrixreps}
\end{equation}
Moreover, for each $\lambda$, $\mA$ and $\mC$ act in $\mH^{(\mA)}_\lambda$ and $\mH^{(\mC)}_\lambda$ as the full matrix algebras of $D_\lambda \times D_\lambda$ and $d_\lambda \times d_\lambda$ complex matrices respectively.
As a consequence, the dimensions of $\mA$ and $\mC$ are simply the dimensions of the subspaces of matrices of the form of Eq.~(\ref{eq:matrixreps}), and we obtain
\begin{equation}
    \dim(\mA) = \sum_\lambda{D_\lambda^2},\;\;\dim(\mC) = \sum_\lambda{d_\lambda^2},\;\;\dim(\mH) = \sum_\lambda{D_\lambda d_\lambda}.
\label{eq:dimensions}
\end{equation}
Note that the representations of Eq.~(\ref{eq:matrixreps}) also imply that any operator $\hh_{\mZ}$ in the center $\mZ$ of these algebras has the matrix representation 
\begin{equation}
    \hh_{\mZ} = \bigoplus_\lambda\left(c_\lambda(\hh_{\mZ})\ (\mathds{1}_{D_\lambda} \otimes \mathds{1}_{d_\lambda})\right),\;\;\;c_\lambda(\hh_{\mZ}) \in \mathbb{C}. 
\label{eq:centerelements}
\end{equation}
The decomposition of Eq.~(\ref{eq:Hilbertdecomp}) also characterizes the Krylov subspaces, i.e., subspaces invariant under time evolution, of the Hamiltonian $H$ of Eq.~(\ref{eq:genhamil}). 
Since the time-evolution unitary $\exp(-i H t)$ is an element of the bond algebra $\mA$, for each $\lambda$ in Eq.~(\ref{eq:Hilbertdecomp}), it only acts on the basis elements of the first factor $\mH^{(\mA)}_\lambda$ while leaving the basis elements of the second factor $\mH^{(\mC)}_\lambda$ invariant. 
Hence, for each $\lambda$, we obtain $d_\lambda$ number of $D_\lambda$-dimensional subspaces that are invariant under time-evolution, which are precisely the Krylov subspaces of Eq.~(\ref{eq:Hilbertsplitting}).
Note that in the above discussion we are free to perform any change of bases in $\mH^{(\mA)}_\lambda$ and $\mH^{(\mC)}_\lambda$, and for $d_\lambda > 1$ (which corresponds to non-Abelian commutants and implies $d_\lambda$-fold degeneracies in the Hamiltonian spectrum, see below) we can have different choices of degenerate Krylov subspaces.
The number of Krylov subspaces $K$ can be expressed only in terms of the dimensions of the irreducible representations of $\mC$:
\begin{equation}
    K = \sum_\lambda{d_\lambda} = \dim(\mM). 
\label{eq:Krylovnumber}
\end{equation}
As we indicate in Eq.~(\ref{eq:Krylovnumber}), the number of Krylov subspaces is simply the dimension of the maximal Abelian subalgebra of $\mC$, which we denote by $\mM$ (there could be multiple choices for $\mM$).
This is evident in the matrix representation of Eq.~(\ref{eq:matrixreps}) in the basis of Eq.~(\ref{eq:Hilbertdecomp}). 
Since $\mM$ is the maximal subspace of operators that are part of $\mC$ and that commute among themselves, using the fact that the maximal Abelian subalgebra of the full matrix algebra is, up to a fixed basis choice, its diagonal subalgebra (i.e., the algebra of all diagonal matrices), we deduce that any operator $\hh \in \mM$ has the matrix representation
\begin{equation}
    \hh_{\mM} = \bigoplus_\lambda\left(\mathds{1}_{D_\lambda} \otimes N^\lambda_{\textrm{diag}}(\hh_{\mM})\right),
\label{eq:abelianmatrixform}
\end{equation}
where $N^\lambda_{\textrm{diag}}(\hh_{\mM})$ is a $d_\lambda \times d_\lambda$ diagonal matrix (and we have implicitly made appropriate fixed basis choice that depends on the $\mM$ used).
The dimension of the subspace of matrices of the form of Eq.~(\ref{eq:abelianmatrixform}) is directly given by Eq.~(\ref{eq:Krylovnumber}).
Using the tensored basis of $\{\ket{u_{\lambda, \alpha}} \otimes \ket{v_{\lambda, \beta}}\}$ given by Eq.~(\ref{eq:Hilbertdecomp}), the Krylov subspaces are uniquely labeled by the states $\{\ket{v_{\lambda, \beta}}\}$, which are simply the eigenvectors of the matrices in Eq.~(\ref{eq:abelianmatrixform}).
Hence, the Krylov subspaces can be uniquely labeled by eigenvalues of a minimal set of generators of $\mM$, which further justifies Eq.~(\ref{eq:Krylovnumber}).
Finally, as evident in Eq.~(\ref{eq:matrixreps}), the existence of these invariant subspaces also implies that elements of the bond algebra $\mA$, hence all Hamiltonians, have a block-diagonal structure in the tensored basis determined by Eq.~(\ref{eq:Hilbertdecomp}).
Particularly, for each $\lambda$ in Eq.~(\ref{eq:Hilbertdecomp}), we obtain $d_\lambda$ identical blocks of dimension $D_\lambda$.
This leads to degeneracies in the full spectrum when the commutant $\mC$ is non-Abelian since it then admits irreducible representations with dimensions $d_\lambda > 1$.
\subsection{Conventional Examples}\label{subsec:conventionalexamples}
A wide spectrum of models usually studied in quantum many-body physics, including those with symmetries, can be described in terms of bond and commutant algebras.  
As we discuss in this section, these range from non-integrable ones without any symmetry to completely solvable ones, with symmetric ones lying between these two extremes.  
\subsubsection{No symmetries}
We first consider Hamiltonians with no symmetries. In this case, the only operator in the commutant $\mC$ is the identity operator $\mathds{1}$, and hence $\dim(\mC) = \gen(\mC) = 1$.
Since the bond algebra $\mA$ is the centralizer of $\mC$, $\mA = \mL(\mH)$, the algebra of all operators on the Hilbert space.  
Due to Eq.~(\ref{eq:dimensions}), this implies that $\lambda$ in Eq.~(\ref{eq:Hilbertdecomp}) takes a single value with $d_\lambda = 1$, $D_\lambda = \dim(\mH)$, and $\dim(\mA) = (\dim(\mH))^2$.
As a consequence of Eq.~(\ref{eq:Krylovnumber}), we obtain $K = 1$, which implies that the system has a single dynamically disconnected Krylov subspace (i.e., the full Hilbert space), as expected for systems without any symmetry. 
\subsubsection{Abelian symmetries}\label{subsubsec:abelian}
We then consider a family of systems with an Abelian symmetry, for example one-dimensional spin-1/2 XXZ models with on-site magnetic fields, given by the family of Hamiltonians
\begin{equation}
    H_{\text{XXZ}} = \sumal{j = 1}{L}{\left[J_j^\perp \left(S^x_j S^x_{j+1} + S^y_j S^y_{j+1} \right) + J_j^z S^z_j S^z_{j+1} \right]} + \sumal{j = 1}{L} h_j S^z_j,
\label{eq:XXZ}
\end{equation}
where $J_j^\perp$'s, $J_j^z$'s, and $h_j$'s are arbitrary coefficients, and we have used periodic boundary conditions ($L + 1 \equiv 1$).
The XXZ model is $U(1)$-symmetric, and the associated conserved quantity is the total spin $S^z_\tot \defn \sum_{j=1}^L{S^z_j}$.
$S^z_\tot$ is part of the commutant algebra $\mC$ corresponding to the bond algebra $\mA$ generated by the terms $\{(S^x_j S^x_{j+1} + S^y_j S^y_{j+1})\}$ and $\{S^z_j\}$ of the XXZ Hamiltonian of Eq.~(\ref{eq:XXZ}), since it commutes with each of them, i.e.
\begin{equation}
    [S^x_j S^x_{j+1} + S^y_j S^y_{j+1}, S^z_{\tot}] = 0,\;\;[S^z_j, S^z_{\tot}] = 0,\;\;1 \leq j \leq L.
\label{eq:U1commutant}
\end{equation}
The commutant algebra $\mC$ is precisely the algebra spanned by all powers of the operator $S^z_\tot$ along with the identity operator $\mathds{1}$.
Using the fact that $\left(S^z_j\right)^2 = \mathds{1}/4$, it is easy to see that $\left(S^z_\tot\right)^{L+1}$ can be expressed in terms of lower powers of $S^z_\tot$, which shows that $\mC$ is spanned by $\{\mathds{1}, S^z_\tot, (S^z_\tot)^2, \cdots, (S^z_\tot)^{L}\}$, and hence $\dim(\mC) = L + 1$.  
Furthermore, since this is an example of an Abelian commutant ($\mC = \mZ$), the irreducible representations of $\mC$ are one-dimensional, i.e., $d_\lambda = 1$ for all $\lambda$ in Eq.~(\ref{eq:Hilbertdecomp}).  
As a consequence of Eq.~(\ref{eq:dimensions}), this means that $\lambda$ runs over $(L+1)$ values, which is consistent with the fact that the total spin for a spin-1/2 system with $L$ spins can only take $(L + 1)$ values (between $-L/2$ and $L/2$). 
The bond algebra $\mA$ admits irreducible representations of dimensions $D_\lambda = \binom{L}{\lambda}$ (which is simply the number of product states with $S^z_\tot = L/2 - \lambda$) such that $\sum_{\lambda=0}^L{D_\lambda d_\lambda} = 2^L = \dim(\mH)$. 
Using Eq.~(\ref{eq:dimensions}), we can also obtain the dimension of the bond algebra to be $\dim(\mA) = \binom{2L}{L}$. 
\subsubsection{Non-Abelian symmetries}\label{subsubsec:nonabelian}
We now illustrate the commutant algebra in a family of systems with a non-Abelian symmetry, for example the one-dimensional spin-1/2 Heisenberg model, given by
\begin{equation}
    H_{\text{Heis}} \defn \sum_j{J_j \vec{S}_{j}\cdot\vec{S}_{j+1}}.   
\label{eq:Heisenberg}
\end{equation}
The Heisenberg Hamiltonian is known to be $SU(2)$-symmetric, and the three generators of $SU(2)$ group  $\{S^x_\tot, S^y_\tot, S^z_\tot\}$ (where $S^\alpha_\tot \defn \sum_{j = 1}^L{S^\alpha_j}$ for $\alpha \in \{x, y, z\}$) are all part of the commutant algebra $\mC$. 
That is, they satisfy
\begin{equation}
    [\vec{S}_j \cdot \vec{S}_{j+1}, \sumal{i = 1}{L}{S^\alpha_i}] = 0,\;\;1 \leq j \leq L,\;\;\alpha \in \{x, y, z\}. 
\label{eq:SU2commutant}
\end{equation}
The full commutant algebra $\mC$ is the associative algebra consisting of all products and their linear combinations of $S^x_\tot$, $S^y_\tot$, and $S^z_\tot$, which is known as the Universal Enveloping Algebra of the Lie algebra $\mathfrak{su}(2)$, denoted by $U(\mathfrak{su}(2))$. 
Since we know that the dimensions of the irreducible representations of $SU(2)$ (hence $U(\mathfrak{su}(2))$) are given by $d_\lambda = 2\lambda + 1$ (corresponding to the spin-$\lambda$ representation) for $0 \leq \lambda \leq L/2$, using Eq.~(\ref{eq:dimensions}) we can show that $\dim(\mC) = \binom{L+3}{3}$.  
The center $\mZ$, which consists of the operators that commute with all operators in $\mC$, is exhausted by the quadratic and higher order Casimir operators such as $\vec{S}_{\tot}^2 \defn (S^x_{\tot})^2 + (S^y_{\tot})^2 + (S^z_{\tot})^2$.
On the other hand, the maximal Abelian subalgebra $\mM$ that uniquely labels all the different sectors is simply a Cartan subalgebra of $SU(2)$. 
This is not unique, and it is generated by $\vec{S}_{\tot}^2$ and one of the $S^\alpha_{\tot}$ for $\alpha \in \{x, y, z\}$.
The corresponding bond algebra $\mA$ can also be understood as follows. 
Up to addition of constants, the Heisenberg Hamiltonian of Eq.~(\ref{eq:Heisenberg}) can be written as $\sum_j{(J_j/2) 
P_{j,j+1}}$, where $P_{j,j+1} \defn 2 \vec{S}_j\cdot \vec{S}_{j+1} + 1/2$ is the \textit{permutation operator} that permutes the spins of $j$ and $j+1$, and hence the bond algebra $\mA$ in this case has a simple form -- it is the group algebra of the symmetric group $S_L$ of $L$ elements with complex coefficients, typically denoted by $\mathbb{C}[S_L]$.   
The dimensions of the irreducible representations of $S_L$ allowed in the spin-1/2 Hilbert space $\mH$ are well-known~\cite{fulton2013representation, bartlett2007reference, readsaleur2007}, and are given by $D_\lambda = \binom{L}{L/2 + \lambda} - \binom{L}{L/2+\lambda+1}$ for even $L$.
Consequently, $\dim(\mA) = \frac{1}{L+1}\binom{2L}{L}$, and $\sum_{\lambda = 0}^{L/2}{D_\lambda d_\lambda} = 2^L = \dim(\mH)$. 
Note that in a tensor product Hilbert space $\mH$, the decomposition of Eq.~(\ref{eq:Hilbertdecomp}) can also be directly understood as a consequence of the fusion rules for $SU(2)$, which leads to the same expressions for $\{D_\lambda\}$ and $\{d_\lambda\}$. 
The non-Abelian commutant here results in degeneracies in the spectrum of the Hamiltonians $H$, since $d_\lambda > 1$ results in multiple identical blocks in the Hamiltonian, as discussed in Sec.~\ref{subsec:defn}.
In particular, there are $d_\lambda = (2\lambda + 1)$ identical blocks (i.e., Krylov subspaces) of dimension $D_\lambda$, which corresponds to the degeneracies of the sectors with quantum numbers spin projection $S^z_{\tot} = -\lambda, -\lambda + 1, \cdots, \lambda - 1, \lambda$ and a total spin $\lambda$ (i.e., with $\vec{S}_{\tot}^2 = \lambda (\lambda + 1)$).
\subsubsection{Solvable models}\label{subsubsec:solvable}
We now turn to completely solvable models with Hamiltonians consisting of commuting terms (i.e., $[\hh_i, \hh_j] = 0$ in Eq.~(\ref{eq:genhamil})).
By construction, the bond algebra corresponding to a family of these models is Abelian ($\mA = \mZ$), and its only irreducible representations are one-dimensional, i.e., $D_\lambda = 1$ for all $\lambda$.
As a consequence of Eqs.~(\ref{eq:dimensions}) and (\ref{eq:Krylovnumber}), the number of Krylov subspaces $K = \dim(\mH)$, which means that all eigenstates are one-dimensional Krylov subspaces, and hence the model is solvable. 
Classic examples of systems that fall into this category are stabilizer code models such as the toric code~\cite{kitaev2003fault, kitaev2010topological} or certain fracton models~\cite{fractonreview, pretko2020fracton}. 
In these cases, both the bond and commutant algebras are group algebras of certain subgroups of the Pauli group (i.e., the group of all Pauli strings under multiplication).
Hence it is typically sufficient to study the group structure of the Pauli strings that span these algebras.
For example, the bond algebra $\mA$ in stabilizer codes by construction is the group algebra of an Abelian stabilizer group $\mS$, a subgroup of the Pauli group.
The commutant algebra $\mC$ is the group algebra of the \textit{group centralizer} of the stabilizer group within the Pauli group, typically denoted by $\mC(\mS)$, and it is a non-Abelian group that consists of all logical operators in the system, including the trivial ones that are part of the stabilizer group $\mS$.
Non-trivial logical operators that are not part of the stabilizer group, e.g., Wilson loops, are part of the quotient group $\mC(\mS)/\mS$.
The (topological) degeneracies in the ground state (and excited states) of stabilizer codes can then be understood either in terms of non-trivial logical operators, or directly as a consequence of the non-Abelian commutant algebra $\mC$.
\subsection{Hilbert Space Fragmentation}\label{subsec:fragmentation}
We now describe Hilbert space fragmentation in the language of bond and commutant algebras. 
As discussed in Sec.~\ref{subsec:defn}, the dynamically disconnected Krylov subspaces of a family of systems can be understood in terms of Eq.~(\ref{eq:Hilbertdecomp}).
We note that the definitions of Eqs.~(\ref{eq:dimensions}) and (\ref{eq:Krylovnumber}), along with the fact that $d_\lambda \geq 1$ for all $\lambda$ impose bounds on the number of Krylov subspaces $K$ in terms of $\dim(\mC)$ and vice-versa:
\begin{equation}
    \sqrt{\dim(\mC)} \leq K \leq \dim(\mC),\;\;K \leq \dim(\mC) \leq K^2. 
\label{eq:Krylovbound}
\end{equation}
These bounds allow us to broadly classify one-dimensional systems into three categories based on the scaling of $\log(\dim(\mC))$ (hence $\dim(\mC)$) with system size $L$.
First, systems where $\log(\dim(\mC))$ is independent of $L$, which occurs in systems with discrete symmetry such as $Z_2$.
Second, systems with $\log(\dim(\mC))$ that scales as $\log L$, which typically occurs in systems with continuous symmetries such as $U(1)$ or $SU(2)$, discussed in Sec.~\ref{subsec:conventionalexamples}. 
These cases are well-known and are typically considered examples of conventional symmetries.
A third possibility is that $\log(\dim(\mC))$ scales linearly with $L$. As a consequence of Eq.~(\ref{eq:Krylovbound}), this is a necessary and sufficient condition for the number of Krylov subspaces $K$ to scale exponentially with $L$, and can be taken to be a definition of Hilbert space fragmentation. 
However, interesting examples of fragmentation are only the systems where $\dim(\mC)$ scales exponentially with system size while the bond algebra $\mA$ is non-Abelian. 
As discussed in Sec.~\ref{subsubsec:solvable}, if the bond algebra $\mA$ is Abelian, $K$ (and hence $\dim(\mC)$) always scales exponentially with system size and the system is completely solvable (hence fragmented in a trivial sense).
In higher dimensional systems, we can similarly define fragmentation as the case when $\log(\dim(\mC))$ scales as a volume-law, i.e., linearly in volume of the system. 
The various scalings of $\dim(\mC)$ in one and two dimensions are summarized in Table~\ref{tab:scalings}, and we mostly focus on fragmented systems in the rest of this work.
Note that higher dimensional systems offer more possibilities for the scaling of $\dim(\mC)$ due to the possibility of subsystem symmetries, but a detailed discussion of all cases is beyond the scope of this work.
\begin{table}[t!]
    \begin{tabular}{|c|c|}
        \hline
        {\bf \textrm{log}}$\boldsymbol{(\dim(\mC))}$  & {\bf Example}\\
        \hline
        $\sim\mathcal{O}(1)$ & Discrete Global Symmetry\\
        $\sim\log L$ & Continuous Global Symmetry\\
        $\sim L$ & Fragmentation\\
        \hline
    \end{tabular}\\
    \vspace{4mm}
    \begin{tabular}{|c|c|}
        \hline
        {\bf \textrm{log}}$\boldsymbol{(\dim(\mC))}$  & {\bf Example}\\
        \hline
        $\sim \mathcal{O}(1)$ & Discrete Global Symmetry\\
        $\sim \log L$ & Continuous Global Symmetry\\
        $\sim L$ & Discrete Subsystem Symmetry\\
        $\sim L\log L$ & Continuous Subsystem Symmetry\\
        $\sim L^2$ & Fragmentation\\
        \hline
    \end{tabular}
    \caption{Classification of systems based on scaling of the dimension of the commutant algebra $\dim(\mC)$ with system size for one dimensional systems of size $L$ (top) and two dimensional systems of size $L \times L$ (bottom).
    }
    \label{tab:scalings}
\end{table}
Several features of fragmentation can also be defined in the language of commutant algebra.
For example, the distinction between strong and weak fragmentation then depends on the dimension of the largest Krylov subspace, which in terms of the decomposition of Eq.~(\ref{eq:Hilbertdecomp}) reads $D_{\max} = \max_\lambda{D_\lambda}$. 
Since strong and weak fragmentation in the literature~\cite{sala2020fragmentation, morningstar2020kineticallyconstrained} has been defined \textit{within} conventional symmetry sectors, in Eq.~(\ref{eq:Hilbertdecomp}) one needs to consider the Hilbert space $\mH$ truncated to states within a particular symmetry sector.\footnote{Indeed, the concept of strong fragmentation becomes trivial when defined w.r.t.\ the full Hilbert space, since the presence of a conventional continuous symmetry such as $U(1)$ or $SU(2)$ is sufficient to ensure $D_{\max}/D \rightarrow 0$ in the thermodynamic limit. Alternately, one could define strong fragmentation as $D_{\max}/D \sim \exp(-c L)$~\cite{chen2021emergent} (or $\exp(-c V)$ in higher dimensions)}.
However, throughout this work we will focus on the full Hilbert space without resolving any conventional symmetries separately. 
Furthermore, frozen eigenstates in fragmented systems are just the one-dimensional representations (singlets) of the algebra $\mA$, and their number is given by $\sum_{\lambda}{d_\lambda \delta_{D_\lambda, 1}}$.
In addition, a further distinction can be made between fragmentation in the product state basis and fragmentation in an entangled basis, depending on whether the Hamiltonian of Eq.~(\ref{eq:genhamil}) is
block diagonal in the product state basis or in an entangled basis.
We refer to the former as ``classical fragmentation" since the same fragmented structure is possible in classical Markov circuits that implements the same transitions as the terms of the bond algebra $\{\hh_j\}$, and to the latter as ``quantum fragmentation." 
Classical fragmentation occurs when all the operators in the commutant $\mC$ are diagonal in the product state basis (e.g., $S^z_\tot$ discussed in Sec.~\ref{subsec:conventionalexamples}).
Hence, $\mC$ is Abelian and admits only one-dimensional irreducible representations (i.e., $d_\lambda = 1$ for all $\lambda$).
The decomposition of Eq.~(\ref{eq:Hilbertdecomp}) is then the same as the Krylov subspace decomposition of Eq.~(\ref{eq:Hilbertsplitting}), and $K = \dim(\mC)$.
While the classical and quantum distinction can also be made for conventional symmetries, we emphasize this for Hilbert space fragmentation since most examples in the literature involve only classical fragmentation. 
In the following sections, we provide various examples of systems with fragmentation, out of which the $t-J_z$ model (Sec.~\ref{sec:tJz}), Pair-Flip model (Sec.~\ref{sec:pairflip}), and spin-1 dipole-conserving models (Sec.~\ref{sec:dipole}) are examples of classical fragmentation, whereas the Temperley-Lieb models (Sec.~\ref{sec:TL}) show quantum fragmentation.
\section{\texorpdfstring{$t-J_z$}{} Model}
\label{sec:tJz}
\subsection{Definition and Symmetries}\label{subsec:tJzmodel}
We illustrate the usefulness of the commutant algebra by explicit construction of conserved quantities for the $t-J_z$ model~\cite{zhang1997tJz, batista2000tJz}.
We consider a general version of the model defined on an arbitrary lattice or graph, given by the family of Hamiltonians 
\begin{gather}
    H_{t-J_z} \defn \sumal{\langle i, j\rangle}{}{[-t_{i, j} \overbrace{\sumal{\sigma \in \{\uparrow, \downarrow\}}{}{\left(\tilde{c}_{i,\sigma} \tilde{c}^\dagger_{j,\sigma} + h.c.\right)}}^{\hT_{i,j}} + J^z_{i,j} \overbrace{S^z_i S^z_{j}}^{\hV_{i,j}}]} \nn \\
    + \sumal{i}{}{\left[h_i S^z_i + g_i (S^z_i)^2\right],}
\label{eq:tJzhamil}
\end{gather}
where $\langle i, j\rangle$ denotes nearest-neighbors, $t_{i,j}$, $J^z_{i,j}$, $h_i$, $g_i$ are arbitrary constants, and we have defined
\begin{gather}
    S^z_j \defn \tilde{c}^\dagger_{j, \uparrow}\tilde{c}_{j, \uparrow} - \tilde{c}^\dagger_{j, \downarrow}\tilde{c}_{j, \downarrow},\nn \\
    \tilde{c}_{j,\sigma} \defn c_{j,\sigma} \left(1 - \cd_{j,-\sigma} c_{j,-\sigma}\right), 
\label{eq:tJzdefns}
\end{gather}
where $-\sigma$ denotes the opposite spin of $\sigma$, and $\cd_{j, \sigma}$ and $c_{j, \sigma}$ are fermionic creation and annihilation operators.
The model is effectively working in the Hilbert space with no double-occupancy at any site and with all fermion moves required to satisfy these constraints, and also exactly maps onto a spin-1 hard core bosonic model via a generalized Jordan-Wigner transformation~\cite{batista2001generalizedJW}. 
Note that we have added the last two terms in Eq.~(\ref{eq:tJzhamil}) in order to break any discrete symmetries of the $t-J_z$ Hamiltonian that we are not interested in.
As we show in App.~\ref{app:diagonalcommutants}, this also ensures that all the operators in the commutant are diagonal in the product state basis.
This $t-J_z$ model as defined in Eq.~(\ref{eq:tJzhamil}) has two obvious $U(1)$ symmetries, which are the separate particle number conservation of $\uparrow$ spins and $\downarrow$ spins:
\begin{equation}
    N^\uparrow \defn \sumal{j}{}{N^\uparrow_j},\;\;\;N^\downarrow \defn \sumal{j}{}{N^\downarrow_j},
\label{eq:tJzsymmetries}
\end{equation}
where we have defined number operators $N^\sigma_j$
\begin{equation}
    N^{\sigma}_j \defn \tilde{c}^\dagger_{j, \sigma} \tilde{c}_{j, \sigma},\;\;\;\sigma \in \{\uparrow, \downarrow\}.
\label{eq:numberdefn}
\end{equation}
Eq.~(\ref{eq:tJzsymmetries}) directly follows from the following commutation relation of the local terms in Eq.~(\ref{eq:tJzhamil}):
\begin{equation}
    \begin{array}{l}
    [N^\sigma_i + N^\sigma_j, \hT_{i,j}] = 0\\
    \vspace{1mm}
    [N^\sigma_i + N^\sigma_j, \hV_{i,j}] = 0\\
    \vspace{1mm}
    [N^\sigma_i, S^z_i] = [N^\sigma_i, (S^z_i)^2] = 0
    \end{array}\;\;\;\text{for}\;\; \sigma \in \{\uparrow, \downarrow\}.
\label{eq:localcomm}
\end{equation}
\subsection{Fragmentation in One Dimension}\label{subsec:tJzfragmentation}
As discussed in Ref.~\cite{sliom2020}, the $t-J_z$ Hamiltonian exhibits Hilbert space fragmentation in one dimension, both with open boundary conditions (OBC) and periodic boundary conditions (PBC).
The transitions implemented by the term $\hT_{i,j}$ can be depicted as 
\begin{equation}
    \ket{\uparrow 0} \leftrightarrow \ket{0 \uparrow},\;\;\;\ket{\downarrow 0} \leftrightarrow \ket{0 \downarrow},
\label{eq:tJztransitions}
\end{equation}
where the two sites are $i$ and $j$, and $\uparrow, \downarrow$, and $0$ denotes the two spins of the fermions and an empty site respectively.
Each Krylov subspace is hence characterized by a pattern of spins $\uparrow$ and $\downarrow$, say from left to right with OBC and anticlockwise along the chain with PBC, which is clearly preserved under the action of the Hamiltonian $H_{t-J_z}$ of Eq.~(\ref{eq:tJzhamil}).
For example, in a system with five sites and open or periodic boundary conditions, the states $\ket{\uparrow \downarrow 0 \downarrow \uparrow}$ and $\ket{\downarrow \uparrow \downarrow 0 \uparrow}$ are dynamically disconnected from one another even though these have the same quantum number under the two $U(1)$ symmetries of the model given in Eq.~(\ref{eq:tJzsymmetries}).
For OBC, the conservation of the pattern of spins results in the formation of exponentially many disconnected subspaces labeled by all possible patterns of $\uparrow$ and $\downarrow$ spins, a total of $\sum_{N=0}^L 2^N = 2^{L+1} - 1$
subspaces for a system of size $L$.
For PBC, it is easy to see that all states with at least one empty site that consist of the same pattern of spins that are equivalent up to a translation along the chain belong to the same Krylov subspace.
Hence the Krylov subspaces in the PBC $t-J_z$ model are labeled by the distinct pattern of spins anticlockwise along the chain that cannot be mapped onto each other by translation.
In addition, both the OBC and PBC $t-J_z$ models have exponentially many one-dimensional Krylov subspaces, i.e., frozen product states, given by configurations with particles on all sites on which $\hT_{i,j}$ vanishes, and are also eigenstates of all the $S^z_{j}$'s (these states are already included in the above OBC Krylov subspace count as $N=L$.)
However, the full set of these frozen states can be completely understood via the $U(1)$ symmetries, since they exhaust the Hilbert space of the quantum number sectors with $N^\uparrow + N^\downarrow = L$.
Of course, the number of Krylov subspaces within a given sector of fixed $N^\uparrow$ and $N^\downarrow$ is lesser, but nevertheless grows exponentially with $L$ for sectors where $N^\uparrow/L$ and $N^\downarrow/L$ are kept finite.
Furthermore, as discussed in Ref.~\cite{sliom2020}, the fragmentation in the OBC $t-J_z$ model within, e.g., the symmetry sector with $N^\uparrow = N^\downarrow = L/4$ (assuming $L$ multiple of $4$ for simplicity) is strong; all
Krylov subspaces in this sector have a dimension of $D_{\max} = \binom{L}{L/2}$, whereas the full Hilbert space for this
symmetry sector has dimension $D = \binom{L}{L/2} \binom{L/2}{L/4}$, hence $D_{\max}/D \rightarrow 0$ as $L \rightarrow \infty$.
\subsection{Commutant Algebra}\label{subsec:tJzcommutant}
The full pattern of the spins is not detected by any local operator, and this shows that there are conserved quantities of $H_{t-J_z}$ other than the charges of the $U(1)$ symmetries $N^\uparrow$ and $N^\downarrow$. 
We can directly understand the extra conserved quantities by observing that
\begin{equation}
    \begin{array}{ll}
    [N^\sigma_i N^\tau_j, \hT_{i,j}] = 0\\
    \vspace{1mm}
    [N^\sigma_i N^\tau_j, \hV_{i,j}] = 0
    \end{array}\;\text{for}\;\;\sigma,\tau \in \{\uparrow, \downarrow\}. 
\label{eq:nncomm}
\end{equation}
For OBC, using Eqs.~(\ref{eq:localcomm}) and (\ref{eq:nncomm}), we can construct a ``quadratic" IoM
\begin{equation}
    N^{\sigma_1 \sigma_2} \defn \sumal{j_1 < j_2}{}{N^{\sigma_1}_{j_1} N^{\sigma_2}_{j_2}}, \;\;\sigma_1, \sigma_2 \in \{\uparrow, \downarrow\},
\label{eq:tJzdouble}
\end{equation}
where $\sum_{j_1 < j_2}(\bullet)$ is shorthand for $\sum_{j_1 = 1}^L{\sum_{j_2 = j_1 + 1}^L(\bullet)}$. 
In Eq.~(\ref{eq:tJzdouble}), $N^{\sigma \sigma}$ for $\sigma \in \{\uparrow, \downarrow\}$ can be expressed in terms of the usual conserved quantities $N^\sigma$ of Eq.~(\ref{eq:tJzsymmetries}) as $N^{\sigma\sigma} = ((N^\sigma)^2 - N^\sigma)/2$, and is not a functionally independent IoM.  
However, it is easy to see that $N^{\sigma_1 \sigma_2}$ for $\sigma_1 \neq \sigma_2$ \textit{cannot} be expressed in terms of products and powers of the local conserved quantities $N^\uparrow$ and $N^{\downarrow}$, and hence are functionally independent IoMs.
Similarly, we can construct families of IoMs for the $H_{t-J_z}$ for OBC as follows 
\begin{equation}
    N^{\sigma_1 \sigma_2 \cdots \sigma_k} = \sumal{j_1 < j_2< \cdots< j_k}{}{N^{\sigma_1}_{j_1} N^{\sigma_2}_{j_2}\cdots N^{\sigma_k}_{j_k}},\;\;\
    \sigma_j \in \{\uparrow,\downarrow\}
\label{eq:tJzconserved}
\end{equation}
where we have used a shorthand notation for the sum, similar to Eq.~(\ref{eq:tJzdouble}), and $0 \leq k \leq L$.  
For $k = 0$, the IoM is defined to be $\mathds{1}$ operator, the $k = 1$ case refers to the usual symmetries of Eq.~(\ref{eq:tJzsymmetries}), and $k = 2$ reduces to the operator of Eq.~(\ref{eq:tJzdouble}). 
For $k = L$, note that the IoM $N^{\sigma_1 \cdots \sigma_L}$ is simply the projector onto a frozen eigenstate with spins on all sites, $\ket{\sigma_1 \cdots \sigma_L}$. 
Several of the IoMs of Eq.~(\ref{eq:tJzconserved}) with two or more indices are functionally independent from the conserved quantities $N^\uparrow$ and $N^\downarrow$, i.e., cannot be expressed as polynomial functions of $N_\uparrow$ and $N_\downarrow$.
Furthermore, as we show in App.~\ref{app:tJzcommutantorth}, the IoMs of Eq.~(\ref{eq:tJzconserved}) are all linearly (although not functionally) independent, and they form a complete basis (although not orthonormal) for the commutant algebra $\Cobc$ for the family of $t-J_z$ models of Eq.~(\ref{eq:tJzhamil}) with OBC. 
Since these IoMs are all diagonal in the product state basis, the $t-J_z$ model is an example of classical fragmentation discussed in Sec.~\ref{subsec:fragmentation}, and the commutant $\Cobc$ is Abelian. 
Furthermore, the dimension of $\Cobc$ is the number of linearly independent operators in Eq.~(\ref{eq:tJzconserved}), which is $\dim(\Cobc) = \sum_{k=0}^L{2^k} = 2^{L+1} - 1$.
This is precisely the number of Krylov subspaces in the $t-J_z$ models, as discussed in the previous subsection and in agreement with the general discussion of Abelian commutants in Sec.~\ref{subsubsec:abelian}.
The commutant algebra $\Cpbc$ for the PBC $t-J_z$ model can be constructed similarly. 
For example, the generalization of the IoM of Eq.~(\ref{eq:tJzdouble}) for PBC reads
\begin{equation}
    N^{[\sigma_1 \sigma_2]} \defn \sumal{m = 1}{2}{\sumal{j_1 < j_2}{}{N^{\sigma_m}_{j_1} N^{\sigma_{m + 1}}_{j_2}}}, 
\label{eq:tJzdoublePBC}
\end{equation}
where we define $\sigma_{m + 2} \equiv \sigma_{m}$ for $1 \leq m \leq 2$, we have used the same shorthand notation for the sum as Eq.~(\ref{eq:tJzdouble}), and have used brackets in the indices to distinguish from the OBC IoMs of Eq.~(\ref{eq:tJzdouble}).  
Similarly, the IoMs of Eq.~(\ref{eq:tJzconserved}) can be generalized to PBC as follows
\begin{equation}
    N^{[\sigma_1 \cdots \sigma_k]} \defn \sumal{m = 1}{k}{\sumal{j_1 < \cdots < j_k}{}{N^{\sigma_{m}}_{j_1} N^{\sigma_{m + 1}}_{j_2}\cdots N^{\sigma_{m+ k -1}}_{j_k}}},\;\;\sigma_\alpha \in \{\uparrow,\downarrow\},
\label{eq:tJzconservedPBC}
\end{equation}
where $1 \leq k \leq L - 1$, we have defined $\sigma_{m+k} \equiv \sigma_m$ for $1 \leq m \leq k$, and we have used shorthand notation similar to Eq.~(\ref{eq:tJzconserved}).
Similar to the OBC case, we can define the $k = 0$ case in Eq.~(\ref{eq:tJzconservedPBC}) to be the $\mathds{1}$ operator, and the $k = 1$ case corresponds to the usual symmetries of Eq.~(\ref{eq:tJzsymmetries}). 
As a consequence of the sum over $m$ in Eq.~(\ref{eq:tJzconservedPBC}), cyclic permutations of the indices denote the same IoMs, i.e., $N^{[\sigma_1 \cdots \sigma_k]} = N^{[\sigma_2 \cdots \sigma_k \sigma_1]} = \cdots = N^{[\sigma_k \sigma_1 \cdots \sigma_{k-1}]}$. 
For $k = L$, additional independent IoMs can be written down, which read 
\begin{equation}
    N^{[\sigma_1 \cdots \sigma_L]} \defn \prodal{j = 1}{L}{N^{\sigma_j}_{j}}.
\label{eq:PBCconservedfull}
\end{equation}
Note that unlike their $k \leq L-1$ counterparts defined in Eq.~(\ref{eq:tJzconservedPBC}), we have chosen the IoMs $N^{[\sigma_1 \cdots \sigma_k]}$ for $k = L$ to not be invariant under cyclic permutations of their indices.

Hence the dimension of $\Cpbc$ is lesser than that of $\Cobc$ by a factor that is polynomial in $L$, but nevertheless it is clear that it grows exponentially with system size $L$, and still constitutes an example of Hilbert space fragmentation.
\subsection{Connections to SLIOMs}\label{subsec:SLIOMconnection}
Reference~\cite{sliom2020} introduced a set of $L$ conserved quantities for the one-dimensional $t-J_z$ model with OBC, dubbed as ``Statistically Localized" Integrals of Motion (SLIOMs), which were shown to uniquely label all the Krylov subspaces of the $t-J_z$ model. 
In this section, we discuss their connection with the IoMs constructed in Sec.~\ref{subsec:tJzcommutant}. 
The (left) SLIOM $\hq^{(\ell)}_l$ for the OBC $t-J_z$ model is simply the spin operator of the $l$-th particle ($\uparrow$ or $\downarrow$) from the left of the chain, and its expression reads
\begin{equation}
    \hq^{(\ell)}_l = \sumal{i = 1}{L}{\mP^{(\ell)}_{l, i} (N^\uparrow_i - N^\downarrow_i)},
\label{eq:tJzSLIOMs}
\end{equation}
where $\mP^{(\ell)}_{l,i}$ is the projector onto configurations where the $l$-th particle from the left is on site $i$.
Although not explicitly discussed in Ref.~\cite{sliom2020}, it is straightforward to show that the SLIOMs $\{\hq^{(\ell)}_l\}$ all commute with \textit{each term} of $H_{t-J_z}$ of Eq.~(\ref{eq:tJzhamil}), i.e.
\begin{equation}
    [\hq^{(\ell)}_l, \hT_{i,j}] = 0,\;\;\;[\hq^{(\ell)}_l, \hV_{i,j}] = 0,
\label{eq:SLIOMcommutant}
\end{equation}
and hence they are all part of the commutant algebra $\Cobc$ spanned by the IoMs of Eq.~(\ref{eq:tJzconserved}). 
This also means that the algebra generated by linear combinations and products the $L$ SLIOMs is completely within $\Cobc$, and as we show explicitly in App.~\ref{app:SLIOMalgebra}, this algebra is precisely $\Cobc$ if the identity operator $\mathds{1}$ is added to the set of generators.
Hence the $L$ SLIOMs along with $\mathds{1}$ are the generators of $\Cobc$, and this suggests that $\gen(\Cobc) \leq L+1$. 
Note that the left SLIOMs of Eq.~(\ref{eq:tJzSLIOMs}) are not the unique set of generators of $\Cobc$. 
As we show in App.~\ref{app:SLIOMalgebra}, a simple different choice that generates the algebra $\Cobc$ are the ``right SLIOMs" $\{\hq^{(r)}_l\}$, which are the spin operators of the $l$-th particle from the right end of the chain, along with $\mathds{1}$. 
Their operator expression is given by
\begin{equation}
    \hq^{(r)}_l = \sumal{i = 1}{L}{\mP^{(r)}_{l, i} (N_i^\uparrow - N_i^\downarrow)},
\label{eq:tJzrightSLIOMs}
\end{equation}
where $\mP^{(r)}_{l,i}$ is the projector onto the $l$-th particle ($\uparrow$ or $\downarrow$) from the right being on site $i$. 
We note that the left and right SLIOMs $\{\hq^{(\ell)}_l\}$ and $\{\hq^{(r)}_l\}$ are only defined for the OBC $t-J_z$ model~\cite{sliom2020}.
For the PBC $t-J_z$ model, it is not clear if there is a smaller set of operators that generate the full commutant $\Cpbc$, and we do not know how to compute $\gen(\Cpbc)$. 
However, as we discuss in Sec.~\ref{sec:Mazur}, for the practical purposes of computing Mazur bounds, in both OBC and PBC, it is convenient to use the expressions for the full commutant $\Cobc$ and $\Cpbc$ respectively, which circumvents the need to determine a smaller set of generators for the commutants. 
\subsection{Higher Dimensions}\label{subsec:tJzhigherdims}
Finally, we briefly discuss the nature of fragmentation in $t-J_z$ models in higher dimensions. 
The local commutation relations in Eqs.~(\ref{eq:localcomm}) and (\ref{eq:nncomm}) hold in any number of dimensions, and the model possesses the two $U(1)$ symmetries $N^\uparrow$ and $N^\downarrow$ of Eq.~(\ref{eq:tJzsymmetries}). 
For simplicity we restrict our discussion to $L \times L$ square lattices with OBC on both sides, although most of the discussion holds more generally. 
The only IoMs that we can construct in that case have the form
\begin{equation}
    N^{\sigma_1 \cdots \sigma_k} \defn \sumal{j_1 \neq j_2 \cdots \neq j_k}{}{N^{\sigma_1}_{j_1} \cdots N^{\sigma_k}_{j_k}},\;\;1 \leq k \leq L^2
\label{eq:higherdimsconserved}
\end{equation}
where the subscripts run over all the sites in the lattice.
However, since there is no restriction in the sum in Eq.~(\ref{eq:higherdimsconserved}), all of the $N^{\sigma_1 \cdots \sigma_k}$ for $k \leq L^2$ can be expressed as a polynomial of $N^\uparrow$ and $N^\downarrow$.
Functionally independent IoMs can appear when $k = L^2$ and Eq.~(\ref{eq:higherdimsconserved}) is replaced by a single product (i.e., no sum) over all sites as
\begin{equation}
    N^{\{\sigma_j\}}_{2D} \defn \prodal{j=1}{L}{N^{\sigma_j}_j},\;\;\sigma_\alpha \in \{\uparrow, \downarrow\},
\label{eq:tJzfrozen}
\end{equation}
where $\{\sigma_j\}$ denotes a configuration of $L^2$ spins on the lattice, and the product runs over all sites of the lattice. 
These IoMs are simply the projectors onto ``frozen" eigenstates $\{\ket{\{\sigma_j\}}\}$, where $\{\sigma_j\}$ is a spin pattern on the lattice such that $N^\uparrow + N^\downarrow = L^2$.
This absence of other independent IoMs is related to the fact that all the ``patterns" of spins on a square lattice with the same number of $N^\uparrow$ and $N^\downarrow$ such that $N^\uparrow + N^\downarrow < L^2$ can be connected to one another by hoppings allowed by $H_{t-J_z}$.
Hence, all the Krylov subspaces other than those determined by the two U(1)'s with $N^\up + N^\downarrow < L^2$ are one-dimensional, and the dimension of the commutant $\mC_{2D}$ for the square lattice $t-J_z$ model is then simply lower bounded by the number of frozen spin configurations, i.e., $\dim(\mC_{2D}) > 2^{L^2}$.
However, we should remark that unlike the $t-J_z$ model in one dimension, the higher dimensional $t-J_z$ model does not exhibit Hilbert space fragmentation within most of the $U(1)$ quantum number sectors defined by the $N^\uparrow$ and $N^\downarrow$.
Similar to the frozen states in the 1d $t-J_z$ model, the frozen states here exhaust the Hilbert spaces of the $U(1)$ quantum number sectors that satisfy $N^\uparrow + N^\downarrow = L^2$.
These form a minority of the $U(1)$ quantum number sectors, and there are no quantum number sectors where generic ``thermal" eigenstates coexist with the frozen eigenstates.
Note that such examples where entire quantum number sectors (of conventional symmetries) are solvable exist in the literature~\cite{vafek2017entanglement}, and are typically not considered to constitute violations of ergodicity.
Nevertheless, from the point of view of the full Hilbert space, since the dimension of the commutant algebra grows exponentially with volume, the 2d $t-J_z$ model formally constitutes an example of fragmentation which is morally no different from many other examples of fragmentation with exponentially many frozen product states.
\section{Pair-Flip Model}
\label{sec:pairflip}
\subsection{Definition and Symmetries}
We now study a less obvious example of classical fragmentation, using a model we call the Pair-Flip (PF) model. 
This is given by the family of Hamiltonians defined on a spin-$(m-1)/2$ (i.e., $m$-level) system as follows
\begin{gather}
    H^{(m)}_{PF} \defn \sumal{\langle i, j\rangle}{}{\sumal{\alpha,\beta = 1}{m}{\overbrace{\left[g^{\alpha,\beta}_{i,j}\left(\ket{\alpha \alpha}\bra{\beta \beta}\right)_{i,j} + h.c.\right]}^{\hF^{\alpha,\beta}_{i,j}}}}\nn \\
    + \sumal{i}{}{\sumal{\alpha = 1}{m}{\lambda^{\alpha}_i\underbrace{\left(\ket{\alpha}\bra{\alpha}\right)_{i}}_{N^\alpha_i}}},
\label{eq:PFHamil}
\end{gather}
where $\langle i, j \rangle$ denotes nearest-neighboring sites, $g^{\alpha,\beta}_{i,j}$ and $\lambda^{\alpha}_{i}$ are arbitrary constants.
Note that we have added the term on the second line to break any discrete symmetries of the model that we are not interested in. As we show in App.~\ref{app:diagonalcommutants}, it also ensures that all the operators in the commutant are diagonal in the product state basis.  
A particular model within the family of Hamiltonians of Eq.~(\ref{eq:PFHamil}) was studied in detail in Ref.~\cite{caha2018pairflip}.
For simplicity, we only focus on bipartite lattices in the following.
For example, for OBC in one-dimension, we have a natural bipartition into even and odd sublattices. 
Similar to the $t-J_z$ model of Sec.~\ref{sec:tJz}, the PF model on any bipartite lattice possesses $U(1)$ conserved quantities given by
\begin{equation}
    N^\alpha \defn \sum_j{(-1)^j N^\alpha_j},\;\;1 \leq \alpha \leq m,
\label{eq:PFsymmetries}
\end{equation}
where the number operator $N^\alpha_j$ is defined in Eq.~(\ref{eq:PFHamil}), and the ``even" and ``odd" sites are on different sublattices.
These follow directly from the local commutation relations
\begin{equation}
    [N^\alpha_i - N^\alpha_j, \hF^{\beta,\gamma}_{i,j}] = 0,\;\;[N^\alpha_i, N^\beta_i] = 0,\;\;\;1 \leq \alpha, \beta, \gamma \leq m. 
\label{eq:PFlocalcommutation}
\end{equation}
Note that not all the conserved quantities of Eq.~(\ref{eq:PFsymmetries}) are independent since $\sum_{\alpha = 1}^m {N_j^\alpha} = \mathds{1}$, hence the PF model has a $U(1)^{m-1}$ conventional symmetry. 
\subsection{Fragmentation in One Dimension}\label{subsec:PFfragmentation1d}
To show that the PF model of Eq.~(\ref{eq:PFHamil}) in one-dimension exhibits Hilbert space fragmentation, we introduce a convenient notation for representing states in the Hilbert space. 
For simplicity, we restrict to OBC in the following. 
We represent the $m$ degrees of freedom per site as $m$ colors, e.g., when $m = 3$, we assign $\ket{+} = \tket{\cfdot{0}{red}}$, $\ket{0} = \tket{\cfdot{0}{olive}}$, $\ket{-} = \tket{\cfdot{0}{blue}}$.
We then use the following procedure to pair the sites using ``dimers."
First, we start from the left of the chain and pair any nearest neighboring sites that have the same color using a dimer of that color.
For example, if $m = 3$,  we allow three colors of dimers
\begin{equation}
    \ket{++} = \tket{\cdimer{0}{0.5}{red}}\;\;\;
    \ket{00} = \tket{\cdimer{0}{0.5}{olive}}\;\;\;
    \ket{--} = \tket{\cdimer{0}{0.5}{blue}}
\end{equation}
Second, we repeat the procedure by focusing only on the remaining unpaired sites, i.e., we ignore all the paired sites and connect any neighboring unpaired sites of the same color with a dimer of that color. 
Finally, we continue this procedure until there are no unpaired sites with neighboring unpaired sites of the same color. 
We refer to these remaining unpaired sites as ``dots." 
This procedure hence maps a product state to a state composed of non-crossing dimer configurations along with some unpaired sites. 
Any product state in the Hilbert space is thus composed of dots (denoted by $\btp\fdot{0}\etp$) and regions of non-crossing dimers (denoted by $\btp\rect{0}{1}\etp$) such that the colors on any adjacent dots (excluding dimer regions) is not the same. 
For example, in the following configuration,
\begin{equation}
    \tket{\fdot{0}\fdot{0.5}\rect{1}{2.5}\fdot{3}\rect{3.5}{4.5}\fdot{5}\fdot{5.5}}
\label{eq:dotdimerexample}
\end{equation}
the $3^{rd}$ dot from the left has a color different from the $2^{nd}$ and $4^{th}$ dots. 
Since the dimer regions denoted by $\btp\rect{0}{1}\etp$ always cover even number of consecutive sites, a system of size $L$ even (resp. odd) has Krylov subspaces with $j$ dots, $0 \leq j \leq L$ and $j$ even (resp. odd).   
Note that distinct configurations of dots and non-crossing dimers do not necessarily represent distinct product states, as evident in the following example
\begin{equation}
    \tket{\cdimer{0}{0.5}{black}\carcdimer{-0.5}{1}{black}} 
    = \tket{\cdimer{-0.5}{0}{black}\cdimer{0.5}{1}{black}},\;\;\;
    \tket{\cfdot{0}{black}\cdimer{0.5}{1}{black}} = \tket{\cdimer{0}{0.5}{black}\cfdot{1}{black}},
\label{eq:configequivalence}
\end{equation}
where black denotes any particular color. 
Here we took a more general perspective (convenient below) that for a given product state we can consider any configuration of dots and dimers satisfying the above properties that correctly represents the state, while the procedure described earlier sweeping from left to right provides one instance of such a pairing configuration.
However, as we now discuss, the Krylov subspaces are uniquely labeled by the pattern of colors of the dots, which can be inferred from any of the pairings. 
As evident from Eq.~(\ref{eq:PFHamil}), the terms $\hF^{\alpha,\beta}_{i,j}$ in  $H^{(m)}_{PF}$ allow for transitions between nearest-neighbor dimers of the same color, and annihilates any configuration of different colors on nearest-neighbors.
For example, when $m = 3$, we can depict the non-vanishing actions of the terms $\hF^{\alpha,\beta}_{i,j}$ as follows
\begin{equation}
    \tket{\cdimer{0}{0.5}{red}} \leftrightarrow \tket{\cdimer{0}{0.5}{olive}} \leftrightarrow \tket{\cdimer{0}{0.5}{blue}},
\label{eq:transitionrules}
\end{equation}
where the two sites represent $i$ and $j$.
Using Eqs.~(\ref{eq:transitionrules}) and (\ref{eq:configequivalence}), we note two important properties of transitions that will help us label the Krylov subspaces of $H^{(m)}_{PF}$:
\begin{enumerate}
    \item A dot of any color can ``hop" over a dimer of any other color via intermediate configurations. For example, 
    \begin{equation}
        \tket{\cfdot{0}{red}\cdimer{0.5}{1}{blue}} \leftrightarrow \tket{\cfdot{0}{red}\cdimer{0.5}{1}{red}} = \tket{\cdimer{0}{0.5}{red}\cfdot{1}{red}} \leftrightarrow \tket{\cdimer{0}{0.5}{blue}\cfdot{1}{red}}
    \label{eq:dothopping}
    \end{equation}
    \item Starting from a configuration with $n$ dimers beside each other, all configurations with $n$ non-crossing dimers of any color can be generated. For example with $n = 2$, we obtain
    \begin{gather}
        \tket{\carcdimer{0}{1.5}{blue}\cdimer{0.5}{1}{red}} \leftrightarrow \tket{\carcdimer{0}{1.5}{blue}\cdimer{0.5}{1}{blue}} \nn \\
        =\tket{\cdimer{0}{0.5}{blue}\cdimer{1}{1.5}{blue}} \leftrightarrow \tket{\cdimer{0}{0.5}{red}\cdimer{1}{1.5}{blue}} \leftrightarrow \tket{\cdimer{0}{0.5}{red}\cdimer{1}{1.5}{red}}\nn \\ =\tket{\carcdimer{0}{1.5}{red}\cdimer{0.5}{1}{red}} \leftrightarrow \tket{\carcdimer{0}{1.5}{red}\cdimer{0.5}{1}{blue}}
    \label{eq:dimerrearrange}
    \end{gather}
\end{enumerate}
Using Eqs.~(\ref{eq:transitionrules}), (\ref{eq:dothopping}), and (\ref{eq:dimerrearrange}), it is easy to see that the pattern of colors of the dots from the left to the right of the chain is unchanged by the action of the Hamiltonian $H^{(m)}_{PF}$, hence these label the different Krylov subspaces. 
The $m(m-1)^{L-1}$ product states that map on to configurations with $L$ dots (hence no dimers) are one-dimensional Krylov subspaces, i.e., they are frozen product states since the action of all the terms $\{\hF^{\alpha,\beta}_{i,j}\}$ vanish on such states. 
However, unlike the $t-J_z$ model, the frozen states here are scattered across various quantum number sectors of the $U(1)$ symmetries, and they do not exhaust the Hilbert space of most of the quantum number sectors they belong to.
The pattern of colors for the PF model is hence the analogue of the pattern of spins for the $t-J_z$ model discussed in Sec.~\ref{sec:tJz}, although the color of the dots in the PF model cannot be deduced by a local operator, unlike the spins in the $t-J_z$ models.  
The counting of the number and dimensions of the Krylov subspaces in the PF model for OBC is fairly complicated and can be extracted from Ref.~\cite{caha2018pairflip}, and we discuss the results in App.~\ref{app:PFcounting}.
Finally, we note that the procedure for mapping product states to a pattern of dots and non-crossing dimers also works for PBC, where we can start the pairing procedure from any site and going around the system until there are no neighboring dots of the same color.
This reveals the fragmentation of the PBC PF model, and the Krylov subspaces are then labeled by the full pattern of dots anticlockwise along the chain, similar to the pattern of spins in the PBC $t-J_z$ model.
In the Krylov subspaces with $L-1$ dots or lesser, all patterns of dots that map on to each other under translation are equivalent, since they can be connected using the rules of Eq.~(\ref{eq:dothopping}) and (\ref{eq:dimerrearrange}).
However, this is not the case for the completely frozen states, i.e., any state with $L$ dots such that neighboring dots do not have the same color still form exponentially many distinct one-dimensional Krylov subspaces.
\subsection{Commutant Algebra}\label{subsec:PFcommutant}
The fact that the pattern of dots is conserved under $H^{(m)}_{PF}$ in one dimension indicates the presence of additional conserved quantities of $H^{(m)}_{PF}$ functionally independent of the $U(1)$ conserved quantities $\{N^\alpha\}$ of Eq.~(\ref{eq:PFsymmetries}). 
In the following, we restrict our discussion to OBC for simplicity. 
Indeed, similar to Eq.~(\ref{eq:nncomm}) in the $t-J_z$, we can construct additional conserved quantities by observing that
\begin{equation}
    \left[N_i^{\sigma} N_j^{\tau}, \hF^{\alpha,\beta}_{i,j}\right] = 0,\;\;\textrm{for}\;\;\sigma \neq \tau,\;\;1\leq \alpha,\beta,\sigma,\tau \leq m. 
\label{eq:PFnncomm}
\end{equation}
Using Eqs.~(\ref{eq:PFsymmetries}) and (\ref{eq:PFnncomm}), we can construct quadratic IoMs for OBC similar to Eq.~(\ref{eq:tJzdouble})
\begin{equation}
    N^{\alpha_1 \alpha_2} \defn \sumal{j_1 < j_2}{}{(-1)^{j_1 + j_2} N_{j_1}^{\alpha_1} N_{j_2}^{\alpha_2}},\;\;\alpha_1 \neq \alpha_2,\;\;1 \leq \alpha_1, \alpha_2 \leq m,
\label{eq:PFNISdouble}
\end{equation}
where we have used a shorthand notation for the sum. 
Furthermore, similar to Eq.~(\ref{eq:tJzconserved}), we can construct families of IoMs of the form
\begin{equation}
    N^{\alpha_1 \alpha_2 \cdots \alpha_k} \defn \sumal{j_1 < j_2 < \cdots < j_k}{}{(-1)^{\sum_{l = 1}^k{j_l}}N_{j_1}^{\alpha_1} N_{j_2}^{\alpha_2} \cdots N_{j_k}^{\alpha_k}},\label{eq:stringops}
\end{equation}
where $0 \leq k \leq L$, and the constraint $\alpha_j \neq \alpha_{j+1}$ is a consequence of Eq.~(\ref{eq:PFnncomm}).
For $k = 0$, this IoM is defined to be the $\mathds{1}$ operator, and the $k = 1$ case refers to the $U(1)$ conserved quantities of Eq.~(\ref{eq:PFsymmetries}). 
However, not all of the IoMs of Eq.~(\ref{eq:stringops}) are linearly independent. 
As we show in App.~\ref{app:PFdependence}, for even (resp. odd) system size $L$, the IoMs of Eq.~(\ref{eq:stringops}) with $k$ odd (resp. even) can be expressed as a linear combination of the ones with $k$ even (resp. odd). 
The total number of linearly independent conserved quantities (i.e. the dimension of the commutant algebra $\textrm{dim}(\mC_{PF})$) is then given by
\begin{gather}
    \textrm{dim}(\mC_{PF}) = \twopartdef{1 + \sumal{p = 1}{L/2}{m (m-1)^{2p-1}}}{L\ \textrm{even}}{\sumal{p = 0}{(L-1)/2}{m (m-1)^{2p}}}{L\ \textrm{odd}}\nn\\
    = \twopartdef{\frac{(m-1)^{L+1} - 1}{m-2}}{m \geq 3}{L+1}{m=2}.
\label{eq:PFcommutantdim}
\end{gather}
This counting is also consistent with the fact that a system with even (resp. odd) size $L$ can only have $j$ even (resp. odd) number of dots, and that Krylov subspaces are uniquely determined by the pattern of dots.\footnote{A direct correspondence can be made as follows:  A Krylov subspace specified by the state of $l$ dots $\alpha_1,\alpha_2,\cdots,\alpha_l$ is annihilated by all IoMs with $k \geq \l$ except $N^{\alpha_1 \alpha_2 \cdots \alpha_l}$, which is easily checked by acting on product state within the Krylov subspace with all dots on the left $l$ sites. (The possibly non-zero values of the IoMs with $k < l$ is not important for making the one-to-one correspondence.)}
As evident from Eq.~(\ref{eq:PFcommutantdim}), $\textrm{dim}(\mC_{PF})$ grows exponentially with $L$ for $m \geq 3$.
Hence, the PF model exhibits Hilbert space fragmentation for $m \geq 3$, according to the definition proposed in Sec.~\ref{subsec:fragmentation}.
Note that such conserved quantities can also be constructed for PBC, and their construction is very similar to the ones for the PBC $t-J_z$ model discussed in Sec.~\ref{subsec:tJzcommutant}, hence we do not illustrate them here. 
Finally, we note that analogues of the SLIOMs for the $t-J_z$ model can also be defined for the PF model.
The $L$ left (resp. right) SLIOMs $\hq_l$ in this case are simply the operators that measure the color of the $l^{th}$ dot from the left (resp. right) side of the chain.  
We expect that these operators along with the identity operator $\mathds{1}$ generate the full commutant algebra of the PF model, and possibly form the minimal set of generators of the commutant algebra.  
However, as we will show in Sec.~\ref{sec:Mazur}, the full commutant algebra is required to accurately capture aspects of the dynamics of the system (particularly, the Mazur bounds of autocorrelation functions), hence we believe that explicit expressions for the SLIOMs are unnecessary.
Although the fragmentation in the PF model closely resembles that in the $t-J_z$ model with ``dots" playing the role of the ``spins," there are some differences in the dynamics of the spins and dots.
At the crudest level, the basic dynamics in the PF model is ``annihilation" of a pair of same-color ``particles" (states) on nearest-neighbor sites and creation of a new pair of same-color particles, while ``dots" are unpairable objects and can ``move'' only when right pairs form nearby; the space between dots is ``alive" with pair-flips.
On the other hand, in the $t-J_z$ model the spins move by themselves to nearby empty sites, and the space between spins is ``dead."
On a more quantitative level, in both these models, given a sector with a fixed number of spins or dots, we can study the distribution of the positions of the dots. 
The spins in the $t-J_z$ model are typically distributed randomly in the bulk of the chain as can be explicitly shown~\cite{sliom2020}.
In the PF model, since the number of dimer configurations in the region between two dots depends on the distance between the dots, we expect a different distribution for the position of the dots.  
This difference between dots and spins is also apparent in the nature of the Mazur bounds for the autocorrelation function of the on-site spin operator in the bulk of the system discussed in Sec.~\ref{sec:Mazur}, which are qualitatively different for the $t-J_z$ and PF model.
Hence we also expect the SLIOMs of the $t-J_z$ and PF models to differ in their localization properties~\cite{sliom2020}.
\subsection{Higher Dimensions}\label{subsec:PFhigherdims}
We now briefly discuss fragmentation in the PF model on higher dimensional lattices, e.g, the square lattice in two dimensions. 
In higher dimensions, product states can be mapped on to configurations of dots and dimers following a procedure similar to the one described in Sec.~\ref{subsec:PFfragmentation1d}.
One-dimensional Krylov subspaces (i.e., frozen product eigenstates) can then be directly constructed by having configurations of all dots such the colors of the dots on any two neighboring sites are different, and if such a configuration exists, conserved quantities associated with them are given by
\begin{equation}
    N^{\vec{\alpha}} = \prodal{j \in G}{}{N^{\alpha_j}_j},
\label{eq:conservedqty2d}
\end{equation}
where $G$ denotes the lattice (or more generally, a graph) and $\alpha_j$ denotes the color of the dot on site $j$ of the lattice. 
The number of such frozen states for the $m$-state PF model is the number of $m$-colorings of $G$, which is given by the so-called Chromatic polynomial of $G$, $p(G, m)$. 
For $m = 2$, it is clear that the number of colorings of a any graph $G$ is always $2$ if $G$ is bipartite, and $0$ if not. 
If $m \geq 3$ and $G$ is a grid graph (i.e., a square lattice with open boundary conditions), $p(G, m)$ is known to grow exponentially with the number of vertices in $m$~\cite{read1988chromatic, oeiscolorings}.
For example, if $m = 3$, the number of such colorings of the square lattice is equal to the partition function of square ice~\cite{liebentropyice1967, liebentropyiceexact1967, baxter3colorings1970}, and the number of frozen configurations on an $L \times L$ lattice asymptotically grows as $W^{L^2}$ where  $W = \left(\frac{4}{3}\right)^{\frac{3}{2}} \approx 1.54$.
These frozen states are one-dimensional Krylov subspaces, and their exponential growth with system size already shows that the PF model on a square lattice shows Hilbert space fragmentation. 
We do not find any Krylov subspaces of larger dimensions, it is likely that all configurations with at least one dimer can be connected using the rules of Eq.~(\ref{eq:transitionrules}).
We have numerically verified this for small system sizes on a square lattice. 
As discussed in Sec.~\ref{subsec:tJzhigherdims}, in the higher dimensional $t-J_z$ model, the frozen states exhaust the Hilbert space of particular conventional quantum number sectors of the $U(1)$ symmetries.
However, the scenario in the higher dimensional PF model is different.
Indeed, the frozen states in the PF model typically do not exhaust the Hilbert space of any of the quantum number sectors of the conventional symmetries, and they coexist with other generic eigenstates belonging to conventional quantum number sectors of the $U(1)$ symmetries.
Hence these frozen product states are anomalous low-entanglement eigenstates within generically thermalizing quantum number sectors, and should be considered as examples of quantum many-body scars.
\section{Temperley-Lieb Models}
\label{sec:TL}
\subsection{Definition and Symmetries}\label{subsec:TLmodel}
We now study an example of a system exhibiting quantum fragmentation, i.e., fragmentation not in the product state basis.
We refer to these systems as Temperley-Lieb (TL) models, given by the family of Hamiltonians defined on a spin-$(m-1)/2$ (i.e., $m$-level) chain with OBC as follows
\begin{equation}
    H^{(m)}_{TL} = \sumal{j=1}{L-1}{J_j \he_{j,j+1}} \defn \sumal{j = 1}{L-1}{J_j \left[\sumal{\alpha,\beta = 1}{m}{\left(\ket{\alpha \alpha}\bra{\beta \beta}\right)_{j,j+1}}\right]}
\label{eq:TLHamil}
\end{equation}
where $J_j$'s are arbitrary constants.
Note that defining (unnormalized) ``singlets" between sites $j$ and $k$ as
\begin{equation}
     \ket{\psi_{\textrm{sing}}}_{j,k} \defn \sum_{\alpha = 1}^{m}{\ket{\alpha \alpha}_{j,k}},
\label{eq:singletdefn}
\end{equation}
the terms $\he_{j,j+1}$ of the Hamiltonian $H^{(m)}_{TL}$ are simply the projectors onto the singlet state between sites $j$ and $j+1$, i.e., $\he_{j,j+1} = (\ket{\psi_{\textrm{sing}}}\bra{\psi_{\textrm{sing}}})_{j,j+1}$.
Hamiltonians in the family of TL models of Eq.~(\ref{eq:TLHamil}) have been previously studied in detail in the literature in various contexts~\cite{barber1989spectrum,batchelor1990spin, readsaleur2007, aufgebauer2010quantum}.
While models similar to $H^{(m)}_{TL}$ can also be defined for PBC and higher dimensions, we will only restrict ourselves to the well studied one-dimensional case with OBC. 
Note that the family of TL models of Eq.~(\ref{eq:TLHamil}) are a part of the family of PF models of Eq.~(\ref{eq:PFHamil}).
Hence $H^{(m)}_{TL}$ possesses all of the conserved quantities of $H^{(m)}_{PF}$, which include the $U(1)$ conserved quantities of Eq.~(\ref{eq:PFsymmetries}), which, in one dimension read
\begin{equation}
    N^\alpha \defn \sum_j{(-1)^j N^\alpha_j},\;\;1 \leq \alpha \leq m,\;\;N^\alpha_j \defn (\ket{\alpha}\bra{\alpha})_j.
\label{eq:PFsymmetriesdiag}
\end{equation}
More generally, additional conserved quantities of $H^{(m)}_{TL}$ can be written as~\cite{readsaleur2007}
\begin{equation}
    M^\beta_\alpha = \sumal{j}{}{(M_j)^\beta_\alpha},\;\;\left(M_j\right)^\beta_\alpha \defn \twopartdef{-\left(\ket{\beta}\bra{\alpha}\right)_j}{j \textrm{~odd}}{\left(\ket{\alpha}\bra{\beta}\right)_j}{j \textrm{~even}},
\label{eq:TLsymmetries}
\end{equation}
where $1 \leq \alpha,\beta \leq m$, and $(M_j)^{\alpha}_{\alpha} = (-1)^j  N_j^\alpha$. 
This directly follows from the local commutation relations
\begin{equation}
    [(M_j)^{\beta}_\alpha + (M_{j+1})^{\beta}_\alpha, \he_{j,j+1}] = 0, \;\; 1 \leq j \leq L -1. 
\label{eq:TLlocalcomm}
\end{equation}
Further, note that there are $m^2-1$ independent $M^{\beta}_\alpha$'s (since $\sum_\alpha{(M_j)^{\alpha}_\alpha} = (-1)^j \mathds{1}$), and it can also be verified that the $\{M^{\beta}_\alpha\}$ are the generators of an $SU(m)$ group~\cite{readsaleur2007}.
Hence the TL models of $H^{(m)}_{TL}$ are $SU(m)$-symmetric. 
Finally, note that when $m = 3$, the TL models $H^{(3)}_{TL}$ can be unitarily transformed into the family of $SU(3)$-symmetric spin-1 biquadratic models~\cite{parkinson1987integrability, parkinson1988biquadratic, barber1989spectrum, Ercolessi2014Analysis}, given by
\begin{equation}
    H_{\textrm{biq}} = U H^{(3)}_{TL} U^\dagger =\sumal{j = 1}{L-1}{J_j \left[(\vec{S}_j \cdot \vec{S}_{j+1})^2 - 1\right]},\;\;U = \prodal{\textrm{$j$\;\;odd}}{}{e^{i \pi S^x_j}},
\label{eq:biquadratic}
\end{equation}
where $\vec{S}_j$ denotes the usual vector of spin-1 operators. 
Similarly, when $m = 2$, the TL models $H^{(2)}_{TL}$ can be unitarily transformed into the family of $SU(2)$-symmetric spin-1/2 Heisenberg models discussed in Sec.~\ref{subsubsec:nonabelian}.
\subsection{Fragmentation in One Dimension}\label{subsec:TLfragmentation}
The dynamically disconnected Krylov subspaces and the fragmentation in the TL models can be understood using a basis of dots and dimers~\cite{readsaleur2007, aufgebauer2010quantum}, which we describe below. 
A ``dimer" between sites $j$ and $k$ is defined to be a singlet configuration $\ket{\psi_{\textrm{sing}}}_{j,k}$ defined in Eq.~(\ref{eq:singletdefn}), 
and is denoted by a line joining the two sites.  
Since the singlet is a maximally entangled state between two spins, no two dimers can end at the same site as a consequence of the monogamy of entanglement.
We construct basis states using a configuration of dimers on the chain, and other unpaired sites in the system (i.e., ones that do not have a dimer ending on them), which we refer to as ``dots."
In particular, any basis state $\ket{\psi}$ with $N$ dimers factorizes as $\ket{\psi} = \ket{\psi_{\textrm{dimer}}} \otimes \ket{\psi_{\textrm{dots}}}$, where $\ket{\psi_{\textrm{dimer}}} \defn \prod_{l=1}^{N}{\ket{\psi_{\textrm{sing}}}_{j_{2l-1},j_{2l}}}$, where $\{j_{2l-1}\}$ and $\{j_{2l}\}$ represent the site indices of the left and right ends of the dimers, such that a dimer connects sites $j_{2l-1}$ and $j_{2l}$.  
We also restrict $\ket{\psi_{\textrm{dimer}}}$ to only have patterns of non-crossing dimers, since it can be shown that any other pattern of dimers can be expressed as a linear combination of configurations of non-crossing dimers~\cite{saito1990noncrossing}. 
Also, we require that no dimers go over any dots.
For a given dimer pattern, we choose an orthogonal basis for the states on the dots $\ket{\psi_{\textrm{dots}}}$ such that they are annihilated by the singlet projector on any adjacent dots (excluding dimer regions). 
For example, we could choose a state on $n$ dots to be a product state $\ket{\psi_{\textrm{dots}}} = \ket{\alpha_1 \alpha_2 \cdots \alpha_n}$ such that $\alpha_j \neq \alpha_{j+1}$, but there are also non-product $\ket{\psi_{\textrm{dots}}}$'s. 
Any state in this basis is hence composed of dots (denoted by $\btp\fdot{0}\etp$) and regions of dimers (denoted by $\btp\rect{0}{1}\etp$), pictorially similar to the basis we used in the PF model in Sec.~\ref{subsec:PFfragmentation1d} (e.g., see Eq.~(\ref{eq:dotdimerexample})).
Such configurations of dots and non-crossing dimers are known to form a complete basis for the full Hilbert space, a fact that has also been used in different contexts~\cite{oguchi1989rvb,veness2017quantum}.
Note that configurations that have the same state $\ket{\psi_{\textrm{dots}}}$ on the dots but differ in the pattern of dimers need not be orthogonal to each other, since it is possible that $\braket{\psi_{\textrm{dimer}}}{\psi'_{\textrm{dimer}}} \neq 0$ for distinct dimer patterns.   
Nevertheless, by construction, configurations with a different number of dimers as well as ones with different states on the dots are orthogonal to each other, while all specified configurations are linearly independent.
Since the dimer regions denoted by $\btp\rect{0}{1}\etp$ always cover even number of consecutive sites, a system of size $L$ even (resp. odd) has Krylov subspaces with $j$ dots, $0 \leq j \leq L$ and $j$ even (resp. odd). %
In the following, we will show that the Krylov subspaces in the TL models are labeled by the state $\ket{\psi_{\textrm{dots}}}$ on the dots. 
To study the Krylov subspaces, we first examine the action of the terms $\{\he_{j,j+1}\}$ of the Hamiltonian on the basis of dimers and dots discussed in the previous paragraph. 
By definition, on configurations of a dimer or dots on sites $j$ and $j+1$, we obtain
\begin{equation}
    \he_{j,j+1}\tket{\cdimer{0}{0.5}{black}} = m \tket{\cdimer{0}{0.5}{black}},\;\;\;\he_{j,j+1}\tket{\cfdot{0}{black}\cfdot{0.5}{black}} = 0.
\label{eq:ejeigenstate}
\end{equation}
Non-vanishing actions allowed by the terms $\he_{j,j+1}$ can be depicted as follows
\begin{gather}
    \he_{j,j+1}\underset{\hspace{1mm}i\hspace{3.5mm}j\hspace{2mm}j+1\hspace{2mm}k}{\tket{\cdimer{-0.5}{0}{black}\cdimer{0.5}{1}{black}}} = {\tket{\cdimer{0}{0.5}{black}\carcdimer{-0.5}{1}{black}}},\nn \\
    \he_{j,j+1}\underset{j\hspace{2mm}j+1\hspace{2mm}k}{\tket{\cfdot{0}{black}\cdimer{0.5}{1}{black}}} = \tket{\cdimer{0}{0.5}{black}\cfdot{1}{black}},
\label{eq:ejscattering}
\end{gather}
where the subscripts label the sites. Note that the sites $i$ and $k$ need not be the neighbors of the sites $j$ or $j+1$.
Note also that in the last equation, the original state at site $j$---which can be any state---is moved to site $k$.
It is easy to see that as a consequence of Eqs.~(\ref{eq:ejeigenstate}) and (\ref{eq:ejscattering}), any configuration of dots and non-crossing dimers maps onto another such configuration with the same number of non-crossing dimers, while retaining the state on the dots. 
Hence, all the basis states with the same state $\ket{\psi_{\textrm{dots}}}$ on the dots can be connected to each other, and such $\ket{\psi_{\textrm{dots}}}$ states label the Krylov subspaces.
Further, all the states with $L$ dots are all frozen eigenstates (i.e., one-dimensional Krylov subspaces) of the TL Hamiltonians, since all the terms of the Hamiltonian act trivially on such states as a consequence of Eq.~(\ref{eq:ejeigenstate}). 
Note that if the state on the dots $\ket{\psi_{\textrm{dots}}}$ is a product state, the Krylov subspaces in the TL models are similar subspaces to those in the PF model (a black dimer here is an equal amplitude superposition of all colored dimers in the PF model).
Indeed, since the family of TL models of Eq.~(\ref{eq:TLHamil}) is a part of the family of PF models of Eq.~(\ref{eq:PFHamil}), the TL models are at least as fragmented as the PF models.
However, in addition, the fragmentation in the TL models also contains cases where $\ket{\psi_{\textrm{dots}}}$ is not a product state, for example, it could consist of a ``triplet'' configuration on two sites like $\ket{\alpha\alpha}_{j,k} - \ket{\beta\beta}_{j,k}$ with $\alpha \neq \beta$. 
Hence, unlike the previous examples, the full fragmentation in the TL models is not evident in the product state basis, and we refer to this type of fragmentation as ``quantum fragmentation."
\subsection{Bond and Commutant Algebras}\label{subsec:TLcommutant}
We now study the fragmentation in the TL models in the language of bond and commutant algebras.
The bond algebra $\mA_{TL}$ in this case is the algebra generated by the terms $\{\he_{j,j+1}\}$ of Eq.~(\ref{eq:TLHamil}). 
For OBC, this algebra is the Temperley-Lieb Algebra with $L-1$ generators (usually denoted by $TL_{L}(q)$), defined by the relations
\begin{equation}
    (e_j)^2 = (q + \qinv) e_j,\;\;\; e_j e_{j \pm 1} e_j = e_j,\;\;\; e_j e_k = e_k e_j
    ,\;\; |j-k| \geq 2.
\label{eq:TLconditions}
\end{equation} 
With some straightforward algebra, it can be verified that Eq.~(\ref{eq:TLconditions}) is satisfied by using $e_j \defn \he_{j, j+1}$ of Eq.~(\ref{eq:TLHamil}), where $q$ is given by 
\begin{equation}
    q + \qinv \defn m\;\;\;\implies q = \frac{m + \sqrt{m^2 - 4}}{2}.
\label{eq:qdefn}
\end{equation}
The commutant of the Temperley-Lieb algebra $TL_L(q)$ has been studied in detail by Read and Saleur in Ref.~\cite{readsaleur2007}.
A large class of operators in the commutant can be obtained by observing the following commutation relation~\cite{readsaleur2007} (assuming $j$ and $k$ belong to different sublattices)
\begin{equation}
    \left[\left(M_j\right)^{\beta_1}_{\alpha_1} \left(M_{k}\right)^{\beta_2}_{\alpha_2}, \he_{j,k}\right]  = 0 \;\;\textrm{if $\alpha_1 \neq \beta_2$ and $\beta_1 \neq \alpha_2$}.
\label{eq:nncommTLuneq}
\end{equation}
Similar to the IoMs of Eqs.~(\ref{eq:tJzdouble}) and (\ref{eq:PFNISdouble}) for the $t-J_z$ and PF models, Eq.~(\ref{eq:nncommTLuneq}) can be used to construct quadratic IoMs
\begin{equation}
    M^{\beta_1, \beta_2}_{\alpha_1, \alpha_2} = \sumal{j_1 < j_2}{}{\left(M_{j_1}\right)^{\beta_1}_{\alpha_1}\left(M_{j_2}\right)^{\beta_2}_{\alpha_2}},\;\;\alpha_1\neq \beta_2,\;\;\beta_1 \neq \alpha_2.
\label{eq:TLJ2conservation}
\end{equation}
Note that unlike the conserved quantities of the $t-J_z$ and PF models discussed in Secs.~\ref{subsec:tJzcommutant} and \ref{subsec:PFcommutant}, the IoMs in Eq.~(\ref{eq:TLJ2conservation}) 
are not diagonal in the product state basis when $\beta_1 \neq \alpha_1$ or $\beta_2 \neq \alpha_2$.
When $\beta_l = \alpha_l$ for $1 \leq l \leq 2$, using Eq.~(\ref{eq:TLsymmetries}), it is easy to see that these reduce to the diagonal IoMs of Eq.~(\ref{eq:PFNISdouble}) in the PF model. 
Further, more non-local IoMs can be constructed as
\begin{equation}
    M^{\beta_1, \cdots, \beta_k}_{\alpha_1,\cdots, \alpha_k} = \sumal{j_1 < \cdots < j_k}{}{\prodal{l = 1}{k}{\left(M_{j_l}\right)^{\beta_l}_{\alpha_l}}},\;\;\alpha_l \neq \beta_{l+1},\;\; \beta_l \neq \alpha_{l+1}, 
\label{eq:TLJkconservation}
\end{equation}
which are defined to equal the identity operator $\mathds{1}$ when $k = 0$, and are the $SU(m)$ generators of Eq.~(\ref{eq:TLsymmetries}) when $k = 1$.
Again, these are not diagonal unless $\beta_l = \alpha_l$ for all $1 \leq l \leq k$, in which case they reduce to the IoMs of the PF model in Eq.~(\ref{eq:stringops}).
This is a consequence of the fact that the TL models are at least as fragmented as the PF models.
Furthermore, the IoMs of Eq.~(\ref{eq:TLJkconservation}) do not all commute with each other, hence the commutant $\mC_{TL}$ is %hence 
non-Abelian.
Moreover, these IoMs of Eq.~(\ref{eq:TLJkconservation}) still do not exhaust the commutant $\mC_{TL}$.
As we discuss in App.~\ref{app:TLcommutants}, additional conserved quantities can be constructed in the cases when $\beta_l = \alpha_{l+1}$ or $\alpha_l = \beta_{l+1}$ in Eq.~(\ref{eq:TLJkconservation}).
Finally, we note that although we know the commutant algebra explicitly, unlike the $t-J_z$ and PF models, we do not know of a way to determine a minimal set of generators analogous to the SLIOMs.
The TL models are some of the few models where the structure of the bond algebra and its representations have been studied extensively.
Given the structure of the bond algebra $\mA_{TL} = TL_L(q)$ and its commutant $\mC_{TL}$, the Hilbert space can be decomposed into representations of $\mA_{TL} \times \mC_{TL}$ according to Eq.~(\ref{eq:Hilbertdecomp}).  
The Krylov subspaces of the TL models are simply the irreducible representations of $TL_{L}(q)$, and the dimensions $\{D_\lambda\}$ and $\{d_\lambda\}$ of the irreducible representations of the bond and commutant algebras for even $L$ are given by~\cite{readsaleur2007}
\begin{equation}
    D_\lambda = \binom{L}{L/2 + \lambda} - \binom{L}{L/2 + \lambda + 1},\;\; d_\lambda = [2\lambda + 1]_q,
\label{eq:TLdimensions}
\end{equation}
where $\lambda$ is an integer $0 \leq \lambda \leq L/2$, and $[\cdot]_q$ denote $q$-deformed integers, defined as $[n]_q \defn (q^n - q^{-n})/(q-q^{-1})$.  
In the description of the Krylov subspaces in terms of dimers and dots, for even $L$ in Eq.~(\ref{eq:TLdimensions}), $2\lambda$ gives the number of dots and $d_\lambda$ gives the corresponding number of distinct $\ket{\psi_{\textrm{dots}}}$.
The degeneracy among the corresponding distinct Krylov subspaces is manifest since the described action of the TL generators is identical in terms of dimers for any $\ket{\psi_{\textrm{dots}}}$.
Using Eqs.~(\ref{eq:dimensions}) and (\ref{eq:TLdimensions}), the dimension of the commutant $\mC_{TL}$ can be shown to scale as~\cite{readsaleur2007}
\begin{equation}
    \textrm{dim}(\mC_{TL}) \sim \frac{q^{2L}}{(1-q^{-2})(1- q^{-4})}\;\;\textrm{for large $L$},
\label{eq:TLcommutantscaling}
\end{equation}
which clearly grows exponentially with $L$ for $q > 1$.
This indicates the presence of Hilbert space fragmentation in the TL models for $m \geq 3$, including in the spin-1 biquadratic model. 
\subsection{Dynamics within Krylov subspaces}\label{subsec:krylovdynamics}
\begin{figure}[t!]
    \centering
    \includegraphics[scale=0.35]{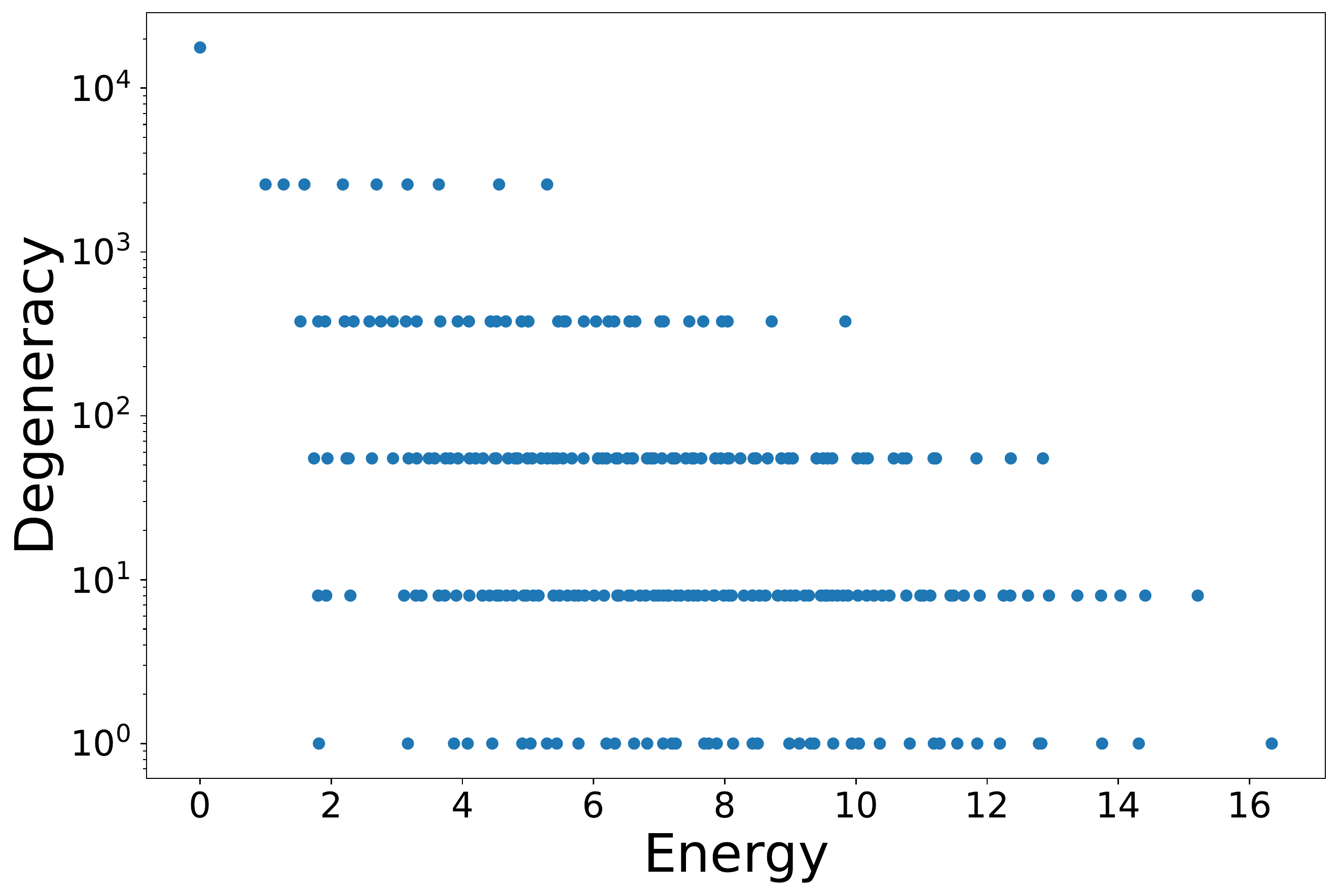}
    \caption{Energy spectrum of the spin-1 biquadratic model of Eq.~(\ref{eq:biquadratic}) with OBC and system size $L = 10$ for a single disorder realization with the $J_j$'s chosen from a uniform distribution in $[0.6, 1.4]$. The horizontal axis shows the energy of the levels and the vertical axis shows the degeneracy of a particular energy level. All the levels at a fixed degeneracy are part of the same Krylov subspace. The number of levels at a fixed degeneracy is $D_\lambda$ (the size of the Krylov subspace), and the corresponding degeneracy is $d_\lambda$ (the number of Krylov subspaces of that size).}
    \label{fig:biquad}
\end{figure}
Hilbert space fragmentation in the TL models leads to several special novel features in the spectrum that are absent in models with classical fragmentation, particularly due to the presence of a non-Abelian commutant.
As discussed in Sec.~\ref{subsec:defn}, the dimensions $\{d_\lambda\}$ of the irreducible representations of the commutant correspond to the degeneracies among the Krylov subspaces with dimension $\{D_\lambda\}$.
In all the cases with classical fragmentation, $d_\lambda = 1$ since the commutant is Abelian. 
However, according to Eq.~(\ref{eq:dimensions}), we know that $d_\lambda > 1$ for $\lambda \neq 0$, and this leads to large degeneracies in the spectra of the TL models.  
For example, the ground state degeneracy in the OBC ferromagnetic spin-1 biquadratic model (i.e., with $J_j > 0$ in Eq.~(\ref{eq:biquadratic})) is known to grow as $F_{2L + 2}$, the $(2L+2)^{th}$ Fibonacci number, which is simply equal to $d_{\lambda = L/2}$, the degeneracy among the Krylov subspaces with $\lambda = L/2$~\cite{readsaleur2007}.   
In Fig.~\ref{fig:biquad}, we plot the energy spectrum and the degeneracies of the energy levels in the spin-1 biquadratic model with disorder. 
As evident there, large degeneracies also extend to typical excited states in the middle of the spectrum, which is a direct consequence of the fragmentation with a non-Abelian commutant.
Hence, we expect the TL models to exhibit highly non-generic dynamics for arbitrary values of $\{J_j\}$ in spite of being non-integrable. 
However, as we now discuss, the dynamics \textit{within} Krylov subspaces is expected to be thermal. 
The TL Hamiltonian $H^{(m)}_{TL}$ of Eq.~(\ref{eq:TLHamil}), when restricted to a particular Krylov subspace, is known to map onto a particular quantum number sector of the spin-1/2 $q$-deformed XXZ (XXZ-q) model with quantum group symmetry $SU(2)_q$, where $q$ is given by Eq.~(\ref{eq:qdefn}).
The dynamics within Krylov subspaces of the TL Hamiltonian can thus be described by the Hamiltonian~\cite{readsaleur2007, aufgebauer2010quantum}
\begin{gather}
    H_\textrm{XXZ-q} =
    -2\sumal{j =1}{L-1}{ J_j\left[\left(S^x_j S^x_{j+1} + S^y_j S^y_{j+1}\right)\right.}\nn \\
    {\left.+ \frac{q + \qinv}{2}\left(S^z_j S^z_{j+1} - \frac{1}{4}\right) + \frac{q - \qinv}{4}\left(S^z_j - S^z_{j+1}\right)\right]},
\label{eq:XXZq}
\end{gather}
where $S^x_j$, $S^y_j$, and $S^z_j$ denote the usual spin-1/2 operators on site $j$.
It can be directly verified that the bond algebra generated by the terms of the Hamiltonian $H_{\textrm{XXZ-q}}$ is the Temperley-Lieb algebra $TL_L(q)$, i.e., if the term acting on sites $j$ and $j+1$ in Eq.~(\ref{eq:XXZq}) is referred to as $e_j$, they satisfy the conditions of Eq.~(\ref{eq:TLconditions}).
Moreover, the picture of dots and dimers discussed in Sec.~\ref{subsec:TLfragmentation} continues to hold, along with the relations of Eqs.~(\ref{eq:ejeigenstate}) and (\ref{eq:ejscattering}), where the dimers are $q$-deformed singlets defined as
\begin{equation}
    \ket{\psi^{(q)}_{\textrm{sing}}}_{j,k} \defn q^{-\frac{1}{2}} \ket{\uparrow \downarrow}_{j,k} - q^{\frac{1}{2}}\ket{\downarrow\uparrow}_{j,k},
\label{eq:qsingletdefn}
\end{equation}
instead of the singlets $\ket{\psi_{\textrm{sing}}}$ of Eq.~(\ref{eq:singletdefn}). 
In the XXZ-q model, it is easy to see that for $2\lambda$ dots there are precisely $2\lambda + 1$ distinct $\ket{\psi_\textrm{dots}}$, which gives the number of degenerate Krylov subspaces in this case, but these are now understood as corresponding to spin-$\lambda$ multiplets of the $SU(2)_q$ symmetry.
The dynamics within Krylov subspaces of the TL models is hence equivalent to the XXZ-q models in corresponding symmetry sectors, with $q$ given by Eq.~(\ref{eq:qdefn}). 
Note that for $q = 1$, the XXZ-q model of Eq.~(\ref{eq:XXZq}) reduces to the Heisenberg model of Eq.~(\ref{eq:Heisenberg}) (up to overall constants).

The XXZ-q models of Eq.~(\ref{eq:XXZq}) are expected to be non-integrable for generic values of $J_j$. 
Reference~\cite{protopopov2020nonabelian} probed the energy level statistics for the disordered Heisenberg model (i.e., the $q = 1$ case of XXZ-q), and found that it exhibits Wigner-Dyson statistics within quantum number sectors of the $SU(2)$ symmetry that correspond to finite energy density, although it is only apparent at very large system sizes.
This suggests that the disordered Heisenberg model thermalizes at large enough system sizes even though it might not appear so for small system sizes.
We numerically observe the same qualitative behavior in the level statistics of the XXZ-q models for $q > 1$ for small system sizes we are able to probe, hence we expect that they too thermalize at large enough system sizes.
This suggests that the TL models also thermalize within exponentially large Krylov subspaces, providing further support for the validity of Krylov-restricted thermalization~\cite{moudgalya2019thermalization}, the property that sufficiently large non-integrable Krylov subspaces in fragmented models thermalize~\cite{yang2019hilbertspace, hahn2021information, khudorozhkov2021hilbert}.
\section{Dipole-Conserving Models}
\label{sec:dipole}
\subsection{Definition and Symmetries}\label{subsec:dipolemodel}
We now turn to models that conserve dipole moment or center-of-mass, which were studied in the context of quantum dynamics in Refs.~\cite{fremling2018dynamics, pai2018localization, sala2020fragmentation, khemani2020localization, moudgalya2020quantum, moudgalya2019thermalization, sliom2020}.
In particular, we focus on the one-dimensional spin-1 dipole-moment conserving models introduced in Ref.~\cite{sala2020fragmentation}, and hence study the family of Hamiltonians
\begin{gather}
    \hdip = \sumal{j}{}{J_j \overbrace{\left[S^-_{j-1} (S^+_j)^2 S^-_{j+1} + h.c.\right]}^{\hP_{[j-1,j+1]}}}\nn \\
    +\sumal{j}{}{\left(h_j S^z_j + g_j (S^z_j)^2\right)},
\label{eq:spin1dip}
\end{gather}
where we have added the last two terms to remove any discrete symmetries, and to ensure for simplicity that all the operators in the commutant are diagonal in the product state basis (see App.~\ref{app:diagonalcommutants}).
The transitions implemented by the terms $\{\hP_{[j-1,j+1]}\}$ can be denoted as
\begin{gather}
    \ket{0 + -} \leftrightarrow \ket{+ - 0},\;\;\; \ket{0 - +} \leftrightarrow \ket{- + 0},\nn \\
    \ket{0 + 0} \leftrightarrow \ket{+ - +},\;\;\; \ket{0 - 0} \leftrightarrow \ket{- + -},
\label{eq:spin1dipoletransitions}
\end{gather}
where $+, 0, -$ respectively denote the spin-1 states with $S^z = +1, 0, -1$, and $\ket{\cdot}$ represents the spin configuration on three consecutive sites $j-1$, $j$, and $j+1$.  
For simplicity, we restrict ourselves to OBC throughout this section. 
The family of models of Eq.~(\ref{eq:spin1dip}) has two obvious conserved quantities, the charge $\hQ$ and the dipole moment $\hD$, given by the operators
\begin{equation}
    \hQ = \sumal{j}{}{S^z_j},\;\;\;\hD = \sumal{j}{}{j S^z_j}. 
\label{eq:chargedipole}
\end{equation}
\subsection{Fragmentation and Commutant Algebra}\label{subsec:invariantstrings}
The Hilbert space fragmentation for the family of models in Eq.~(\ref{eq:spin1dip}) was pointed out in Refs.~\cite{sala2020fragmentation, khemani2020localization}, where they noted the existence of exponentially many Krylov subspaces, and systematically constructed and counted the frozen product eigenstates (i.e., one-dimensional Krylov subspaces).
In addition to frozen eigenstates and Krylov subspaces similar to the other examples we studied, the dipole-conserving models also possess Krylov subspaces with frozen regions in the chain or ``blockades," which dynamically disconnect parts of the system. 
For example, all the states in such Krylov subspaces factorize as $\ket{\psi} = \ket{L^{(\ell)}} \otimes \ket{B^{(b)}} \otimes \ket{R^{(r)}}$, where $\ket{B^{(b)}}$ is the frozen region spanning $b$ sites that dynamically disconnects the sites to its left and right (this is the same for all the states within the Krylov subspace), and $\ket{L^{(\ell)}}$ and $\ket{R^{(r)}}$ are the wavefunctions on the $\ell$ and $r$ sites to the left and right of the blockade respectively.
There are no analogues of these blockades in the $t-J_z$, PF, or the TL models.
A simple example of a blockaded Krylov subspace in $H_{\dip}$ contains states of the form $\ket{\cdots + 0 \cdots 0\ \fbox{+ +}\ 0 \cdots 0 + \cdots}$, where it is easy to see that the boxed blockade configuration ($\ket{B^{(b)}} = \ket{++}$) can never be changed under the transitions of Eq.~(\ref{eq:spin1dipoletransitions}).
The bond algebra $\mA_{\dip}$ corresponding to the family of Hamiltonians of Eq.~(\ref{eq:spin1dip}) is the algebra generated by the operators $\{\hP_{[j-1, j+1]}\}$ and $\{S^z_j, (S^z_j)^2\}$. 
The inclusion of the latter set of terms ensures that the corresponding commutant algebra $\mC_{\dip}$ only consists of operators diagonal in the product state basis (see App.~\ref{app:diagonalcommutants}), and this model exhibits classical fragmentation.
However, unlike the $t-J_z$ and PF models, we are not able to find simple expressions for a basis of $\mC_{\dip}$ in terms of simple operators such as $S^z_j$. 
Nevertheless, since $\mC_{\dip}$ is abelian, the projectors onto the Krylov subspaces form an orthogonal basis of the commutant algebra.
For example, for any Krylov subspace $\mK_\alpha$  spanned by an orthogonal set of states $\left\{\ket{\psi_{\alpha\beta}}\right\}$, consider the projector $\Pi_{\mK_\alpha} \defn \sum_\beta{\ket{\psi_{\alpha\beta}}\bra{\psi_{\alpha\beta}}}$.
Since $\Pi_{\mK_\alpha}$ is diagonal in the product state basis, it commutes with all other diagonal operator in that basis, hence $[\Pi_{\mK_\alpha}, S^z_j] = [\Pi_{\mK_\alpha}, (S^z_j)^2] = 0$ for all $j$ and $\alpha$. 
Furthermore, for any $\ket{\psi_{\alpha\beta}} \in \mK_\alpha$, $\hP_{[j-1, j+1]}\ket{\psi_{\alpha\beta}} \in \mK_\alpha$, hence it is easy to see that $[\Pi_{\mK_\alpha}, \hP_{[j-1,j+1]}] = 0$ for all $j$.
Since we know that all the operators in the commutant are diagonal, the Krylov subspace projectors $\Pi_{\mK_\alpha}$ form an orthogonal basis for the commutant $\mC_{\dip}$.
The operators in the commutant can then be directly understood from the structure of the states in the Krylov subspaces.
The structure of the Krylov subspaces was worked out in Ref.~\cite{sliom2020}, which constructed SLIOMs for the dipole-conserving models $\hdip$ of Eq.~(\ref{eq:spin1dip}) that uniquely label the Krylov subspaces.
They first identified that patterns of ``domain-walls," systematically determined for any product state, were conserved under the actions of Eq.~(\ref{eq:spin1dipoletransitions}).
In addition to the pattern of domain walls, the dipole moments between the domain-walls were also independently conserved.
However, this dipole moment conservation makes it hard to associate Krylov subspaces to invariant ``patterns," similar to the pattern of spins in the $t-J_z$ model, the pattern of dots in the PF model, or the state on the dots in the TL model, and one might wonder if such a pattern exists.
Nevertheless, in App.~\ref{app:canonicalconfigs}, we show that such patterns indeed exist, when the model is described in a language of appropriately defined ``dots" and ``links."
This makes the Krylov subspaces much more apparent, and allows us to use the transitions of Eq.~(\ref{eq:spin1dipoletransitions}) in the language of dots and links (see Eq.~(\ref{eq:hdiptransitionsdiag})) to bring all the states in a Krylov subspace to a unique canonical form that characterizes the Krylov subspace.
We further show that each canonical configuration can be uniquely mapped onto a tiling pattern of a chain of length $L$ using three objects $\btp\fdot{0}\etp$, $\btp\udot{0}\etp$, $\btp\dimer{0}{0.5}\etp$ (see Eq.~(\ref{eq:tilingmap})), which enables us to compute the exact number of Krylov subspaces as a function of system size (i.e., the number of canonical configurations).
In particular, for OBC on a system size $L$, we find that (see Eq.~(\ref{eq:commutantdimdipole}))
\begin{equation}
    \textrm{dim}(\mC_{\textrm{dip}}) = 2P_{L+1} - 1 \sim (\sqrt{2} + 1)^L\;\;\textrm{for large $L$},
\label{eq:dipoledim}
\end{equation}
where $P_{L+1}$ is the $(L+1)^{th}$ Pell number.
The exponential growth of the dimension signifies the presence of Hilbert space fragmentation. 
Although the explicit expressions for the operators in the commutant are rather obscure, with the above picture in hand, in Sec.~\ref{subsubsec:dipolemazur} we will be able to calculate contribution from a large class of such projectors to a Mazur bound for spin autocorrelations, showing analytically that blockades lead to effective spin localization even at infinite temperature.
Although we have not shown this explicitly, it is likely that  the SLIOMs constructed in Ref.~\cite{sliom2020} along with the $\mathds{1}$ operator form a set of generators for the full commutant algebra, similar to the SLIOMs in the $t-J_z$ model discussed in Sec.~\ref{subsec:SLIOMconnection}.
Indeed, if each Krylov subspace is described by a distinct set of eigenvalues of SLIOMs, we can use the SLIOMs to construct a product of projectors onto a space with the appropriate eigenvalue for each SLIOM, which is related to the projector onto that Krylov subspace.
On a different note, we also believe that the commutant $\mC_{\dip}$ can be straightforwardly generalized to the family of spin-1/2 dipole-conserving ``pair-hopping models"~\cite{moudgalya2020quantum, moudgalya2019thermalization}, since as shown in  Ref.~\cite{moudgalya2019thermalization}, the Krylov subspaces there closely resemble the ones in the spin-1 model of Eq.~(\ref{eq:spin1dip}). 

\section{Mazur Bounds}\label{sec:Mazur}
\subsection{Definition}\label{subsec:Mazurdefn}
The effect of conserved quantities on the dynamics of isolated quantum systems can be quantified using Mazur bounds~\cite{mazurbound1969, dhar2020revisiting} on the long-time average of dynamical autocorrelation functions under time-evolution.
Given a system with Hamiltonian $H$ and conserved quantities $\{I_\alpha\}$, the autocorrelation function of an observable $A$ can be bounded as
\begin{gather}
    C_A
    \defn \lim_{\tau \rightarrow \infty} \frac{1}{\tau} \int_0^\tau{dt\ \langle A(t) A(0)\rangle}\nn \\
    \geq \sumal{\alpha, \beta}{}{\opbraket{A}{I_\alpha} \left(K^{-1} \right)_{\alpha\beta}\opbraket{I_\beta}{A}} \defn M_{A},
\label{eq:mazurboundnotorth}
\end{gather}
where we have defined
\begin{gather}
    A(t) = e^{i H t} A e^{-i H t},\;\;\; \opbraket{A}{B} \defn \langle A^\dagger B \rangle \defn \frac{1}{D}\textrm{Tr}\left(A^\dagger B\right),\nn \\
    K_{\alpha\beta} \defn \opbraket{I_\alpha}{I_\beta},
\label{eq:mazurdefns}
\end{gather}
where $D$ is the Hilbert space dimension, $\langle \bullet \rangle$ denotes the infinite-temperature expectation value, and $K^{-1}$ is the inverse of the correlation matrix $K$.
We have also introduced a braket notation in operator space, i.e., $\opket{A}$ denotes an operator $A$ and $\opbraket{A}{B}$ denotes the overlap between two operators $A$ and $B$ as defined in Eq.~(\ref{eq:mazurdefns}), and we will be using these two notations interchangeably throughout this section. 
Note that the correlation matrix can be diagonalized by working with an orthogonal basis of conserved quantities, i.e., appropriate linear combinations of $\{I_\alpha\}$, which we refer to as $\{Q_\alpha\}$, that satisfy $\opbraket{Q_\alpha}{Q_\beta} \sim \delta_{\alpha,\beta}$. 
The Mazur bound of Eq.~(\ref{eq:mazurboundnotorth}) can then be expressed in terms of $\{Q_\alpha\}$ as follows
\begin{equation}
    C_{A} \geq \sumal{\alpha}{}{\frac{\opbraket{A}{Q_\alpha} \opbraket{Q_\alpha}{A}}{\opbraket{Q_\alpha}{Q_\alpha}}} = M_{A}.
\label{eq:mazurbound}
\end{equation}
Denoting eigenstates of $H$ with energies $\{E_\alpha\}$ as $\{\ket{e_\alpha}\}$, a trivial example of the Mazur bound can be obtained directly by choosing $\{Q_\alpha\}$ in the RHS of Eq.~(\ref{eq:mazurbound}) to be the set of eigenstate projectors $\{\ket{e_\alpha}\bra{e_\alpha}\}$
\begin{equation}
    M_{A} = \frac{1}{D}\sumal{\alpha}{}{|\bra{e_\alpha} A \ket{e_\alpha}|^2}.
\label{eq:projectormazur}
\end{equation}
Meanwhile, the LHS of Eq.~(\ref{eq:mazurbound}), which can be expressed as an expectation value in the diagonal ensemble: 
\begin{gather}
    C_{A} = \lim_{\tau \rightarrow \infty} \frac{1}{D \tau} \int_0^\tau{dt\ e^{i (E_\alpha - E_\beta)t} |\bra{e_\alpha} A \ket{e_\beta}|^2} \nn \\
    = \frac{1}{D}\sumal{\alpha}{}{|\bra{e_\alpha} A\ket{e_\alpha}|^2} + \frac{1}{D}\sumal{\alpha \neq \beta, E_\alpha = E_\beta}{}{|\bra{e_\alpha} A \ket{e_\beta}|^2} \geq M_{A}.\nn \\
\end{gather}
However, if the spectrum of the Hamiltonian $\{E_\alpha\}$ is non-degenerate, the inequality saturates, and this is known as the Suzuki equality~\cite{suzukiequality1971, dhar2020revisiting}.
Note that Eq.~(\ref{eq:mazurbound}) can also be extended to finite temperature autocorrelation functions but in this work we only focus on the infinite-temperature case. 
In typical applications of the Mazur bound, for example in the study of integrable systems, the only conserved quantities that are considered in Eq.~(\ref{eq:mazurbound}) are local or quasilocal conserved quantities~\cite{prosen2013quasilocalfamilies, prosen2014quasilocal, ilievski2016quasilocal, zadnik2016quasilocal, doyon2017pseudolocal}. 
However, when considering the dynamics of some Hamiltonian in the family of Eq.~(\ref{eq:genhamil}), given that the commutant algebra contains many different linearly independent conserved quantities, as well as exponentially many of them in fragmented systems, it is not a priori clear which of these contribute the most to the Mazur bound.
Hence we study the Mazur bound by considering the full commutant algebra in Eq.~(\ref{eq:mazurbound}); this also helps quantify the relative importance of the various conserved quantities in the commutant via their contribution to the bounds for observables of interest.
In the following, we focus on autocorrelation functions of local (on-site or nearest-neighboring terms) observables $A$.
Without loss of generality, we choose observables to be traceless ($\langle A \rangle = 0 = \textrm{Tr}(A)$), although it is sometimes more convenient to subtract out the contribution of the traceful part ($\langle A \rangle^2$) later, which is equivalent to studying the Mazur bound for the ``connected" autocorrelation function. 
Further, we choose $\{I_\alpha\}$ to be any linear basis for the commutant algebra $\mC$, and $\{Q_\alpha\}$ to be an orthogonal basis for $\mC$.
\subsection{Conventional Symmetries}\label{subsec:Mazurconventional}
We start with the Mazur bound in the case of conventional symmetries discussed in Sec.~\ref{subsec:conventionalexamples}.
The answer is straightforward for systems without any symmetry, the only conserved quantity in the commutant is $\mathds{1}$, and the bound of Eq.~(\ref{eq:mazurbound}) reduces to $M_{A} = (\textrm{Tr}(A)/D)^2 = 0$.
Hence the autocorrelation function in such systems at late times typically decays to zero. 
We then consider systems with $U(1)$ symmetry, such as the spin-1/2 XXZ model of Eq.~(\ref{eq:XXZ}), and focus on the Mazur bound for the autocorrelation function of the operator $S^z_j$, the spin operator on site $j$.  
As discussed in Sec.~\ref{subsubsec:abelian}, the commutant algebra $\mC$ is spanned by the operators $\{\mathds{1}, S^z_{\tot}, \cdots, (S^z_{\tot})^{L}\}$. 
This is not an orthogonal basis for $\mC$, since $\opbraket{(S^z_{\tot})^m}{(S^z_{\tot})^n} = \textrm{Tr}((S^z_{\tot})^{m+n}) \neq 0$ for all $m$ and $n$. 
Using the expression $S^z_{\tot} = \sum_{j=1}^L{S^z_j}$, an orthogonal basis $\{Q^z_0, \cdots, Q^z_L\}$ for $\mC$ reads
\begin{equation}
    Q^z_n = \sumal{j_1 \neq j_2 \neq \cdots \neq j_n = 1}{L}{S^z_{j_1} S^z_{j_2} \cdots S^z_{j_n}},\;\;0 \leq n \leq L, 
\label{eq:U1commutantbasis}
\end{equation}
where $Q^z_0 = \mathds{1}$ and $Q^z_1 = S^z_{\tot}$. 
Applying the Mazur bound of Eq.~(\ref{eq:mazurbound}) for the observable $A = S^z_j$,  we obtain 
\begin{equation}
	C_{S^z_j} \geq M_{S^z_j} = \sumal{n = 0}{L}{\frac{\opbraket{S^z_j}{Q^z_n}^2}{\opbraket{Q^z_n}{Q^z_n}}} = \frac{\opbraket{S^z_j}{Q^z_1}^2}{\opbraket{Q^z_1}{Q^z_1}} = \frac{1}{4 L}, 
\label{eq:U1mazurbound}
\end{equation}
where we have used $\opbraket{S^z_j}{Q^z_n} = \delta_{n, 1}/4$,  and $\opbraket{Q^z_1}{Q^z_1} = L/4$. 
As evident from Eq.~(\ref{eq:U1mazurbound}), the only conserved quantity in the commutant that contributes is $Q^z_1 = S^z_{\tot}$,  hence it is sufficient to only use the local conserved quantity for the Mazur bound.  
However,  this is not sufficient for multi-site observables such as  $A = S^z_j S^z_{j+1}$,  for which the Mazur bound of Eq.~(\ref{eq:mazurbound}) reads
\begin{equation}
M_{S^z_j S^z_{j+1}} = \sumal{n = 0}{L}{\frac{\opbraket{S^z_j S^z_{j+1}}{Q^z_n}^2}{\opbraket{Q^z_n}{Q^z_n}}} = \frac{\opbraket{S^z_j S^z_{j+1}}{Q^z_2}^2}{\opbraket{Q^z_2}{Q^z_2}} = \frac{1}{8 L (L-1)},
\label{eq:U1mazurboundZZ}
\end{equation}
where we have used $\opbraket{S^z_j S^z_{j+1}}{Q^z_n} = \delta_{n, 2}/8$, and $\opbraket{Q^z_2}{Q^z_2} = L (L-1)/8$.
Generically,  all operators in the commutant need to be considered in order to obtain a tight Mazur bound, although local operators can be sufficient for certain observables of interest. 
A similar analysis can be extended to the $SU(2)$-symmetric systems discussed in Sec.~\ref{subsubsec:nonabelian} such as the spin-1/2 Heisenberg model of Eq.~(\ref{eq:Heisenberg}).  
Following the discussion there, it is clear that an (overcomplete) orthogonal basis for the commutant is constructed from operators that have the form
\begin{equation}
	Q^{\alpha_1 \cdots \alpha_n}_n = \sumal{j_1 \neq \cdots \neq j_n = 1}{L}{S^{\alpha_1}_{j_1} \cdots S^{\alpha_n}_{j_n}},\;\;0 \leq n \leq L,\;\;\alpha_l \in \{x, y, z\}, 
\label{eq:SU2commutantbasis}
\end{equation}
where distinct basis elements are characterized by the number of $x$'s, $y$'s, and $z$'s among $\alpha_\ell$'s.
To obtain tight Mazur bounds for observables $S^z_j$ and $S^z_j S^z_{j+1}$, it is sufficient to identify the operators in the commutant that have a non-vanishing overlaps with them, and the basis of Eq.~(\ref{eq:SU2commutantbasis}) makes it clear that the only such operators are $Q^z_1 = S^z_{\tot}$ and $Q^z_2$ of Eq.~(\ref{eq:U1commutantbasis}) respectively.
This leads to the same Mazur bounds as the $U(1)$ case in Eqs.~(\ref{eq:U1mazurbound}) and (\ref{eq:U1mazurboundZZ}).  
The larger commutant in the $SU(2)$ case implies that similar bounds also hold for arbitrary spin components $S_j^\alpha$ (hence arbitrary on-site traceless observables) and arbitrary products $S_j^\alpha S_k^\beta$, including with different spin components.
For example, using the operators $Q^{\alpha\beta}_2$ with $\alpha \neq \beta$ in Eq.~(\ref{eq:SU2commutantbasis}), we obtain $M_{S_j^\alpha  S_k^\beta} = 1/[16 L(L-1)]$ if $\alpha \neq \beta$.
Note that these Mazur bounds are valid as bounds valid for {\it arbitrary} XXZ and Heisenberg models of Eqs.~(\ref{eq:XXZ}) and (\ref{eq:Heisenberg}) respectively.
The translation invariant XXZ and Heisenberg models are integrable, and consist of additional quasilocal conserved quantities~\cite{prosen2013quasilocalfamilies, prosen2014quasilocal} not part of the commutant algebra, resulting in larger Mazur bounds. 
\subsection{Fragmented Systems}\label{subsec:fragmentedMazur}
\begin{figure}[t!]
    \centering
    \includegraphics[scale=0.48]{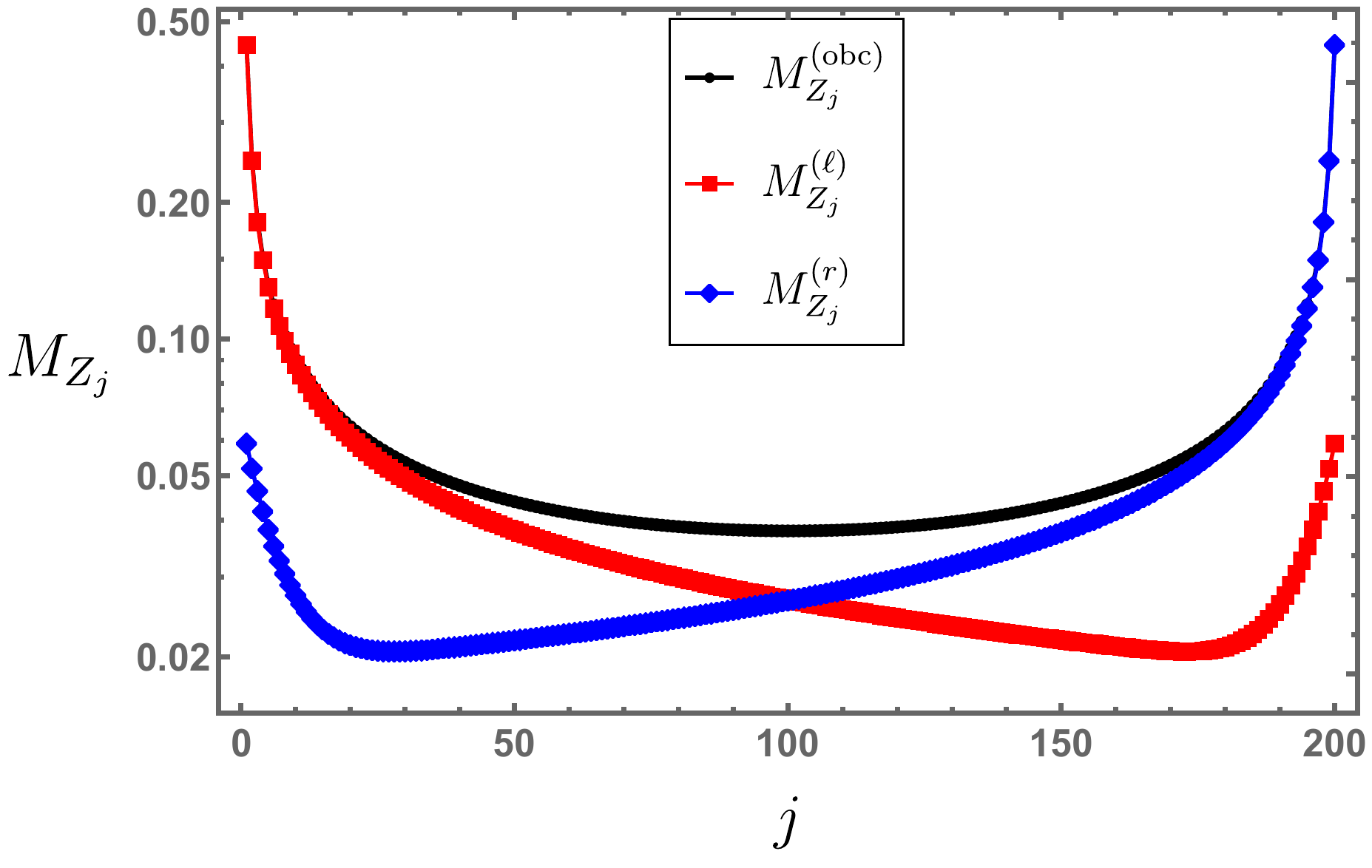}
    \caption{(Color online) Mazur bounds for the autocorrelation function of the spin operator $Z_j$ in the one-dimensional $t-J_z$ model with OBC of system size $L = 200$. The bounds, defined in Eq.~(\ref{eq:mazurqtys}), are obtained by considering the left SLIOMs ($M^{(\ell)}_{Z_j}$, red), right SLIOMs ($M^{(r)}_{Z_j}$, blue), or the full commutant algebra $\Cobc$ ($M^{(\textrm{obc})}_{Z_j}$, black) as the only conserved quantities. The bounds $M^{(\ell)}_{Z_j}$ and $M^{(r)}_{Z_j}$ agree with the bound $M^{(\textrm{obc})}_{Z_j}$ close to the left and right edges of the chain respectively, but deviate everywhere else.}
    \label{fig:tJzmazur}
\end{figure}
We now discuss various aspects of Mazur bounds in systems exhibiting Hilbert space fragmentation.
\subsubsection{$t-J_z$ model}
\label{subsubsec:tJzMazur}
We first illustrate the Mazur bound for the one-dimensional $t-J_z$ model discussed in Sec.~\ref{sec:tJz}.
Only considering the $U(1)$ conserved quantity $Z_{\tot} \defn N^\uparrow - N^\downarrow$, where $N^\uparrow$ and $N^\downarrow$ are defined in Eq.~(\ref{eq:tJzsymmetries}), the Mazur bound for the on-site spin operator defined as $Z_j \defn N^\uparrow_j - N^\downarrow_j$ reads
\begin{equation}
    M^{(U(1))}_{Z_j} \defn \frac{\opbraket{Z_j}{Z_{\tot}}^2}{\opbraket{Z_{\tot}}{Z_{\tot}}} = \frac{2}{3L},
\label{eq:U1mazurtJz}
\end{equation}
where we have used $\opbraket{Z_j}{Z_{\tot}} = 2/3$, and $\opbraket{Z_{\tot}}{Z_{\tot}} = 2 L/3$.
For Mazur bounds of products of spin operators such as $Z_j Z_{j+1}$, the contribution of $Z_{\tot}$ vanishes since $\opbraket{Z_j Z_{j+1}}{Z_{\tot}} = 0$.
However, the rest of the $U(1)$ commutant (i.e., the algebra generated by operators $N^\uparrow$ and $N^\downarrow$) has a non-vanishing contribution to the Mazur bound, which can be computed by working with a partial set of orthogonal conserved quantities
\begin{equation}
    Q^Z_n \defn \sumal{j_1 \neq j_2 \neq \cdots \neq j_n = 1}{L}{Z_{j_1} Z_{j_2} \cdots Z_{j_n}},\;\;0 \leq n \leq L, 
\label{eq:tJzconservedall}
\end{equation}
where $Q^Z_0 = \mathds{1}$ and $Q^Z_1 = Z_{\tot}$.
The Mazur bound for the operator $Z_j Z_{j+1}$ then reads
\begin{equation}
    M^{(U(1))}_{Z_j Z_{j+1}} = \frac{\opbraket{Z_j Z_{j+1}}{Q^Z_2}^2}{\opbraket{Q^Z_2}{Q^Z_2}} = \frac{8}{9L(L-1)} \sim \frac{1}{L^2}
\label{eq:U1mazurZZ}
\end{equation}
where we have used $\opbraket{Z_j Z_{j+1}}{Q^Z_2} = 8/9$ and $\opbraket{Q^Z_2}{Q^Z_2} = 8 L(L-1)/9$.
Reference~\cite{sliom2020} numerically observed that Eq.~(\ref{eq:U1mazurtJz}) is not a tight bound for the late-time spin autocorrelation function in the OBC $t-J_z$ model, which was attributed to the presence of additional non-local conserved quantities, particularly $L$ (left) SLIOMs $\{\hq^{(\ell)}_l\}$.
They further numerically studied the Mazur bounds after including the SLIOM contribution, and showed that it led to a tighter bound closer to the true behavior of the spin autocorrelation function.  
However, as discussed in Sec.~\ref{subsec:SLIOMconnection}, the SLIOMs are not the only conserved quantities of the $t-J_z$ model, and there could be additional contributions to the Mazur bounds from other operators such as products of SLIOMs, more generally from all operators in the commutant $\Cobc$.   %
Indeed, as we discuss in the following, the left or right SLIOMs by themselves do not provide tight Mazur bounds for all of the spin operators $Z_j$, and we obtain a better bound by considering the full commutant algebra $\Cobc$. 
The Mazur bound of Eq.~(\ref{eq:mazurboundnotorth}) using the full commutant $\Cobc$ can be computed by choosing $\{I_\alpha\}$ to be the non-orthogonal IoMs $\{N^{\sigma_1 \cdots \sigma_k}\}$ (Eq.~(\ref{eq:tJzconserved})) or $\{N^{[\sigma_1 \cdots \sigma_k]}\}$ (Eq.~(\ref{eq:tJzconservedPBC})) for OBC or PBC respectively. 
However, it is convenient to work in terms of orthogonal IoMs $\{Q_\alpha\}$ for the $t-J_z$ model studied in App.~\ref{app:tJzcommutantorth} (see Eqs.~(\ref{eq:worddefn}) and (\ref{eq:worddefnPBC}) for OBC and PBC respectively), and use the Mazur bound of Eq.~(\ref{eq:mazurbound}).
Hence we explore the different contributions to the Mazur bound $M^{(\ell)}_{Z_j}$, $M^{(r)}_{Z_j}$, and $M^{(\textrm{obc})}_{Z_j}$, defined as
\begin{gather}
    M^{(\ell)}_{Z_j} \defn \sumal{l = 1}{L}{\frac{\opbraket{Z_j}{\hq^{(\ell)}_l}^2}{\opbraket{\hq^{(\ell)}_l}{\hq^{(\ell)}_l}}},\;\;M^{(r)}_{Z_j} \defn \sumal{l = 1}{L}{\frac{\opbraket{Z_j}{\hq^{(r)}_l}^2}{\opbraket{\hq^{(r)}_l}{\hq^{(r)}_l}}}\nn \\
    M^{(\textrm{obc})}_{Z_j} \defn \sumal{\alpha = 1}{2^{L+1}-1}{\frac{\opbraket{Z_j}{Q_\alpha}^2}{\opbraket{Q_\alpha}{Q_\alpha}}},
\label{eq:mazurqtys}
\end{gather}
which are the Mazur bounds for the autocorrelation functions of local spin operators $\{Z_j\}$ obtained by considering the left SLIOMs $\{\hq^{(\ell)}_l\}$, the right SLIOMs $\{\hq^{(r)}_l\}$, and the full commutant $\Cobc$ respectively. 
Exact expressions for the bounds $M^{(\ell)}_{Z_j}$ and $M^{(r)}_{Z_j}$ can be obtained by working in an orthogonal basis for the commutant algebra discussed in App.~\ref{subsec:SLIOMMazur}, and they read (see Eq.~(\ref{eq:mazursliom}))
\begin{gather}
    M^{(\ell)}_{Z_j} =  \sumal{l = 1}{L}{\frac{\langle Z_j \hq^{(\ell)}_l\rangle^2}{\langle \hq^{(\ell)}_l \hq^{(\ell)}_l\rangle}} = \frac{1}{3^{2j}}\sumal{l = 1}{j}{\frac{2^l \binom{j -1}{l-1}^2}{\sumal{\alpha = l}{L}{\frac{1}{3^\alpha}\binom{\alpha - 1}{l-1}}}}, \nn \\
    M^{(r)}_{Z_j} = \sumal{l = 1}{L}{\frac{\langle Z_j \hq^{(r)}_l\rangle^2}{\langle \hq^{(r)}_l \hq^{(r)}_l\rangle}} = M^{(\ell)}_{Z_{L-j+1}}. %\frac{1}{3^{2(L-j+1)}}\sumal{l = 1}{L-j+1}{\frac{2^l \binom{L - j}{l-1}^2}{\sumal{\alpha = l}{L}{\frac{1}{3^\alpha}\binom{\alpha - 1}{l-1}}}}. 
\label{eq:mazursliommain}
\end{gather}
Note that the bound $M^{(\ell)}_{Z_j}$ corresponding to the left SLIOMs is precisely equal to the bound derived in Eq.~(8) of Ref.~\cite{sliom2020}. 
As pointed out there, the expression for the bound $M^{(\ell)}_1$ corresponding to the autocorrelation function of the edge spin operator $Z_1$ simplifies,
\begin{equation}
    M^{(\ell)}_1 = \frac{1}{3^2}\times \frac{2}{\sumal{\alpha = 1}{L}{3^{-\alpha}}} = \frac{4}{9\left(1 - 3^{-L}\right)}, 
\label{eq:mazuredge}
\end{equation}
which shows that the autocorrelation function for the edge spin equilibriates to a constant value. 
However, while we physically expect true Mazur bounds to be invariant under a reflection of the chain ($j \leftrightarrow L - j + 1$), neither of $M^{(\ell)}_{Z_j}$ or $M^{(r)}_{Z_j}$ are, which suggests that these are not tight Mazur bounds.\footnote{Note that $M^{(\ell)}_{Z_j} + M^{(r)}_{Z_j}$, while reflection-invariant, is not a valid Mazur bound since the left and right SLIOMs are not orthogonal to each other (see discussion in Sec.~\ref{subsec:Mazurdefn}).}
In Fig.~\ref{fig:tJzmazur}, we plot the Mazur bounds $M^{(\ell)}_{Z_j}$ and $M^{(r)}_{Z_j}$ as a function of $j$ for a fixed system size $L$. 
It clearly shows that the bounds obtained using the left and right SLIOMs are different although they are conserved quantities for the same family of models.  
We then compute the Mazur bound $M^{(\textrm{obc})}_{Z_j}$ using all the conserved quantities in the full commutant algebra $\Cobc$. 
As shown in App.~\ref{subsec:commutantMazur}, the exact expression reads (see Eq.~(\ref{eq:mazurboundfull}))
\begin{equation}
M^{(\textrm{obc})}_{Z_j} =  \sumal{\alpha = 0}{j-1}{\ \sumal{\beta = 0}{L-j}{\ \frac{2^{\alpha + \beta + 1}\binom{j-1}{\alpha}^2 \binom{L-j}{\beta}^2}{3^L \binom{L}{\alpha + \beta + 1}}}}.
\label{eq:mazurboundfullmain}
\end{equation}
This bound has several advantages compared to the bounds $M^{(\ell)}_{Z_j}$ and $M^{(r)}_{Z_j}$ obtained using only the SLIOMs. 
First, the expression is invariant under a reflection of the chain (i.e., $M^{(\textrm{obc})}_{Z_j} = M^{(\textrm{obc})}_{L-j + 1}$) consistent with physical expectations.  
Second, the expression also allows for a saddle point approximation for large $L$ in the bulk of the chain, and the continuum approximation of $M^{(\textrm{obc})}_{Z_j}$ reads
\begin{equation}
     M^{(\textrm{obc})}(x) =  \frac{1}{3}\sqrt{\frac{2}{\pi L x(1-x)}} \sim \frac{1}{\sqrt{L}}\;\;\textrm{for large}\;L, 
\label{eq:mazursaddlemain}
\end{equation}
where $x \defn j/L$, and $0 < x < 1$. 
Eq.~(\ref{eq:mazursaddlemain}) proves the numerical observations in Ref.~\cite{sliom2020} that the spin autocorrelation functions decay as $~1/\sqrt{L}$ in the bulk of the chain. 
In Fig.~\ref{fig:tJzmazur}, we plot the three Mazur bounds of Eq.~(\ref{eq:mazurqtys}) as a function of $j$. 
As evident there, the left and right SLIOMs only accurately capture the Mazur bound close to the left and right edges of the chain respectively, whereas using the full commutant yields a better bound everywhere else in the chain. 
The Mazur bound for the spin operators $Z_j$ computed using the full commutant $\Cobc$ of the OBC $t-J_z$ model can be generalized to the PBC $t-J_z$ model with the commutant $\Cpbc$.
As shown in App.~\ref{subsec:PBCcommutantMazur}, an exact expression for the corresponding Mazur bound reads (see Eq.~(\ref{eq:mazurboundPBCfull}))
\begin{equation}
    M^{(\textrm{pbc})}_{Z_j} \defn \sumal{\alpha}{}{\frac{\opbraket{Z_j}{Q_\alpha}^2}{\opbraket{Q_\alpha}{Q_\alpha}}} = \frac{2}{3L} + \left(1 - \frac{1}{L}\right)\left(\frac{2}{3}\right)^L,
\label{eq:PBCmazur}
\end{equation}
where $\{Q_\alpha\}$ is an orthogonal basis for the commutant $\Cpbc$. 
Unlike the OBC case, the knowledge of the commutant algebra in the PBC does not drastically enhance the Mazur bound $M^{(U(1))}_{Z_j}$ of Eq.~(\ref{eq:U1mazurtJz}), since the second term in Eq.~(\ref{eq:PBCmazur}) decays exponentially with system size.
However, considering the Mazur bound for the operator $Z_j Z_{j+1}$, we find a different scenario. 
In App.~\ref{subsec:PBCcommutantMazur}, we compute the Mazur bound $M^{(\textrm{pbc})}_{Z_j Z_{j+1}}$ for this operator using the full commutant algebra $\Cpbc$, and we obtain (see Eq.~(\ref{eq:ZZmazurpbc}))
\begin{equation}
    M^{(\textrm{pbc})}_{Z_j Z_{j+1}} = \frac{8}{27 (L-1)} - \frac{4}{27 L (L-1)} + \mathcal{O}(e^{-L}) \sim \frac{1}{L}.
\label{eq:ZZpbc}
\end{equation}
Hence $M^{(\textrm{pbc})}_{Z_j Z_{j+1}}$ is clearly larger than $M^{(U(1))}_{Z_j Z_{j+1}}$ of Eq.~(\ref{eq:U1mazurZZ}), the bound obtained by only considering the part of the commutant generated by the $U(1)$ conserved quantities. 
Hence fragmentation in the PBC $t-J_z$ model leads to a larger saturation of the autocorrelation of appropriate operators, in spite of the absence of SLIOMs~\cite{sliom2020}. 
\subsubsection{Spin-1 dipole-conserving model}\label{subsubsec:dipolemazur}
We now apply Mazur bounds to the autocorrelation function of the on-site spin operator (which we denote here by $Z_j$) in the spin-1 dipole-conserving model of Eq.~(\ref{eq:spin1dip}). 
As discussed in Sec.~\ref{subsec:invariantstrings}, since the dipole-conserving model exhibits classical fragmentation, an orthogonal basis for the commutant algebra $\mC_{\dip}$ is simply the projectors onto Krylov subspaces $\Pi_{\mK_\alpha}$.
We can express $\Pi_{\mK_\alpha} = \sumal{\beta}{}{\ket{\psi_{\alpha\beta}}\bra{\psi_{\alpha\beta}}}$, where $\ket{\psi_{\alpha\beta}}$'s are the product states that span $\mK_\alpha$.
The detailed structure of states in the Krylov subspaces is discussed in App.~\ref{app:canonicalconfigs}.
Using Eq.~(\ref{eq:mazurbound}), the Mazur bound for the autocorrelation function of $Z_j$ using the conserved quantities $\{\Pi_{\mK_\alpha}\}$ reads
\begin{gather}
    M^{(\textrm{dip})}_{Z_j} = \frac{1}{3^L}\sumal{\alpha}{}{\frac{[\Tr(Z_j \Pi_{\mK_\alpha})]^2}{\Tr(\Pi_{\mK_\alpha}^2)}} = \sumal{\alpha}{}{\frac{(Z^{(\mK_\alpha)}_j)^2}{3^L D_{\mK_\alpha}}}, \nn \\
    Z^{(\mK_\alpha)}_j \defn \sumal{\beta = 1}{D_{\mK_\alpha}}{\bra{\psi_{\alpha\beta}} Z_j \ket{\psi_{\alpha\beta}}},
\label{eq:dipfullmazur}
\end{gather}
where $Z^{(\mK_\alpha)}_j$ is the sum of the spin value on site $j$ in all the product states in the Krylov subspace $\mK_\alpha$.
As discussed in Sec.~\ref{subsec:invariantstrings}, models conserving dipole moment have certain special subspaces that contain ``blockades" that dynamically disconnect different parts of the system. 
The contribution of a large class of such blockaded subspaces to the Mazur bound of the local spin operator $Z_j$ can be exactly evaluated, as we discuss below.
This also reveals their crucial role in ``localization" of local $Z_j$ operator (i.e., the saturation of the autocorrelation function to a finite value) numerically observed in Refs.~\cite{pai2018localization, sala2020fragmentation, khemani2020localization}, and analytically argued for in Ref.~\cite{sliom2020}.
Restricting ourselves to the Krylov subspaces $\{\mK_{j, \alpha}\}$ where site $j$ is part of a frozen region and the non-fluctuating spin $Z_j = \pm 1$ (since the contribution of such subspaces with $Z_j = 0$ vanishes), we obtain $Z^{(\mK_{j, \alpha})}_j = \pm D_{\mK_{j, \alpha}}$, hence the full Mazur bound of Eq.~(\ref{eq:dipfullmazur}) can be lower bounded by the blockaded subspace contribution as
\begin{gather}
    M^{(\textrm{dip})}_{Z_j} \geq M^{(\textrm{block})}_{Z_j} = \sumal{\alpha}{}{\frac{(Z^{(\mK_{j, \alpha})}_j)^2}{3^L D_{\mK_{j, \alpha}}}} = \frac{1}{3^L}\sumal{\alpha}{}{D_{\mK_{j, \alpha}}}. 
\label{eq:dipblockmazur}
\end{gather}
Hence it is sufficient to simply compute the total dimension of such blockaded Krylov subspaces. 
We also note that the RHS of Eq.~(\ref{eq:dipblockmazur}) is simply the probability that the site $j$ is frozen.
Equipped with the detailed understanding of the Krylov subspaces in $H_{\dip}$, we are able to compute the total dimension of such subspaces in App.~\ref{app:dipolemazur}.
To do so, we exploit the decoupling of all the states in the blockaded Krylov subspaces into $\ket{\psi} = \ket{L^{(\ell)}} \otimes \ket{B^{(b)}} \otimes \ket{R^{(r)}}$, hence we are able to easily count the dimension of all subspaces with a given blockade configuration $\ket{B^{(b)}}$.
We then sum over contributions from inequivalent choices of the blockade configurations that avoid overcounting, and in the thermodynamic limit we obtain (see Eq.~(\ref{eq:blockadeexact}))
\begin{equation}
    M^{(\textrm{block})}_{Z_j} = \frac{2}{15} \approx 0.1333.
\label{eq:blockadeexactmain}
\end{equation}
We have also verified that this is precisely the probability that the site $j$ is frozen in the thermodynamic limit, which can also be computed using a slightly different approach.
While this partial bound is consistent with results presented in Ref.~\cite{sala2020fragmentation}, it does not fully capture the Mazur bound numerically computed there, which is $\sim 0.24$. 
We believe this is due to significant contributions from the subspaces where $j$ is a part of a small ``active region" sandwiched between two blockades,  and such Krylov subspaces
are not included in Eq.~(\ref{eq:blockadeexactmain}). 
In particular, we have in mind subspaces that comprise of states such as 
\begin{equation}
    \overset{\hspace{6mm}j}{\ket{\cdots\ \fbox{++}\ 0 + 0\ \fbox{++}\ \cdots}},\;\;\;\overset{\hspace{6mm}j}{\ket{\cdots\ \fbox{++}\ + - +\ \fbox{++}\ \cdots}},   
\end{equation} 
where the configurations within the boxes are blockades (ensured by additional requirements on the states marked with dots, see discussion in Sec.~\ref{sec:dipole}). 
The quantity $Z^{(\mK_\alpha)}_j \neq 0$ within such subspaces (e.g., $Z^{(\mK_\alpha)}_j = D_{\mK_{\alpha}}/2$ in the above specific example), and there are exponentially many such subspaces (with combined dimension being a finite fraction of the total Hilbert space dimension), which
can lead to a large contribution according to Eq.~(\ref{eq:dipfullmazur}). 
Although we do not attempt it here, the contributions from those subspaces can in principle  be computed using techniques similar to those used in App.~\ref{app:dipolemazur}.  
\subsubsection{PF models}\label{subsubsec:PFMazur}
\begin{figure}[t!]
\includegraphics[scale = 1]{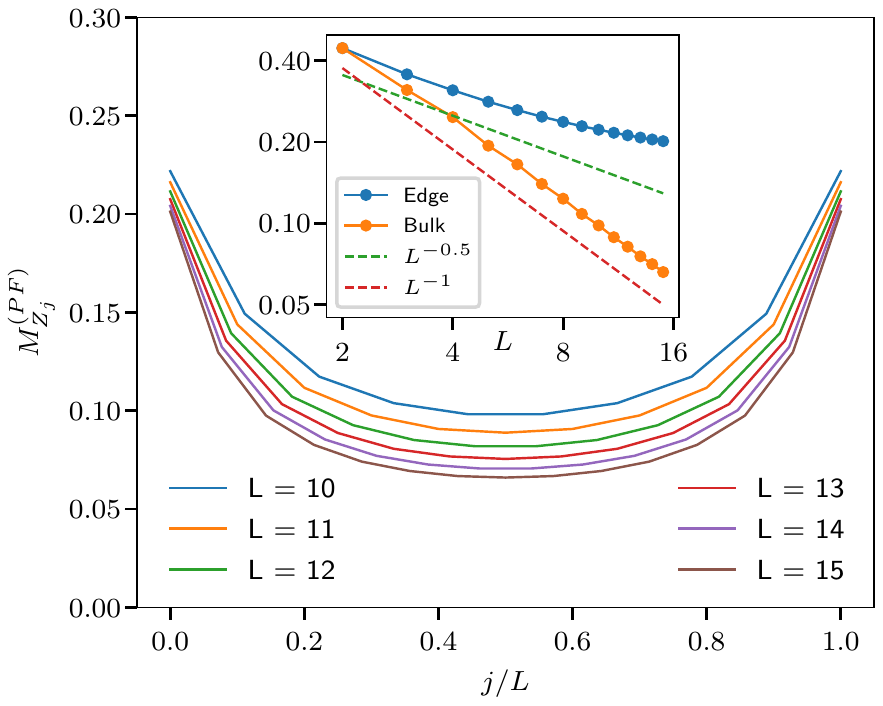}
\caption{(Color online) Mazur bounds $M^{(\textrm{PF})}_{Z_j}$ for the on-site spin operator $Z_j$ in the OBC PF model of Eq.~(\ref{eq:PFHamil}) for $m = 3$.
Fragmentation leads to a non-uniform profile of the Mazur bound across the chain, and apparent localization close to the edges of the chain.
(Inset) Log-log plot of the scaling of $M^{(\textrm{PF})}_{Z_j}$ on the edge ($j = 1$) and in the bulk ($j = L/2$) as a function of system size $L$, along with lines that depict $\sim 1/\sqrt{L}$ and $\sim 1/L$ scalings. 
While the boundary appears to saturate to $\sim 0.2$, the bulk decays as $\sim 1/L$, suggesting a qualitatively different behavior than the OBC $t-J_z$ model.
These results are also valid lower bounds for the Mazur bound in the OBC TL models of Eq.~(\ref{eq:TLHamil}), which are at least as fragmented as the PF model, see Sec.~\ref{subsubsec:TLMazur} for a discussion.
}
\label{fig:PFMazur}
\end{figure}
Note that we were only able to analytically compute the Mazur bounds in the $t-J_z$ model and the spin-1 dipole-conserving model since we were able to explicitly construct a manageable
orthogonal basis (a subset in the latter case) for the operators in the commutant algebra.
In the PF and TL models, although we have explicit expressions for all the operators in the commutant, they are not orthogonal to each other and nor are we able to analytically compute their correlation matrix in Eq.~(\ref{eq:mazurdefns}).
In the PF model, the projectors onto the Krylov subspaces would be a choice of an orthogonal basis of the commutant, but we have not been able to use them to obtain analytical results for the Mazur bounds.
Nevertheless, these projectors are rather simple to construct numerically in the PF model, where the fragmentation happens in the product state basis. 
We can then use these projectors to numerically compute the exact Mazur bound $M^{(\textrm{PF})}_{Z_j}$ for the on-site spin operator $Z_j$ using an expression similar to Eq.~(\ref{eq:dipfullmazur}).
Note that this is equivalent to computing the Mazur bounds of Eq.~(\ref{eq:mazurboundnotorth}) using the IoMs of Eq.~(\ref{eq:stringops}), although constructing and inverting the correlation matrix is computationally more expensive. 
In Fig.~\ref{fig:PFMazur}, we plot this bound $M^{(\textrm{PF})}_{Z_j}$ in the OBC PF model with $m = 3$ as a function of $j$ for various system sizes $L$, where we can view this as a spin-1 system with standard definition $Z_j \equiv \ketbra{+} - \ketbra{-}$.
Similar to the bound in the $t-J_z$ model shown in Fig.~\ref{fig:tJzmazur}, we find that the fragmentation in the PF model leads to a spatial dependence of the bound $M^{(\textrm{PF})}_{Z_j}$, with higher bounds closer to the edges of the chain.
As shown in the inset of Fig.~\ref{fig:PFMazur}, the Mazur bound for the edge spin at large system sizes appears to approach a non-zero value of $\sim 0.2$, suggesting localization at the edges similar to the $t-J_z$ model, although to a smaller numerical value. 
The Mazur bound in the bulk of the chain appears to be rather accurately $\approx 1/L$ at large system sizes, somewhat larger than the total contribution of the two $U(1)$ symmetries of the $m = 3$ PF model, which can be shown to be $2/(3L)$ (similar to Eq.~(\ref{eq:U1mazurtJz})). 
This is qualitatively different from the bulk Mazur bounds in the OBC $t-J_z$ model, where the nature of the scaling with system size is enhanced to $\sim 1/\sqrt{L}$, see Eq.~(\ref{eq:mazursaddlemain}).  
These results are consistent with the expectation that the dynamics of ``dots" in the PF model is significantly different from the spins in the $t-J_z$ model, as discussed in Sec.~\ref{subsec:PFcommutant}; however, we do not have a good physical understanding of the observed qualitative behavior and leave this as an open question.
(The difference in the dynamics is likely less important for the edge spin operator, since in the PF model inside a Krylov subspace only the very first dot---whose state does not vary---can ``visit'' the edge site $j=1$, just like only the very first spin can visit the edge site in the $t-J_z$ model, which appears to be sufficient to develop non-zero autocorrelation.)
\subsubsection{TL models}\label{subsubsec:TLMazur}
The Mazur bound of $M^{(\textrm{PF})}_{Z_j}$ is also a lower bound for the Mazur bound $M^{(\textrm{TL})}_{Z_j}$ of the on-site spin operator in the TL models of Eq.~(\ref{eq:TLHamil}), since all the IoMs of the PF model are also IoMs of the TL models, as discussed in Sec.~\ref{subsec:TLcommutant}.
Hence, we expect edge localization of the spin operator in the TL models as well.
However, a more precise calculation or numerics is more challenging since the Krylov subspaces are specified in terms of a non-orthogonal dimer basis, and computing projectors onto the Krylov subspaces involves an explicit computation of overlap matrix between the dimer basis states and its inverse.   
Furthermore, since the commutant $\mC_{TL}$ is non-Abelian, the projectors onto the Krylov subspaces do not span the full commutant, rather they only span its maximal Abelian subalgebra. 
That is, given two degenerate Krylov subspaces $\mK_\alpha$ and $\mK_{\alpha'}$ (here labeled by different wavefunctions on the ``dots"),
we can chose orthogonal bases $\{\ket{\psi_{\alpha\beta}}\}$ and $\{\ket{\psi_{\alpha'\beta}}\}$ for them such that the operators $\{\he_{j,j+1}\}$ have identical matrix forms in these bases.
Hence, operators such as $\Pi_{\mK_\alpha\mK_{\alpha'}} \defn \sum_\beta{\ket{\psi_{\alpha\beta}}\bra{\psi_{\alpha'\beta}}}$ for degenerate Krylov subspaces $\mK_\alpha$ and $\mK_{\alpha'}$ are also part of the commutant, in addition to projectors onto the Krylov subspaces, which read $\Pi_{\mK_\alpha} \defn \sum_\beta{\ket{\psi_{\alpha\beta}}\bra{\psi_{\alpha\beta}}}$.
This suggests that the expression for the Mazur bound for operators such as $Z_j$ also involves computing matrix elements such as $\bra{\psi_{\alpha\beta}} Z_j \ket{\psi_{\alpha'\beta'}}$ between basis states of different Krylov subspaces, which further complicates the computation. 
We have hence not been able to perform analytical Mazur bound calculations for spin operators in the TL model, and this leaves open the question of whether the bounds shown in Fig.~\ref{fig:PFMazur} are further enhanced in the TL models that lead to qualitatively different behavior, and we defer this study for future work.
Nevertheless, we are able to analytically obtain such an exact Mazur bound for the autocorrelation function of the edge energy operator $\he_{1,2}$.
We achieve this by utilizing the fact that the energy terms act within Krylov subspaces and have simple expressions in the ``dots and dimers'' representation, cf.~Eqs.~(\ref{eq:ejeigenstate})-(\ref{eq:ejscattering}).
Specifically, we focus on local energy operators $\he_{j,j+1}$ and note that it is sufficient to consider the  projectors $\{\Pi_{\mK_\alpha}\}$ in the computation of the Mazur bound since $\Tr(\he_{j,j+1}  \Pi_{\mK_\alpha \mK_{\alpha'}} ) = 0$ for two degenerate Krylov subspaces $\mK_\alpha$ and $\mK_{\alpha'}$, where $\Pi_{\mK_\alpha}$ and $\Pi_{\mK_\alpha \mK_{\alpha'}}$ are defined in the previous paragraph.
Hence, we arrive at an analog of Eq.~(\ref{eq:dipfullmazur}) 
\begin{align}
M^{(\textrm{TL})}_{\he_{j,j+1}} &= \frac{1}{m^L} \sum_\alpha \frac{[\Tr(\he_{j,j+1} \Pi_{\mK_\alpha})]^2}{\Tr(\Pi_{\mK_\alpha}^2)} \nn\\
&= \frac{m^2}{m^L} \sum_\alpha \frac{[N_{\mK_\alpha}^{\text{dimer@}(j,j+1)}]^2}{D_{\mK_\alpha}},
\label{eq:Mej}
\end{align}
where $\alpha$ runs over distinct Krylov subspaces labeled by the number of dots and distinct $\ket{\psi_\text{dots}}$'s, cf.~Sec.~\ref{subsec:TLfragmentation}, $D_{\mK_\alpha}$ is the dimension of the Krylov subspace $\mK_\alpha$, and $N_{\mK_\alpha}^{\text{dimer@}(j,j+1)}$ is the number of configurations in $\mK_\alpha$ where the sites $j$ and $j+1$ are connected by a dimer. 
In Eq.~(\ref{eq:Mej}), we have in mind a Krylov subspace $\mK_\alpha$ that is spanned by a number of distinct pictures of dots and dimers, and we can associate a basis vector with each such picture.
While this basis is not orthogonal, we can still use it to evaluate the trace, which simplifies here since $\he_{j,j+1}$ acting on any such picture produces a single other such picture, according to Eqs.~(\ref{eq:ejeigenstate}) and (\ref{eq:ejscattering}).
However, the output and input pictures coincide only if the sites $j,j+1$ are connected by a dimer, and the amplitude is $m$ in each such case, cf.~Eq.~(\ref{eq:ejeigenstate}).
Each such case gives a nonzero diagonal matrix element that enters $\Tr(\he_{j,j+1} \Pi_{\mK_\alpha})$, which can then be evaluated to $m$ times $N^{\text{dimer@}(j,j+1)}_{\mK_\alpha}$, the number of pictures where the sites $j,j+1$ form a dimer.\footnote{Note that the above simple diagonal matrix elements in the matrix representation of $\he_{j,j+1}$ in this basis are not equal to expectation values in the basis states since the basis is not orthogonal.}
Furthermore, since $\he_{j,j+1}$ is not traceless, i.e., has non-zero expectation value at infinite temperature, we are only interested in its connected autocorrelation function, and the corresponding ``connected" Mazur bound is given by
\begin{equation}
M^\text{conn}_{\he_{j,j+1}} \defn M^{(\textrm{TL})}_{\he_{j,j+1}} - (\langle \he_{j,j+1} \rangle_{T=\infty})^2 = M^{(\textrm{TL})}_{\he_{j,j+1}} - \frac{1}{m^2} ~.
\label{eq:Mconndefn}
\end{equation}
While we do not know how to count $N^{\text{dimer@}(j,j+1)}_{\mK_\alpha}$ in Eq.~(\ref{eq:Mej}) for general $j$, as we show in App.~\ref{app:TLmazur}, the counting simplifies at the edge for $j = 1$. 
This allows us to obtain the exact result for the connected Mazur bound at the edge as (see Eq.~(\ref{eq:Mconnapp}))
\begin{equation}
M^\text{conn}_{\he_{1,2}} = \frac{1}{L-1} %\left[
\left(
1 - \frac{4}{m^2} + \frac{6}{m^2 L} %\right] 
\right)
~.
\label{eq:Mconnmain}
\end{equation}
For $m>2$, the Mazur bound $M^{\text{conn}}_{\he_{1,2}}$ decays as $\sim 1/L$, a slower decay than $\sim 1/L^2$ expected from the global $SU(m)$ symmetry (this follows from arguments similar to two-site operators in the case of $SU(2)$ discussed in Sec.~\ref{subsec:Mazurconventional}, since the bond energy is a two-site operator).
Hence, the autocorrelation function of the edge energy operator provides a clear dynamical signature of the Hilbert space fragmentation in the TL models.
Note that for $m=2$, Eq.~(\ref{eq:Mconnmain}) simplifies to $3/[2L(L-1)] \sim L^{-2}$, since the $m=2$ TL chain only has a global $SU(2)$ symmetry and does not exhibit fragmentation [this formula agrees with the result for $M_{S_j^\alpha S_k^\alpha}$ for $SU(2)$-symmetric spin-1/2 case discussed in Sec.~\ref{subsec:Mazurconventional}, see~Eq.~(\ref{eq:U1mazurbound}), since $U \he_{j,j+1} U^\dagger = 1/2 - 2 \vec{S}_j \cdot \vec{S}_{j+1}$ where $U$ is a sublattice rotation similar to Eq.~(\ref{eq:biquadratic}).].
Finally, we note that despite the enhancement of the Mazur bound for the edge energy operator in the TL ($m > 2$) models,  it does not exhibit ``localization," in contrast to the edge spin operator.
On the other hand, in the $t-J_z$ model, we expect the edge energy operators to be also ``localized," while we have not shown this explicitly in Sec.~\ref{subsubsec:tJzMazur}, this can be checked for various Hamiltonian terms in Eq.~(\ref{eq:tJzhamil}) taken near the edge.
We think that the physics of this difference between the TL and $t-J_z$ model is that there is a symmetry distinction between local energy and local spin operators in the former, while there is no such distinction in the latter, e.g., the local field term in the $t-J_z$ Hamiltonian in Eq.~(\ref{eq:tJzhamil}) is proportional to the spin operator $Z_j$.
While the symmetry distinction in the TL model is clear already from the $SU(m)$ symmetry, it would be interesting to better understand the interplay with the fragmentation phenomenon, and if there may be other more subtle distinctions among various observables.
\subsubsection{Contribution of frozen eigenstates}
Note that in all of the Mazur bounds discussed in fragmented systems, the contribution of the frozen eigenstates is exponentially small in the system size, and this can be shown on general grounds.  
Indeed, denoting the normalized frozen states by $\{\ket{\psi_f}\}$, the projectors onto them $\{\ket{\psi_f}\bra{\psi_f}\}$ are  mutually orthogonal elements of the commutant algebra.
For any Hermitian operator $\hO$, their Mazur bound contribution is given by
\begin{equation}
    M^{\textrm{(frozen)}}_{\hO} = \frac{1}{D}\sumal{f}{}{\bra{\psi_f} \hO \ket{\psi_f}^2} \leq  \frac{N_f \norm{\hO}^2 }{D},
\label{eq:Mazurfrozen}
\end{equation}
where $N_f$ is the number of frozen states, $\norm{\hO}$ is the operator norm of $\hO$, and $D$ is the Hilbert space dimension.
Restricting to fragmented systems and strictly local operators, we have $D \sim d^L$, $N_f \sim \phi^L$ with some $\phi < d$, and $\norm{\hO} \sim \mathcal{O}(1)$, hence $M^{\textrm{(frozen)}}_{\hO} \sim (\phi/d)^L$, which is exponentially small.
This argument holds irrespective of spatial dimension or the particular system, hence we do not expect frozen eigenstates to play any significant role in the enhancement of the Mazur bounds.
\section{Summary and Outlook}
\label{sec:conclusion}
In this work, we studied dynamical properties of families of Hamiltonians using their \textit{commutant algebra}, the algebra of all operators (IoMs) that commute with the entire family of Hamiltonians. 
Focusing on Hamiltonians of the form of Eq.~(\ref{eq:genhamil}), the commutant algebra is the centralizer of the \textit{bond algebra} generated by the local terms of the Hamiltonian, i.e., the algebra of operators that commute with each term of the Hamiltonian.
The bond and commutant algebra language can be used to clearly define IoMs of a family of Hamiltonians, since there is no clear definition of IoMs for a single Hamiltonian in a finite-dimensional Hilbert space (any Hamiltonian has exponentially many conserved quantities -- the eigenstate projectors).
In Sec.~\ref{sec:commutants}, we illustrated this for Hamiltonians with conventional symmetries such as $U(1)$ or $SU(2)$. 
Further, we showed that the number of dynamically disconnected Krylov subspaces of a family of Hamiltonians scales in the same way with system size as the dimension of the commutant [i.e., distinguishing exponential vs polynomial scaling,  see precise Eq.~(\ref{eq:Krylovbound})], hence allowing us to naturally classify systems into various categories based on the character of this scaling.
As we summarize in Tab.~\ref{tab:scalings}, this dimension is either constant or scales polynomially with system size in systems exhibiting conventional discrete or continuous global symmetries.  
On the other hand, in systems exhibiting Hilbert space fragmentation, this dimension scales exponentially with system size, and this provides a precise definition of Hilbert space fragmentation, something that has been absent in the literature so far. 
Hence systems with Hilbert space fragmentation can be thought to lie somewhere between systems with conventional symmetries and completely solvable systems.
Furthermore, the fragmentation is a property of the bond algebra, and the same Krylov subspaces exist for any Hamiltonian or Floquet circuit built using the operators in this algebra.   
We illustrate this definition by constructing the complete basis for the commutant algebra in several models exhibiting fragmentation -- the $t-J_z$, Pair-Flip (PF), Temperley-Lieb (TL), and spin-1 dipole-conserving models in Secs.~\ref{sec:tJz}, \ref{sec:pairflip}, \ref{sec:TL}, and \ref{sec:dipole} respectively, and we show that the dimension of the commutant scales exponentially with system size.  
The commutant algebra formalism also clearly distinguishes between two types fragmentation, depending on whether the fragmentation occurs in the product state basis or in an entangled basis, which we refer to as ``classical" and ``quantum" fragmentation respectively. 
In the former case, all operators in the commutant are diagonal in the product state basis, and this type of fragmentation can also occur in constrained classical systems with the same set of transition rules~\cite{morningstar2020kineticallyconstrained}, where the phenomenon is an example of ``reducibility"~\cite{ritort2003glassy}. 
Three of the examples we discuss exhibit classical fragmentation, the $t-J_z$, PF, and dipole-conserving models, while the TL models (which includes the well-known spin-1 biquadratic model) exhibit quantum fragmentation. 
Using this language, we also clarify the distinction between the full commutant and the minimal set of operators required to uniquely label its Krylov subspaces, which helps us understand the relation to SLIOMs in certain fragmented systems~\cite{sliom2020}.
Specifically, the Krylov subspaces of any system can be uniquely labeled by the eigenvalues under the minimal set of generators of the maximal Abelian subalgebra of the commutant.
In systems with conventional symmetries, this minimal set consists of a few simple conserved quantities [such as $S^z_{\tot}$ for the $U(1)$ symmetry or $\vec{S}_{\tot}^2$, $S^z_{\tot}$ for the $SU(2)$ symmetry], hence the language of commutant algebra is not necessary.
In certain systems exhibiting fragmentation, non-local operators referred to as SLIOMs~\cite{sliom2020} form one such minimal generating set, although there are multiple such choices of non-local operators.
However, in all the models we study, the full commutant algebra is easier to construct than the SLIOMs, and the expressions or the existence of the latter is not evident in many systems, e.g., the PF and TL models.
Finally, in Sec.~\ref{sec:Mazur}, we studied the contribution of the full commutant algebra to the Mazur bounds for the autocorrelation function of local operators, both in the case of conventional symmetries as well as fragmented systems.
In the case of conventional symmetries, it is typically sufficient to consider the contribution of the minimal set of local conserved quantities that generate the full commutant, although there can be additional (perhaps subleading) contributions from the non-local operators in the commutant algebra.
In all the fragmented systems we discuss, we found significant enhancement of the Mazur bounds due to the contribution of the exponentially many operators in the commutant.
Focusing on the Mazur bounds for the on-site spin operator in various systems: (i) In the $t-J_z$ model, we obtain exact results everywhere in the system, analytically recovering the numerical results of Ref.~\cite{sliom2020}. 
(ii) In the spin-1 dipole-conserving system with OBC, we analytically show the spin localization in the bulk of the system.
(iii) In the PF and TL models, we numerically illustrate the behavior of the Mazur bound enhancement everywhere in the chain, finding localization near the edge but qualitatively weaker enhancement in the bulk compared to the $t - J_z$ model (the TL model results are only lower-bounded by the PF ones and are not definitive for the bulk behavior).
In addition, in the $t-J_z$ model with PBC, we show the enhancement of the Mazur bound for two-site operators, going beyond results in Ref.~\cite{sliom2020}.
Also, in the TL model, we analytically show the enhancement of the Mazur bound for the edge energy operator.
We believe that our exact results on the commutant of these systems will also be useful in understanding other dynamical phenomena in fragmented systems, such as the fracton Casimir effect in dipole-conserving models~\cite{feng2021hilbert}.
The presented Mazur bound calculations also highlight two aspects of fragmentation.
First, we show that the SLIOMs do not accurately capture the full Mazur bound in fragmented systems (explicitly demonstrated in the OBC $t-J_z$ model), and our results suggest that their existence is not necessary for the enhancement of Mazur bounds (e.g., in the PBC $t-J_z$ model).
This points to the need of going beyond constructing the minimal set of generators such as the SLIOMs in fragmented systems.
Second, while the commutant often contains exponentially many IoMs that are simply the projectors onto frozen eigenstates that are product states, their contribution to the Mazur bound is exponentially small, hence the frozen eigenstates play a minimal role in the enhanced Mazur bounds.
This shows that such inert/frozen eigenstates, which are the focus of a lot of the literature on fragmentation, might not be the most important aspect of fragmentation. 
On the other hand, as shown in the spin-1 dipole-conserving model, blockades that dynamically disconnect parts of the system are crucial for operator localization.
Looking forward, it would be interesting to find new examples of classical and quantum fragmentation that fit into this language, and apply and verify that this formalism applies to the several other examples of Hilbert space fragmentation in the literature~\cite{detomasi2019dynamics, yang2019hilbertspace, lee2021frustration, hahn2021information, langlett2021hilbert, mukherjee2021minimal, mukherjee2021constraint, khudorozhkov2021hilbert, richter2021anomalous}.
An efficient and systematic procedure of determining the commutant corresponding to a family of systems, possibly along the lines of existing works on conserved quantity detection~\cite{kim2015slowest,lin2017explicit, chertkov2020engineering}, would be useful in looking for fragmentation, and might also help in finding the associated non-obvious IoMs. 
These examples can also be used to better understand the finer classification of dynamical phenomena within each of the coarse classes defined by the dimension of the commutant in Tab.~\ref{tab:scalings}. 
For example, even within systems with fragmentation, there could be further distinctions in the properties of the Krylov subspaces.
Some of these properties such as weak vs strong fragmentation can be understood in terms of the full spectrum of the numbers $\{D_\lambda\}$ and $\{d_\lambda\}$ in Eq.~(\ref{eq:Hilbertdecomp}) (as opposed to the dimension of the commutant which only involves $\{d_\lambda\}$).  
On the other hand, there are properties such as Mazur bound saturation or operator localization that require a detailed knowledge of the operators in the commutant.  
In addition, there are systems that exhibit similar features as fragmentation but with polynomially many Krylov subspaces, as shown in a recent work~\cite{richter2021anomalous} on the Motzkin spin chain.
Although such systems would not be fragmented in our definition, it is important to precisely understand how they differ from systems with conventional continuous global symmetries.\footnote{The Krylov subspaces in the Motzkin chain work Ref.~\cite{richter2021anomalous} can be labeled by two conserved numbers, $N_( - N_)$ and $N_( + N_) - 2N_{(\cdot)}$, where `$($' stands for the `$u$' state and `$)$' for the `$d$' state in their notation, while $N_{(\cdot)}$ counts the number of syntactically matched pairs of parentheses in the product state basis. The former can be thought of as a conventional $U(1)$ symmetry, while the latter is a non-local $U(1)$ symmetry.}
While we have focused on families of Hamiltonians of the form of Eq.~(\ref{eq:genhamil}), where the bond algebra is generated from strictly local terms, a minimal generalization consists of adding a few operators that are the sums of strictly local operators (e.g., a uniform magnetic field) to the generators of the bond algebra.
As we will discuss in an upcoming work~\cite{scarsinprep}, studying such generalized algebras yields insights into many examples of quantum many-body scars.
It would also be interesting to study more general families of Hamiltonians with additional symmetries (e.g., translation, lattice symmetries), or symmetries in other types of systems such as open quantum systems~\cite{albert2014symmetries, buca2012note, essler2020integrability, robertson2021exact} or quantum maps~\cite{backer2003numerical, esposti2005quantum, moudgalya2019operator}, and explore whether such systems can be fruitfully understood in this language of commutant algebras.
Further, throughout this work we have restricted ourselves to finite Hilbert space dimensions, where the bond and commutant algebras are essentially algebras of matrices of the form of Eq.~(\ref{eq:matrixreps}).
It is important to make this formalism more rigorous in the thermodynamic limit, where many subtle issues such as the topology of operator space arise.
One way of working with the conserved quantities directly in the thermodynamic limit might utilize their simple Matrix Product Operator representations shown in App.~\ref{app:MPOforms}.
Finally, while the commutant algebra language demystifies many aspects of symmetries and fragmentation, it also warrants a closer examination of a number of fundamental questions.
While the scaling of the dimension of the commutant with system size in Table~\ref{tab:scalings} can be defined to be the distinction between conventional discrete/continuous global symmetries and fragmentation, one still needs to verify if all symmetries considered ``conventional" (e.g., Lie group symmetries) obey this property. 
Conventional symmetries are typically unitary operators with an on-site action, which leads to conserved quantities that are local operators. 
If this is the case, do all families of systems with \textit{only} a finite number of on-site symmetries have a commutant with dimension that scales polynomially with system size?
Addressing such questions is also important to provide a precise definition of ergodicity and its breaking in quantum many-body systems. 
Ergodicity in a quantum Hamiltonian is usually tested by studying the properties of the spectrum within quantum number sectors after resolving the symmetries of the system, where typically only the well-known conventional symmetries are considered~\cite{rigol2008thermalization}. 
However, the commutant algebra language treats the conventional conserved quantities and the non-local conserved quantities on equal footing, and it is not a priori clear which symmetries should be resolved~\cite{mondainicomment2018, shiraishimorireply2018} before testing for ergodicity. 
In spite of the fact that fragmented systems usually have large Krylov subspaces that thermalize, they are considered to be examples of ergodicity breaking due to the many Krylov subspaces of small dimension that remain after resolving only the conventional symmetries (and also the very existence of the Krylov subspaces not labeled by the conventional symmetries is usually referred to as ETH-breaking, e.g., being responsible for unconventional thermalization dynamics from simple initial states).
On the other hand, conventional symmetries themselves also possess some number of subspaces of small dimension [e.g., the ferromagnetic multiplet in $SU(2)$ symmetric systems], which are not considered to be ergodicity breaking.
This discrepancy in terminology opens up the question of how to precisely define ergodicity breaking, and whether local on-site conserved quantities are fundamentally the only ones that need to be ``resolved."
We defer detailed explorations of these fundamental questions for future work. 
\section*{Acknowledgements}
We thank Berislav Buca, Anushya Chandran, Arpit Dua, Duncan Haldane, Paul Fendley, Chris Laumann, Cheng-Ju Lin, Daniel Mark, Alan Morningstar, Tibor Rakovszky, Pablo Sala, and Shivaji Sondhi for useful discussions.
S.M. thanks Abhinav Prem, Rahul Nandkishore, Nicolas Regnault and B. Andrei Bernevig for related previous collaborations. 
This work was supported by the Walter Burke Institute for Theoretical Physics at Caltech; the Institute for Quantum Information and Matter, an NSF Physics Frontiers Center (NSF Grant PHY-1733907); and the National Science Foundation through grant DMR-2001186.
A part of this work was completed at the Aspen Center for Physics, which is supported by National Science Foundation grant PHY-1607611.
This work was partially supported by a grant from the Simons Foundation.
\bibliography{refs}

\begin{thebibliography}{114}%
\makeatletter
\providecommand \@ifxundefined [1]{%
 \@ifx{#1\undefined}
}%
\providecommand \@ifnum [1]{%
 \ifnum #1\expandafter \@firstoftwo
 \else \expandafter \@secondoftwo
 \fi
}%
\providecommand \@ifx [1]{%
 \ifx #1\expandafter \@firstoftwo
 \else \expandafter \@secondoftwo
 \fi
}%
\providecommand \natexlab [1]{#1}%
\providecommand \enquote  [1]{``#1''}%
\providecommand \bibnamefont  [1]{#1}%
\providecommand \bibfnamefont [1]{#1}%
\providecommand \citenamefont [1]{#1}%
\providecommand \href@noop [0]{\@secondoftwo}%
\providecommand \href [0]{\begingroup \@sanitize@url \@href}%
\providecommand \@href[1]{\@@startlink{#1}\@@href}%
\providecommand \@@href[1]{\endgroup#1\@@endlink}%
\providecommand \@sanitize@url [0]{\catcode `\\12\catcode `\$12\catcode
  `\&12\catcode `\#12\catcode `\^12\catcode `\_12\catcode `\%12\relax}%
\providecommand \@@startlink[1]{}%
\providecommand \@@endlink[0]{}%
\providecommand \url  [0]{\begingroup\@sanitize@url \@url }%
\providecommand \@url [1]{\endgroup\@href {#1}{\urlprefix }}%
\providecommand \urlprefix  [0]{URL }%
\providecommand \Eprint [0]{\href }%
\providecommand \doibase [0]{https://doi.org/}%
\providecommand \selectlanguage [0]{\@gobble}%
\providecommand \bibinfo  [0]{\@secondoftwo}%
\providecommand \bibfield  [0]{\@secondoftwo}%
\providecommand \translation [1]{[#1]}%
\providecommand \BibitemOpen [0]{}%
\providecommand \bibitemStop [0]{}%
\providecommand \bibitemNoStop [0]{.\EOS\space}%
\providecommand \EOS [0]{\spacefactor3000\relax}%
\providecommand \BibitemShut  [1]{\csname bibitem#1\endcsname}%
\let\auto@bib@innerbib\@empty
%</preamble>
\bibitem [{\citenamefont {Srednicki}(1994)}]{srednicki1994chaos}%
  \BibitemOpen
  \bibfield  {author} {\bibinfo {author} {\bibfnamefont {M.}~\bibnamefont
  {Srednicki}},\ }\bibfield  {title} {\bibinfo {title} {Chaos and quantum
  thermalization},\ }\href {https://doi.org/10.1103/PhysRevE.50.888} {\bibfield
   {journal} {\bibinfo  {journal} {Phys. Rev. E}\ }\textbf {\bibinfo {volume}
  {50}},\ \bibinfo {pages} {888} (\bibinfo {year} {1994})}\BibitemShut
  {NoStop}%
\bibitem [{\citenamefont {Deutsch}(1991)}]{deutsch1991quantum}%
  \BibitemOpen
  \bibfield  {author} {\bibinfo {author} {\bibfnamefont {J.~M.}\ \bibnamefont
  {Deutsch}},\ }\bibfield  {title} {\bibinfo {title} {Quantum statistical
  mechanics in a closed system},\ }\href
  {https://doi.org/10.1103/PhysRevA.43.2046} {\bibfield  {journal} {\bibinfo
  {journal} {Phys. Rev. A}\ }\textbf {\bibinfo {volume} {43}},\ \bibinfo
  {pages} {2046} (\bibinfo {year} {1991})}\BibitemShut {NoStop}%
\bibitem [{\citenamefont {Shiraishi}\ and\ \citenamefont
  {Mori}(2018)}]{shiraishimorireply2018}%
  \BibitemOpen
  \bibfield  {author} {\bibinfo {author} {\bibfnamefont {N.}~\bibnamefont
  {Shiraishi}}\ and\ \bibinfo {author} {\bibfnamefont {T.}~\bibnamefont
  {Mori}},\ }\bibfield  {title} {\bibinfo {title} {Shiraishi and {M}ori
  reply},\ }\href {https://doi.org/10.1103/PhysRevLett.121.038902} {\bibfield
  {journal} {\bibinfo  {journal} {Phys. Rev. Lett.}\ }\textbf {\bibinfo
  {volume} {121}},\ \bibinfo {pages} {038902} (\bibinfo {year}
  {2018})}\BibitemShut {NoStop}%
\bibitem [{\citenamefont {{Rigol}}\ \emph {et~al.}(2008)\citenamefont
  {{Rigol}}, \citenamefont {{Dunjko}},\ and\ \citenamefont
  {{Olshanii}}}]{rigol2008thermalization}%
  \BibitemOpen
  \bibfield  {author} {\bibinfo {author} {\bibfnamefont {M.}~\bibnamefont
  {{Rigol}}}, \bibinfo {author} {\bibfnamefont {V.}~\bibnamefont {{Dunjko}}},\
  and\ \bibinfo {author} {\bibfnamefont {M.}~\bibnamefont {{Olshanii}}},\
  }\bibfield  {title} {\bibinfo {title} {{Thermalization and its mechanism for
  generic isolated quantum systems}},\ }\href
  {https://doi.org/10.1038/nature06838} {\bibfield  {journal} {\bibinfo
  {journal} {Nature}\ }\textbf {\bibinfo {volume} {452}},\ \bibinfo {pages}
  {854} (\bibinfo {year} {2008})}\BibitemShut {NoStop}%
\bibitem [{\citenamefont {Polkovnikov}\ \emph {et~al.}(2011)\citenamefont
  {Polkovnikov}, \citenamefont {Sengupta}, \citenamefont {Silva},\ and\
  \citenamefont {Vengalattore}}]{polkovnikov2011colloquium}%
  \BibitemOpen
  \bibfield  {author} {\bibinfo {author} {\bibfnamefont {A.}~\bibnamefont
  {Polkovnikov}}, \bibinfo {author} {\bibfnamefont {K.}~\bibnamefont
  {Sengupta}}, \bibinfo {author} {\bibfnamefont {A.}~\bibnamefont {Silva}},\
  and\ \bibinfo {author} {\bibfnamefont {M.}~\bibnamefont {Vengalattore}},\
  }\bibfield  {title} {\bibinfo {title} {Colloquium: Nonequilibrium dynamics of
  closed interacting quantum systems},\ }\href
  {https://doi.org/10.1103/RevModPhys.83.863} {\bibfield  {journal} {\bibinfo
  {journal} {Reviews of Modern Physics}\ }\textbf {\bibinfo {volume} {83}},\
  \bibinfo {pages} {863} (\bibinfo {year} {2011})}\BibitemShut {NoStop}%
\bibitem [{\citenamefont {{D'Alessio}}\ \emph {et~al.}(2016)\citenamefont
  {{D'Alessio}}, \citenamefont {{Kafri}}, \citenamefont {{Polkovnikov}},\ and\
  \citenamefont {{Rigol}}}]{d2016quantum}%
  \BibitemOpen
  \bibfield  {author} {\bibinfo {author} {\bibfnamefont {L.}~\bibnamefont
  {{D'Alessio}}}, \bibinfo {author} {\bibfnamefont {Y.}~\bibnamefont
  {{Kafri}}}, \bibinfo {author} {\bibfnamefont {A.}~\bibnamefont
  {{Polkovnikov}}},\ and\ \bibinfo {author} {\bibfnamefont {M.}~\bibnamefont
  {{Rigol}}},\ }\bibfield  {title} {\bibinfo {title} {{From quantum chaos and
  eigenstate thermalization to statistical mechanics and thermodynamics}},\
  }\href {https://doi.org/10.1080/00018732.2016.1198134} {\bibfield  {journal}
  {\bibinfo  {journal} {Advances in Physics}\ }\textbf {\bibinfo {volume}
  {65}},\ \bibinfo {pages} {239} (\bibinfo {year} {2016})}\BibitemShut
  {NoStop}%
\bibitem [{\citenamefont {\ifmmode~\check{S}\else \v{S}\fi{}untajs}\ \emph
  {et~al.}(2020)\citenamefont {\ifmmode~\check{S}\else \v{S}\fi{}untajs},
  \citenamefont {Bon\ifmmode~\check{c}\else \v{c}\fi{}a}, \citenamefont
  {Prosen},\ and\ \citenamefont {Vidmar}}]{suntajs2020quantum}%
  \BibitemOpen
  \bibfield  {author} {\bibinfo {author} {\bibfnamefont {J.}~\bibnamefont
  {\ifmmode~\check{S}\else \v{S}\fi{}untajs}}, \bibinfo {author} {\bibfnamefont
  {J.}~\bibnamefont {Bon\ifmmode~\check{c}\else \v{c}\fi{}a}}, \bibinfo
  {author} {\bibfnamefont {T.}~\bibnamefont {Prosen}},\ and\ \bibinfo {author}
  {\bibfnamefont {L.}~\bibnamefont {Vidmar}},\ }\bibfield  {title} {\bibinfo
  {title} {Quantum chaos challenges many-body localization},\ }\href
  {https://doi.org/10.1103/PhysRevE.102.062144} {\bibfield  {journal} {\bibinfo
   {journal} {Phys. Rev. E}\ }\textbf {\bibinfo {volume} {102}},\ \bibinfo
  {pages} {062144} (\bibinfo {year} {2020})}\BibitemShut {NoStop}%
\bibitem [{\citenamefont {Abanin}\ \emph {et~al.}(2021)\citenamefont {Abanin},
  \citenamefont {Bardarson}, \citenamefont {{De Tomasi}}, \citenamefont
  {Gopalakrishnan}, \citenamefont {Khemani}, \citenamefont {Parameswaran},
  \citenamefont {Pollmann}, \citenamefont {Potter}, \citenamefont {Serbyn},\
  and\ \citenamefont {Vasseur}}]{abanin2021distinguishing}%
  \BibitemOpen
  \bibfield  {author} {\bibinfo {author} {\bibfnamefont {D.}~\bibnamefont
  {Abanin}}, \bibinfo {author} {\bibfnamefont {J.}~\bibnamefont {Bardarson}},
  \bibinfo {author} {\bibfnamefont {G.}~\bibnamefont {{De Tomasi}}}, \bibinfo
  {author} {\bibfnamefont {S.}~\bibnamefont {Gopalakrishnan}}, \bibinfo
  {author} {\bibfnamefont {V.}~\bibnamefont {Khemani}}, \bibinfo {author}
  {\bibfnamefont {S.}~\bibnamefont {Parameswaran}}, \bibinfo {author}
  {\bibfnamefont {F.}~\bibnamefont {Pollmann}}, \bibinfo {author}
  {\bibfnamefont {A.}~\bibnamefont {Potter}}, \bibinfo {author} {\bibfnamefont
  {M.}~\bibnamefont {Serbyn}},\ and\ \bibinfo {author} {\bibfnamefont
  {R.}~\bibnamefont {Vasseur}},\ }\bibfield  {title} {\bibinfo {title}
  {Distinguishing localization from chaos: Challenges in finite-size systems},\
  }\href {https://doi.org/https://doi.org/10.1016/j.aop.2021.168415} {\bibfield
   {journal} {\bibinfo  {journal} {Annals of Physics}\ }\textbf {\bibinfo
  {volume} {427}},\ \bibinfo {pages} {168415} (\bibinfo {year}
  {2021})}\BibitemShut {NoStop}%
\bibitem [{\citenamefont {{Bernien}}\ \emph {et~al.}(2017)\citenamefont
  {{Bernien}}, \citenamefont {{Schwartz}}, \citenamefont {{Keesling}},
  \citenamefont {{Levine}}, \citenamefont {{Omran}}, \citenamefont {{Pichler}},
  \citenamefont {{Choi}}, \citenamefont {{Zibrov}}, \citenamefont {{Endres}},
  \citenamefont {{Greiner}}, \citenamefont {{Vuleti{\'c}}},\ and\ \citenamefont
  {{Lukin}}}]{bernien2017probing}%
  \BibitemOpen
  \bibfield  {author} {\bibinfo {author} {\bibfnamefont {H.}~\bibnamefont
  {{Bernien}}}, \bibinfo {author} {\bibfnamefont {S.}~\bibnamefont
  {{Schwartz}}}, \bibinfo {author} {\bibfnamefont {A.}~\bibnamefont
  {{Keesling}}}, \bibinfo {author} {\bibfnamefont {H.}~\bibnamefont
  {{Levine}}}, \bibinfo {author} {\bibfnamefont {A.}~\bibnamefont {{Omran}}},
  \bibinfo {author} {\bibfnamefont {H.}~\bibnamefont {{Pichler}}}, \bibinfo
  {author} {\bibfnamefont {S.}~\bibnamefont {{Choi}}}, \bibinfo {author}
  {\bibfnamefont {A.~S.}\ \bibnamefont {{Zibrov}}}, \bibinfo {author}
  {\bibfnamefont {M.}~\bibnamefont {{Endres}}}, \bibinfo {author}
  {\bibfnamefont {M.}~\bibnamefont {{Greiner}}}, \bibinfo {author}
  {\bibfnamefont {V.}~\bibnamefont {{Vuleti{\'c}}}},\ and\ \bibinfo {author}
  {\bibfnamefont {M.~D.}\ \bibnamefont {{Lukin}}},\ }\bibfield  {title}
  {\bibinfo {title} {{Probing many-body dynamics on a 51-atom quantum
  simulator}},\ }\href {https://doi.org/10.1038/nature24622} {\bibfield
  {journal} {\bibinfo  {journal} {Nature}\ }\textbf {\bibinfo {volume} {551}},\
  \bibinfo {pages} {579} (\bibinfo {year} {2017})}\BibitemShut {NoStop}%
\bibitem [{\citenamefont {Turner}\ \emph {et~al.}(2018)\citenamefont {Turner},
  \citenamefont {Michailidis}, \citenamefont {Abanin}, \citenamefont {Serbyn},\
  and\ \citenamefont {Papic}}]{turner2017quantum}%
  \BibitemOpen
  \bibfield  {author} {\bibinfo {author} {\bibfnamefont {C.}~\bibnamefont
  {Turner}}, \bibinfo {author} {\bibfnamefont {A.}~\bibnamefont {Michailidis}},
  \bibinfo {author} {\bibfnamefont {D.}~\bibnamefont {Abanin}}, \bibinfo
  {author} {\bibfnamefont {M.}~\bibnamefont {Serbyn}},\ and\ \bibinfo {author}
  {\bibfnamefont {Z.}~\bibnamefont {Papic}},\ }\bibfield  {title} {\bibinfo
  {title} {Weak ergodicity breaking from quantum many-body scars},\ }\href
  {https://doi.org/10.1038/s41567-018-0137-5} {\bibfield  {journal} {\bibinfo
  {journal} {Nature Physics}\ }\textbf {\bibinfo {volume} {14}},\ \bibinfo
  {pages} {745} (\bibinfo {year} {2018})}\BibitemShut {NoStop}%
\bibitem [{\citenamefont {Serbyn}\ \emph {et~al.}(2021)\citenamefont {Serbyn},
  \citenamefont {Abanin},\ and\ \citenamefont
  {Papi{\'{c}}}}]{serbyn2020review}%
  \BibitemOpen
  \bibfield  {author} {\bibinfo {author} {\bibfnamefont {M.}~\bibnamefont
  {Serbyn}}, \bibinfo {author} {\bibfnamefont {D.~A.}\ \bibnamefont {Abanin}},\
  and\ \bibinfo {author} {\bibfnamefont {Z.}~\bibnamefont {Papi{\'{c}}}},\
  }\bibfield  {title} {\bibinfo {title} {Quantum many-body scars and weak
  breaking of ergodicity},\ }\href {https://doi.org/10.1038/s41567-021-01230-2}
  {\bibfield  {journal} {\bibinfo  {journal} {Nature Physics}\ }\textbf
  {\bibinfo {volume} {17}},\ \bibinfo {pages} {675} (\bibinfo {year}
  {2021})}\BibitemShut {NoStop}%
\bibitem [{\citenamefont {Moudgalya}\ \emph
  {et~al.}(2018{\natexlab{a}})\citenamefont {Moudgalya}, \citenamefont
  {Rachel}, \citenamefont {Bernevig},\ and\ \citenamefont
  {Regnault}}]{moudgalya2018a}%
  \BibitemOpen
  \bibfield  {author} {\bibinfo {author} {\bibfnamefont {S.}~\bibnamefont
  {Moudgalya}}, \bibinfo {author} {\bibfnamefont {S.}~\bibnamefont {Rachel}},
  \bibinfo {author} {\bibfnamefont {B.~A.}\ \bibnamefont {Bernevig}},\ and\
  \bibinfo {author} {\bibfnamefont {N.}~\bibnamefont {Regnault}},\ }\bibfield
  {title} {\bibinfo {title} {Exact excited states of nonintegrable models},\
  }\href {https://doi.org/10.1103/PhysRevB.98.235155} {\bibfield  {journal}
  {\bibinfo  {journal} {Phys. Rev. B}\ }\textbf {\bibinfo {volume} {98}},\
  \bibinfo {pages} {235155} (\bibinfo {year} {2018}{\natexlab{a}})}\BibitemShut
  {NoStop}%
\bibitem [{\citenamefont {Moudgalya}\ \emph
  {et~al.}(2018{\natexlab{b}})\citenamefont {Moudgalya}, \citenamefont
  {Regnault},\ and\ \citenamefont {Bernevig}}]{moudgalya2018b}%
  \BibitemOpen
  \bibfield  {author} {\bibinfo {author} {\bibfnamefont {S.}~\bibnamefont
  {Moudgalya}}, \bibinfo {author} {\bibfnamefont {N.}~\bibnamefont
  {Regnault}},\ and\ \bibinfo {author} {\bibfnamefont {B.~A.}\ \bibnamefont
  {Bernevig}},\ }\bibfield  {title} {\bibinfo {title} {Entanglement of exact
  excited states of {Affleck}-{Kennedy}-{Lieb}-{Tasaki} models: Exact results,
  many-body scars, and violation of the strong eigenstate thermalization
  hypothesis},\ }\href {https://doi.org/10.1103/PhysRevB.98.235156} {\bibfield
  {journal} {\bibinfo  {journal} {Phys. Rev. B}\ }\textbf {\bibinfo {volume}
  {98}},\ \bibinfo {pages} {235156} (\bibinfo {year}
  {2018}{\natexlab{b}})}\BibitemShut {NoStop}%
\bibitem [{\citenamefont {Choi}\ \emph {et~al.}(2019)\citenamefont {Choi},
  \citenamefont {Turner}, \citenamefont {Pichler}, \citenamefont {Ho},
  \citenamefont {Michailidis}, \citenamefont {Papi\ifmmode~\acute{c}\else
  \'{c}\fi{}}, \citenamefont {Serbyn}, \citenamefont {Lukin},\ and\
  \citenamefont {Abanin}}]{choi2018emergent}%
  \BibitemOpen
  \bibfield  {author} {\bibinfo {author} {\bibfnamefont {S.}~\bibnamefont
  {Choi}}, \bibinfo {author} {\bibfnamefont {C.~J.}\ \bibnamefont {Turner}},
  \bibinfo {author} {\bibfnamefont {H.}~\bibnamefont {Pichler}}, \bibinfo
  {author} {\bibfnamefont {W.~W.}\ \bibnamefont {Ho}}, \bibinfo {author}
  {\bibfnamefont {A.~A.}\ \bibnamefont {Michailidis}}, \bibinfo {author}
  {\bibfnamefont {Z.}~\bibnamefont {Papi\ifmmode~\acute{c}\else \'{c}\fi{}}},
  \bibinfo {author} {\bibfnamefont {M.}~\bibnamefont {Serbyn}}, \bibinfo
  {author} {\bibfnamefont {M.~D.}\ \bibnamefont {Lukin}},\ and\ \bibinfo
  {author} {\bibfnamefont {D.~A.}\ \bibnamefont {Abanin}},\ }\bibfield  {title}
  {\bibinfo {title} {Emergent {SU}(2) dynamics and perfect quantum many-body
  scars},\ }\href {https://doi.org/10.1103/PhysRevLett.122.220603} {\bibfield
  {journal} {\bibinfo  {journal} {Phys. Rev. Lett.}\ }\textbf {\bibinfo
  {volume} {122}},\ \bibinfo {pages} {220603} (\bibinfo {year}
  {2019})}\BibitemShut {NoStop}%
\bibitem [{\citenamefont {Schecter}\ and\ \citenamefont
  {Iadecola}(2019)}]{schecter2019weak}%
  \BibitemOpen
  \bibfield  {author} {\bibinfo {author} {\bibfnamefont {M.}~\bibnamefont
  {Schecter}}\ and\ \bibinfo {author} {\bibfnamefont {T.}~\bibnamefont
  {Iadecola}},\ }\bibfield  {title} {\bibinfo {title} {Weak ergodicity breaking
  and quantum many-body scars in spin-1 {XY} magnets},\ }\href
  {https://doi.org/10.1103/PhysRevLett.123.147201} {\bibfield  {journal}
  {\bibinfo  {journal} {Phys. Rev. Lett.}\ }\textbf {\bibinfo {volume} {123}},\
  \bibinfo {pages} {147201} (\bibinfo {year} {2019})}\BibitemShut {NoStop}%
\bibitem [{\citenamefont {Iadecola}\ and\ \citenamefont
  {Schecter}(2020)}]{iadecola2020quantum}%
  \BibitemOpen
  \bibfield  {author} {\bibinfo {author} {\bibfnamefont {T.}~\bibnamefont
  {Iadecola}}\ and\ \bibinfo {author} {\bibfnamefont {M.}~\bibnamefont
  {Schecter}},\ }\bibfield  {title} {\bibinfo {title} {Quantum many-body scar
  states with emergent kinetic constraints and finite-entanglement revivals},\
  }\href {https://doi.org/10.1103/PhysRevB.101.024306} {\bibfield  {journal}
  {\bibinfo  {journal} {Phys. Rev. B}\ }\textbf {\bibinfo {volume} {101}},\
  \bibinfo {pages} {024306} (\bibinfo {year} {2020})}\BibitemShut {NoStop}%
\bibitem [{\citenamefont {Shibata}\ \emph {et~al.}(2020)\citenamefont
  {Shibata}, \citenamefont {Yoshioka},\ and\ \citenamefont
  {Katsura}}]{shibata2020onsager}%
  \BibitemOpen
  \bibfield  {author} {\bibinfo {author} {\bibfnamefont {N.}~\bibnamefont
  {Shibata}}, \bibinfo {author} {\bibfnamefont {N.}~\bibnamefont {Yoshioka}},\
  and\ \bibinfo {author} {\bibfnamefont {H.}~\bibnamefont {Katsura}},\
  }\bibfield  {title} {\bibinfo {title} {Onsager's scars in disordered spin
  chains},\ }\href {https://doi.org/10.1103/PhysRevLett.124.180604} {\bibfield
  {journal} {\bibinfo  {journal} {Phys. Rev. Lett.}\ }\textbf {\bibinfo
  {volume} {124}},\ \bibinfo {pages} {180604} (\bibinfo {year}
  {2020})}\BibitemShut {NoStop}%
\bibitem [{\citenamefont {Mark}\ \emph {et~al.}(2020)\citenamefont {Mark},
  \citenamefont {Lin},\ and\ \citenamefont {Motrunich}}]{mark2020unified}%
  \BibitemOpen
  \bibfield  {author} {\bibinfo {author} {\bibfnamefont {D.~K.}\ \bibnamefont
  {Mark}}, \bibinfo {author} {\bibfnamefont {C.-J.}\ \bibnamefont {Lin}},\ and\
  \bibinfo {author} {\bibfnamefont {O.~I.}\ \bibnamefont {Motrunich}},\
  }\bibfield  {title} {\bibinfo {title} {Unified structure for exact towers of
  scar states in the {A}ffleck-{K}ennedy-{L}ieb-{T}asaki and other models},\
  }\href {https://doi.org/10.1103/PhysRevB.101.195131} {\bibfield  {journal}
  {\bibinfo  {journal} {Phys. Rev. B}\ }\textbf {\bibinfo {volume} {101}},\
  \bibinfo {pages} {195131} (\bibinfo {year} {2020})}\BibitemShut {NoStop}%
\bibitem [{\citenamefont {Moudgalya}\ \emph
  {et~al.}(2020{\natexlab{a}})\citenamefont {Moudgalya}, \citenamefont
  {O'Brien}, \citenamefont {Bernevig}, \citenamefont {Fendley},\ and\
  \citenamefont {Regnault}}]{moudgalya2020large}%
  \BibitemOpen
  \bibfield  {author} {\bibinfo {author} {\bibfnamefont {S.}~\bibnamefont
  {Moudgalya}}, \bibinfo {author} {\bibfnamefont {E.}~\bibnamefont {O'Brien}},
  \bibinfo {author} {\bibfnamefont {B.~A.}\ \bibnamefont {Bernevig}}, \bibinfo
  {author} {\bibfnamefont {P.}~\bibnamefont {Fendley}},\ and\ \bibinfo {author}
  {\bibfnamefont {N.}~\bibnamefont {Regnault}},\ }\bibfield  {title} {\bibinfo
  {title} {Large classes of quantum scarred {H}amiltonians from matrix product
  states},\ }\href {https://doi.org/10.1103/PhysRevB.102.085120} {\bibfield
  {journal} {\bibinfo  {journal} {Phys. Rev. B}\ }\textbf {\bibinfo {volume}
  {102}},\ \bibinfo {pages} {085120} (\bibinfo {year}
  {2020}{\natexlab{a}})}\BibitemShut {NoStop}%
\bibitem [{\citenamefont {Moudgalya}\ \emph
  {et~al.}(2020{\natexlab{b}})\citenamefont {Moudgalya}, \citenamefont
  {Regnault},\ and\ \citenamefont {Bernevig}}]{moudgalya2020eta}%
  \BibitemOpen
  \bibfield  {author} {\bibinfo {author} {\bibfnamefont {S.}~\bibnamefont
  {Moudgalya}}, \bibinfo {author} {\bibfnamefont {N.}~\bibnamefont
  {Regnault}},\ and\ \bibinfo {author} {\bibfnamefont {B.~A.}\ \bibnamefont
  {Bernevig}},\ }\bibfield  {title} {\bibinfo {title}
  {$\ensuremath{\eta}$-pairing in {H}ubbard models: From spectrum generating
  algebras to quantum many-body scars},\ }\href
  {https://doi.org/10.1103/PhysRevB.102.085140} {\bibfield  {journal} {\bibinfo
   {journal} {Phys. Rev. B}\ }\textbf {\bibinfo {volume} {102}},\ \bibinfo
  {pages} {085140} (\bibinfo {year} {2020}{\natexlab{b}})}\BibitemShut
  {NoStop}%
\bibitem [{\citenamefont {Mark}\ and\ \citenamefont
  {Motrunich}(2020)}]{mark2020eta}%
  \BibitemOpen
  \bibfield  {author} {\bibinfo {author} {\bibfnamefont {D.~K.}\ \bibnamefont
  {Mark}}\ and\ \bibinfo {author} {\bibfnamefont {O.~I.}\ \bibnamefont
  {Motrunich}},\ }\bibfield  {title} {\bibinfo {title}
  {$\ensuremath{\eta}$-pairing states as true scars in an extended hubbard
  model},\ }\href {https://doi.org/10.1103/PhysRevB.102.075132} {\bibfield
  {journal} {\bibinfo  {journal} {Phys. Rev. B}\ }\textbf {\bibinfo {volume}
  {102}},\ \bibinfo {pages} {075132} (\bibinfo {year} {2020})}\BibitemShut
  {NoStop}%
\bibitem [{\citenamefont {Pakrouski}\ \emph {et~al.}(2020)\citenamefont
  {Pakrouski}, \citenamefont {Pallegar}, \citenamefont {Popov},\ and\
  \citenamefont {Klebanov}}]{pakrouski2020many}%
  \BibitemOpen
  \bibfield  {author} {\bibinfo {author} {\bibfnamefont {K.}~\bibnamefont
  {Pakrouski}}, \bibinfo {author} {\bibfnamefont {P.~N.}\ \bibnamefont
  {Pallegar}}, \bibinfo {author} {\bibfnamefont {F.~K.}\ \bibnamefont
  {Popov}},\ and\ \bibinfo {author} {\bibfnamefont {I.~R.}\ \bibnamefont
  {Klebanov}},\ }\bibfield  {title} {\bibinfo {title} {Many-body scars as a
  group invariant sector of {H}ilbert space},\ }\href
  {https://doi.org/10.1103/PhysRevLett.125.230602} {\bibfield  {journal}
  {\bibinfo  {journal} {Phys. Rev. Lett.}\ }\textbf {\bibinfo {volume} {125}},\
  \bibinfo {pages} {230602} (\bibinfo {year} {2020})}\BibitemShut {NoStop}%
\bibitem [{\citenamefont {Ren}\ \emph {et~al.}(2021)\citenamefont {Ren},
  \citenamefont {Liang},\ and\ \citenamefont {Fang}}]{ren2020quasisymmetry}%
  \BibitemOpen
  \bibfield  {author} {\bibinfo {author} {\bibfnamefont {J.}~\bibnamefont
  {Ren}}, \bibinfo {author} {\bibfnamefont {C.}~\bibnamefont {Liang}},\ and\
  \bibinfo {author} {\bibfnamefont {C.}~\bibnamefont {Fang}},\ }\bibfield
  {title} {\bibinfo {title} {Quasisymmetry groups and many-body scar
  dynamics},\ }\href {https://doi.org/10.1103/PhysRevLett.126.120604}
  {\bibfield  {journal} {\bibinfo  {journal} {Phys. Rev. Lett.}\ }\textbf
  {\bibinfo {volume} {126}},\ \bibinfo {pages} {120604} (\bibinfo {year}
  {2021})}\BibitemShut {NoStop}%
\bibitem [{\citenamefont {O'Dea}\ \emph {et~al.}(2020)\citenamefont {O'Dea},
  \citenamefont {Burnell}, \citenamefont {Chandran},\ and\ \citenamefont
  {Khemani}}]{odea2020from}%
  \BibitemOpen
  \bibfield  {author} {\bibinfo {author} {\bibfnamefont {N.}~\bibnamefont
  {O'Dea}}, \bibinfo {author} {\bibfnamefont {F.}~\bibnamefont {Burnell}},
  \bibinfo {author} {\bibfnamefont {A.}~\bibnamefont {Chandran}},\ and\
  \bibinfo {author} {\bibfnamefont {V.}~\bibnamefont {Khemani}},\ }\bibfield
  {title} {\bibinfo {title} {From tunnels to towers: Quantum scars from {L}ie
  algebras and $q$-deformed {L}ie algebras},\ }\href
  {https://doi.org/10.1103/PhysRevResearch.2.043305} {\bibfield  {journal}
  {\bibinfo  {journal} {Phys. Rev. Research}\ }\textbf {\bibinfo {volume}
  {2}},\ \bibinfo {pages} {043305} (\bibinfo {year} {2020})}\BibitemShut
  {NoStop}%
\bibitem [{\citenamefont {Shiraishi}\ and\ \citenamefont
  {Mori}(2017)}]{mori2017eth}%
  \BibitemOpen
  \bibfield  {author} {\bibinfo {author} {\bibfnamefont {N.}~\bibnamefont
  {Shiraishi}}\ and\ \bibinfo {author} {\bibfnamefont {T.}~\bibnamefont
  {Mori}},\ }\bibfield  {title} {\bibinfo {title} {Systematic construction of
  counterexamples to the eigenstate thermalization hypothesis},\ }\href
  {https://doi.org/10.1103/PhysRevLett.119.030601} {\bibfield  {journal}
  {\bibinfo  {journal} {Phys. Rev. Lett.}\ }\textbf {\bibinfo {volume} {119}},\
  \bibinfo {pages} {030601} (\bibinfo {year} {2017})}\BibitemShut {NoStop}%
\bibitem [{\citenamefont {Pai}\ \emph {et~al.}(2019)\citenamefont {Pai},
  \citenamefont {Pretko},\ and\ \citenamefont
  {Nandkishore}}]{pai2018localization}%
  \BibitemOpen
  \bibfield  {author} {\bibinfo {author} {\bibfnamefont {S.}~\bibnamefont
  {Pai}}, \bibinfo {author} {\bibfnamefont {M.}~\bibnamefont {Pretko}},\ and\
  \bibinfo {author} {\bibfnamefont {R.~M.}\ \bibnamefont {Nandkishore}},\
  }\bibfield  {title} {\bibinfo {title} {Localization in fractonic random
  circuits},\ }\href {https://doi.org/10.1103/PhysRevX.9.021003} {\bibfield
  {journal} {\bibinfo  {journal} {Phys. Rev. X}\ }\textbf {\bibinfo {volume}
  {9}},\ \bibinfo {pages} {021003} (\bibinfo {year} {2019})}\BibitemShut
  {NoStop}%
\bibitem [{\citenamefont {Sala}\ \emph {et~al.}(2020)\citenamefont {Sala},
  \citenamefont {Rakovszky}, \citenamefont {Verresen}, \citenamefont {Knap},\
  and\ \citenamefont {Pollmann}}]{sala2020fragmentation}%
  \BibitemOpen
  \bibfield  {author} {\bibinfo {author} {\bibfnamefont {P.}~\bibnamefont
  {Sala}}, \bibinfo {author} {\bibfnamefont {T.}~\bibnamefont {Rakovszky}},
  \bibinfo {author} {\bibfnamefont {R.}~\bibnamefont {Verresen}}, \bibinfo
  {author} {\bibfnamefont {M.}~\bibnamefont {Knap}},\ and\ \bibinfo {author}
  {\bibfnamefont {F.}~\bibnamefont {Pollmann}},\ }\bibfield  {title} {\bibinfo
  {title} {Ergodicity breaking arising from {H}ilbert space fragmentation in
  dipole-conserving {H}amiltonians},\ }\href
  {https://doi.org/10.1103/PhysRevX.10.011047} {\bibfield  {journal} {\bibinfo
  {journal} {Phys. Rev. X}\ }\textbf {\bibinfo {volume} {10}},\ \bibinfo
  {pages} {011047} (\bibinfo {year} {2020})}\BibitemShut {NoStop}%
\bibitem [{\citenamefont {Khemani}\ \emph {et~al.}(2020)\citenamefont
  {Khemani}, \citenamefont {Hermele},\ and\ \citenamefont
  {Nandkishore}}]{khemani2020localization}%
  \BibitemOpen
  \bibfield  {author} {\bibinfo {author} {\bibfnamefont {V.}~\bibnamefont
  {Khemani}}, \bibinfo {author} {\bibfnamefont {M.}~\bibnamefont {Hermele}},\
  and\ \bibinfo {author} {\bibfnamefont {R.}~\bibnamefont {Nandkishore}},\
  }\bibfield  {title} {\bibinfo {title} {Localization from {H}ilbert space
  shattering: From theory to physical realizations},\ }\href
  {https://doi.org/10.1103/PhysRevB.101.174204} {\bibfield  {journal} {\bibinfo
   {journal} {Phys. Rev. B}\ }\textbf {\bibinfo {volume} {101}},\ \bibinfo
  {pages} {174204} (\bibinfo {year} {2020})}\BibitemShut {NoStop}%
\bibitem [{\citenamefont {Ritort}\ and\ \citenamefont
  {Sollich}(2003)}]{ritort2003glassy}%
  \BibitemOpen
  \bibfield  {author} {\bibinfo {author} {\bibfnamefont {F.}~\bibnamefont
  {Ritort}}\ and\ \bibinfo {author} {\bibfnamefont {P.}~\bibnamefont
  {Sollich}},\ }\bibfield  {title} {\bibinfo {title} {Glassy dynamics of
  kinetically constrained models},\ }\href
  {https://doi.org/10.1080/0001873031000093582} {\bibfield  {journal} {\bibinfo
   {journal} {Advances in Physics}\ }\textbf {\bibinfo {volume} {52}},\
  \bibinfo {pages} {219} (\bibinfo {year} {2003})}\BibitemShut {NoStop}%
\bibitem [{\citenamefont {Bergholtz}\ and\ \citenamefont
  {Karlhede}(2005)}]{bergholtz2005half}%
  \BibitemOpen
  \bibfield  {author} {\bibinfo {author} {\bibfnamefont {E.~J.}\ \bibnamefont
  {Bergholtz}}\ and\ \bibinfo {author} {\bibfnamefont {A.}~\bibnamefont
  {Karlhede}},\ }\bibfield  {title} {\bibinfo {title} {Half-filled lowest
  {L}andau level on a thin torus},\ }\href@noop {} {\bibfield  {journal}
  {\bibinfo  {journal} {Phys. Rev. Lett.}\ }\textbf {\bibinfo {volume} {94}},\
  \bibinfo {pages} {026802} (\bibinfo {year} {2005})}\BibitemShut {NoStop}%
\bibitem [{\citenamefont {{Bergholtz}}\ and\ \citenamefont
  {{Karlhede}}(2006)}]{bergholtz2006one}%
  \BibitemOpen
  \bibfield  {author} {\bibinfo {author} {\bibfnamefont {E.~J.}\ \bibnamefont
  {{Bergholtz}}}\ and\ \bibinfo {author} {\bibfnamefont {A.}~\bibnamefont
  {{Karlhede}}},\ }\bibfield  {title} {\bibinfo {title} {{`One-dimensional'
  theory of the quantum {H}all system}},\ }\href
  {https://doi.org/10.1088/1742-5468/2006/04/L04001} {\bibfield  {journal}
  {\bibinfo  {journal} {Journal of Statistical Mechanics: Theory and
  Experiment}\ }\textbf {\bibinfo {volume} {2006}},\ \bibinfo {pages} {L04001}
  (\bibinfo {year} {2006})}\BibitemShut {NoStop}%
\bibitem [{\citenamefont {Olmos}\ \emph {et~al.}(2010)\citenamefont {Olmos},
  \citenamefont {Müller},\ and\ \citenamefont
  {Lesanovsky}}]{olmos2010thermalization}%
  \BibitemOpen
  \bibfield  {author} {\bibinfo {author} {\bibfnamefont {B.}~\bibnamefont
  {Olmos}}, \bibinfo {author} {\bibfnamefont {M.}~\bibnamefont {Müller}},\
  and\ \bibinfo {author} {\bibfnamefont {I.}~\bibnamefont {Lesanovsky}},\
  }\bibfield  {title} {\bibinfo {title} {Thermalization of a strongly
  interacting 1{D} {R}ydberg lattice gas},\ }\href
  {https://doi.org/10.1088/1367-2630/12/1/013024} {\bibfield  {journal}
  {\bibinfo  {journal} {New Journal of Physics}\ }\textbf {\bibinfo {volume}
  {12}},\ \bibinfo {pages} {013024} (\bibinfo {year} {2010})}\BibitemShut
  {NoStop}%
\bibitem [{\citenamefont {Sikora}\ \emph {et~al.}(2011)\citenamefont {Sikora},
  \citenamefont {Shannon}, \citenamefont {Pollmann}, \citenamefont {Penc},\
  and\ \citenamefont {Fulde}}]{sikora2011extended}%
  \BibitemOpen
  \bibfield  {author} {\bibinfo {author} {\bibfnamefont {O.}~\bibnamefont
  {Sikora}}, \bibinfo {author} {\bibfnamefont {N.}~\bibnamefont {Shannon}},
  \bibinfo {author} {\bibfnamefont {F.}~\bibnamefont {Pollmann}}, \bibinfo
  {author} {\bibfnamefont {K.}~\bibnamefont {Penc}},\ and\ \bibinfo {author}
  {\bibfnamefont {P.}~\bibnamefont {Fulde}},\ }\bibfield  {title} {\bibinfo
  {title} {Extended quantum {U}(1)-liquid phase in a three-dimensional quantum
  dimer model},\ }\href {https://doi.org/10.1103/PhysRevB.84.115129} {\bibfield
   {journal} {\bibinfo  {journal} {Phys. Rev. B}\ }\textbf {\bibinfo {volume}
  {84}},\ \bibinfo {pages} {115129} (\bibinfo {year} {2011})}\BibitemShut
  {NoStop}%
\bibitem [{\citenamefont {Nakamura}\ \emph {et~al.}(2012)\citenamefont
  {Nakamura}, \citenamefont {Wang},\ and\ \citenamefont
  {Bergholtz}}]{nakamura2012exactly}%
  \BibitemOpen
  \bibfield  {author} {\bibinfo {author} {\bibfnamefont {M.}~\bibnamefont
  {Nakamura}}, \bibinfo {author} {\bibfnamefont {Z.-Y.}\ \bibnamefont {Wang}},\
  and\ \bibinfo {author} {\bibfnamefont {E.~J.}\ \bibnamefont {Bergholtz}},\
  }\bibfield  {title} {\bibinfo {title} {Exactly solvable fermion chain
  describing a $\ensuremath{\nu}=1/3$ fractional quantum hall state},\ }\href
  {https://doi.org/10.1103/PhysRevLett.109.016401} {\bibfield  {journal}
  {\bibinfo  {journal} {Phys. Rev. Lett.}\ }\textbf {\bibinfo {volume} {109}},\
  \bibinfo {pages} {016401} (\bibinfo {year} {2012})}\BibitemShut {NoStop}%
\bibitem [{\citenamefont {van Horssen}\ \emph {et~al.}(2015)\citenamefont {van
  Horssen}, \citenamefont {Levi},\ and\ \citenamefont
  {Garrahan}}]{horssen2015dynamics}%
  \BibitemOpen
  \bibfield  {author} {\bibinfo {author} {\bibfnamefont {M.}~\bibnamefont {van
  Horssen}}, \bibinfo {author} {\bibfnamefont {E.}~\bibnamefont {Levi}},\ and\
  \bibinfo {author} {\bibfnamefont {J.~P.}\ \bibnamefont {Garrahan}},\
  }\bibfield  {title} {\bibinfo {title} {Dynamics of many-body localization in
  a translation-invariant quantum glass model},\ }\href
  {https://doi.org/10.1103/PhysRevB.92.100305} {\bibfield  {journal} {\bibinfo
  {journal} {Phys. Rev. B}\ }\textbf {\bibinfo {volume} {92}},\ \bibinfo
  {pages} {100305} (\bibinfo {year} {2015})}\BibitemShut {NoStop}%
\bibitem [{\citenamefont {Lan}\ \emph {et~al.}(2018)\citenamefont {Lan},
  \citenamefont {van Horssen}, \citenamefont {Powell},\ and\ \citenamefont
  {Garrahan}}]{lan2018quantum}%
  \BibitemOpen
  \bibfield  {author} {\bibinfo {author} {\bibfnamefont {Z.}~\bibnamefont
  {Lan}}, \bibinfo {author} {\bibfnamefont {M.}~\bibnamefont {van Horssen}},
  \bibinfo {author} {\bibfnamefont {S.}~\bibnamefont {Powell}},\ and\ \bibinfo
  {author} {\bibfnamefont {J.~P.}\ \bibnamefont {Garrahan}},\ }\bibfield
  {title} {\bibinfo {title} {Quantum slow relaxation and metastability due to
  dynamical constraints},\ }\href
  {https://doi.org/10.1103/PhysRevLett.121.040603} {\bibfield  {journal}
  {\bibinfo  {journal} {Phys. Rev. Lett.}\ }\textbf {\bibinfo {volume} {121}},\
  \bibinfo {pages} {040603} (\bibinfo {year} {2018})}\BibitemShut {NoStop}%
\bibitem [{\citenamefont {Gopalakrishnan}\ and\ \citenamefont
  {Zakirov}(2018)}]{gopalakrishnan2018facilitated}%
  \BibitemOpen
  \bibfield  {author} {\bibinfo {author} {\bibfnamefont {S.}~\bibnamefont
  {Gopalakrishnan}}\ and\ \bibinfo {author} {\bibfnamefont {B.}~\bibnamefont
  {Zakirov}},\ }\bibfield  {title} {\bibinfo {title} {Facilitated quantum
  cellular automata as simple models with non-thermal eigenstates and
  dynamics},\ }\href {https://doi.org/10.1088/2058-9565/aad759} {\bibfield
  {journal} {\bibinfo  {journal} {Quantum Science and Technology}\ }\textbf
  {\bibinfo {volume} {3}},\ \bibinfo {pages} {044004} (\bibinfo {year}
  {2018})}\BibitemShut {NoStop}%
\bibitem [{\citenamefont {Moudgalya}\ \emph {et~al.}()\citenamefont
  {Moudgalya}, \citenamefont {Prem}, \citenamefont {Nandkishore}, \citenamefont
  {Regnault},\ and\ \citenamefont {Bernevig}}]{moudgalya2019thermalization}%
  \BibitemOpen
  \bibfield  {author} {\bibinfo {author} {\bibfnamefont {S.}~\bibnamefont
  {Moudgalya}}, \bibinfo {author} {\bibfnamefont {A.}~\bibnamefont {Prem}},
  \bibinfo {author} {\bibfnamefont {R.}~\bibnamefont {Nandkishore}}, \bibinfo
  {author} {\bibfnamefont {N.}~\bibnamefont {Regnault}},\ and\ \bibinfo
  {author} {\bibfnamefont {B.~A.}\ \bibnamefont {Bernevig}},\ }\bibinfo {title}
  {{Thermalization and Its Absence within Krylov Subspaces of a Constrained
  Hamiltonian}},\ in\ \href {https://doi.org/10.1142/9789811231711_0009} {\emph
  {\bibinfo {booktitle} {Memorial Volume for Shoucheng Zhang}}},\
  Chap.~\bibinfo {chapter} {7}, pp.\ \bibinfo {pages} {147--209}\BibitemShut
  {NoStop}%
\bibitem [{\citenamefont {Moudgalya}\ \emph
  {et~al.}(2020{\natexlab{c}})\citenamefont {Moudgalya}, \citenamefont
  {Bernevig},\ and\ \citenamefont {Regnault}}]{moudgalya2020quantum}%
  \BibitemOpen
  \bibfield  {author} {\bibinfo {author} {\bibfnamefont {S.}~\bibnamefont
  {Moudgalya}}, \bibinfo {author} {\bibfnamefont {B.~A.}\ \bibnamefont
  {Bernevig}},\ and\ \bibinfo {author} {\bibfnamefont {N.}~\bibnamefont
  {Regnault}},\ }\bibfield  {title} {\bibinfo {title} {Quantum many-body scars
  in a {L}andau level on a thin torus},\ }\href
  {https://doi.org/10.1103/PhysRevB.102.195150} {\bibfield  {journal} {\bibinfo
   {journal} {Phys. Rev. B}\ }\textbf {\bibinfo {volume} {102}},\ \bibinfo
  {pages} {195150} (\bibinfo {year} {2020}{\natexlab{c}})}\BibitemShut
  {NoStop}%
\bibitem [{\citenamefont {Yang}\ \emph {et~al.}(2020)\citenamefont {Yang},
  \citenamefont {Liu}, \citenamefont {Gorshkov},\ and\ \citenamefont
  {Iadecola}}]{yang2019hilbertspace}%
  \BibitemOpen
  \bibfield  {author} {\bibinfo {author} {\bibfnamefont {Z.-C.}\ \bibnamefont
  {Yang}}, \bibinfo {author} {\bibfnamefont {F.}~\bibnamefont {Liu}}, \bibinfo
  {author} {\bibfnamefont {A.~V.}\ \bibnamefont {Gorshkov}},\ and\ \bibinfo
  {author} {\bibfnamefont {T.}~\bibnamefont {Iadecola}},\ }\bibfield  {title}
  {\bibinfo {title} {Hilbert-space fragmentation from strict confinement},\
  }\href {https://doi.org/10.1103/PhysRevLett.124.207602} {\bibfield  {journal}
  {\bibinfo  {journal} {Phys. Rev. Lett.}\ }\textbf {\bibinfo {volume} {124}},\
  \bibinfo {pages} {207602} (\bibinfo {year} {2020})}\BibitemShut {NoStop}%
\bibitem [{\citenamefont {Hahn}\ \emph {et~al.}(2021)\citenamefont {Hahn},
  \citenamefont {McClarty},\ and\ \citenamefont {Luitz}}]{hahn2021information}%
  \BibitemOpen
  \bibfield  {author} {\bibinfo {author} {\bibfnamefont {D.}~\bibnamefont
  {Hahn}}, \bibinfo {author} {\bibfnamefont {P.~A.}\ \bibnamefont {McClarty}},\
  and\ \bibinfo {author} {\bibfnamefont {D.~J.}\ \bibnamefont {Luitz}},\
  }\href@noop {} {\bibinfo {title} {Information dynamics in a model with
  {H}ilbert space fragmentation}} (\bibinfo {year} {2021}),\ \Eprint
  {https://arxiv.org/abs/2104.00692} {arXiv:2104.00692 [cond-mat.stat-mech]}
  \BibitemShut {NoStop}%
\bibitem [{\citenamefont {De~Tomasi}\ \emph {et~al.}(2019)\citenamefont
  {De~Tomasi}, \citenamefont {Hetterich}, \citenamefont {Sala},\ and\
  \citenamefont {Pollmann}}]{detomasi2019dynamics}%
  \BibitemOpen
  \bibfield  {author} {\bibinfo {author} {\bibfnamefont {G.}~\bibnamefont
  {De~Tomasi}}, \bibinfo {author} {\bibfnamefont {D.}~\bibnamefont
  {Hetterich}}, \bibinfo {author} {\bibfnamefont {P.}~\bibnamefont {Sala}},\
  and\ \bibinfo {author} {\bibfnamefont {F.}~\bibnamefont {Pollmann}},\
  }\bibfield  {title} {\bibinfo {title} {Dynamics of strongly interacting
  systems: From fock-space fragmentation to many-body localization},\ }\href
  {https://doi.org/10.1103/PhysRevB.100.214313} {\bibfield  {journal} {\bibinfo
   {journal} {Phys. Rev. B}\ }\textbf {\bibinfo {volume} {100}},\ \bibinfo
  {pages} {214313} (\bibinfo {year} {2019})}\BibitemShut {NoStop}%
\bibitem [{\citenamefont {Herviou}\ \emph {et~al.}(2021)\citenamefont
  {Herviou}, \citenamefont {Bardarson},\ and\ \citenamefont
  {Regnault}}]{herviou2021mbl}%
  \BibitemOpen
  \bibfield  {author} {\bibinfo {author} {\bibfnamefont {L.}~\bibnamefont
  {Herviou}}, \bibinfo {author} {\bibfnamefont {J.~H.}\ \bibnamefont
  {Bardarson}},\ and\ \bibinfo {author} {\bibfnamefont {N.}~\bibnamefont
  {Regnault}},\ }\bibfield  {title} {\bibinfo {title} {Many-body localization
  in a fragmented {H}ilbert space},\ }\href
  {https://doi.org/10.1103/PhysRevB.103.134207} {\bibfield  {journal} {\bibinfo
   {journal} {Phys. Rev. B}\ }\textbf {\bibinfo {volume} {103}},\ \bibinfo
  {pages} {134207} (\bibinfo {year} {2021})}\BibitemShut {NoStop}%
\bibitem [{\citenamefont {Rakovszky}\ \emph {et~al.}(2020)\citenamefont
  {Rakovszky}, \citenamefont {Sala}, \citenamefont {Verresen}, \citenamefont
  {Knap},\ and\ \citenamefont {Pollmann}}]{sliom2020}%
  \BibitemOpen
  \bibfield  {author} {\bibinfo {author} {\bibfnamefont {T.}~\bibnamefont
  {Rakovszky}}, \bibinfo {author} {\bibfnamefont {P.}~\bibnamefont {Sala}},
  \bibinfo {author} {\bibfnamefont {R.}~\bibnamefont {Verresen}}, \bibinfo
  {author} {\bibfnamefont {M.}~\bibnamefont {Knap}},\ and\ \bibinfo {author}
  {\bibfnamefont {F.}~\bibnamefont {Pollmann}},\ }\bibfield  {title} {\bibinfo
  {title} {Statistical localization: From strong fragmentation to strong edge
  modes},\ }\href {https://doi.org/10.1103/PhysRevB.101.125126} {\bibfield
  {journal} {\bibinfo  {journal} {Phys. Rev. B}\ }\textbf {\bibinfo {volume}
  {101}},\ \bibinfo {pages} {125126} (\bibinfo {year} {2020})}\BibitemShut
  {NoStop}%
\bibitem [{\citenamefont {Lee}\ \emph {et~al.}(2021)\citenamefont {Lee},
  \citenamefont {Pal},\ and\ \citenamefont {Changlani}}]{lee2021frustration}%
  \BibitemOpen
  \bibfield  {author} {\bibinfo {author} {\bibfnamefont {K.}~\bibnamefont
  {Lee}}, \bibinfo {author} {\bibfnamefont {A.}~\bibnamefont {Pal}},\ and\
  \bibinfo {author} {\bibfnamefont {H.~J.}\ \bibnamefont {Changlani}},\
  }\bibfield  {title} {\bibinfo {title} {Frustration-induced emergent {H}ilbert
  space fragmentation},\ }\href {https://doi.org/10.1103/PhysRevB.103.235133}
  {\bibfield  {journal} {\bibinfo  {journal} {Phys. Rev. B}\ }\textbf {\bibinfo
  {volume} {103}},\ \bibinfo {pages} {235133} (\bibinfo {year}
  {2021})}\BibitemShut {NoStop}%
\bibitem [{\citenamefont {Langlett}\ and\ \citenamefont
  {Xu}(2021)}]{langlett2021hilbert}%
  \BibitemOpen
  \bibfield  {author} {\bibinfo {author} {\bibfnamefont {C.~M.}\ \bibnamefont
  {Langlett}}\ and\ \bibinfo {author} {\bibfnamefont {S.}~\bibnamefont {Xu}},\
  }\bibfield  {title} {\bibinfo {title} {Hilbert space fragmentation and exact
  scars of generalized {F}redkin spin chains},\ }\href
  {https://doi.org/10.1103/PhysRevB.103.L220304} {\bibfield  {journal}
  {\bibinfo  {journal} {Phys. Rev. B}\ }\textbf {\bibinfo {volume} {103}},\
  \bibinfo {pages} {L220304} (\bibinfo {year} {2021})}\BibitemShut {NoStop}%
\bibitem [{\citenamefont {Mukherjee}\ \emph
  {et~al.}(2021{\natexlab{a}})\citenamefont {Mukherjee}, \citenamefont {Cai},\
  and\ \citenamefont {Liu}}]{mukherjee2021constraint}%
  \BibitemOpen
  \bibfield  {author} {\bibinfo {author} {\bibfnamefont {B.}~\bibnamefont
  {Mukherjee}}, \bibinfo {author} {\bibfnamefont {Z.}~\bibnamefont {Cai}},\
  and\ \bibinfo {author} {\bibfnamefont {W.~V.}\ \bibnamefont {Liu}},\
  }\bibfield  {title} {\bibinfo {title} {Constraint-induced breaking and
  restoration of ergodicity in spin-1 {PXP} models},\ }\href
  {https://doi.org/10.1103/PhysRevResearch.3.033201} {\bibfield  {journal}
  {\bibinfo  {journal} {Phys. Rev. Research}\ }\textbf {\bibinfo {volume}
  {3}},\ \bibinfo {pages} {033201} (\bibinfo {year}
  {2021}{\natexlab{a}})}\BibitemShut {NoStop}%
\bibitem [{\citenamefont {Mukherjee}\ \emph
  {et~al.}(2021{\natexlab{b}})\citenamefont {Mukherjee}, \citenamefont
  {Banerjee}, \citenamefont {Sengupta},\ and\ \citenamefont
  {Sen}}]{mukherjee2021minimal}%
  \BibitemOpen
  \bibfield  {author} {\bibinfo {author} {\bibfnamefont {B.}~\bibnamefont
  {Mukherjee}}, \bibinfo {author} {\bibfnamefont {D.}~\bibnamefont {Banerjee}},
  \bibinfo {author} {\bibfnamefont {K.}~\bibnamefont {Sengupta}},\ and\
  \bibinfo {author} {\bibfnamefont {A.}~\bibnamefont {Sen}},\ }\bibfield
  {title} {\bibinfo {title} {A minimal model for {H}ilbert space fragmentation
  with local constraints},\ }\href@noop {} {\bibfield  {journal} {\bibinfo
  {journal} {arXiv e-prints}\ } (\bibinfo {year} {2021}{\natexlab{b}})},\
  \Eprint {https://arxiv.org/abs/2106.14897} {arXiv:2106.14897
  [cond-mat.str-el]} \BibitemShut {NoStop}%
\bibitem [{\citenamefont {Khudorozhkov}\ \emph {et~al.}(2021)\citenamefont
  {Khudorozhkov}, \citenamefont {Tiwari}, \citenamefont {Chamon},\ and\
  \citenamefont {Neupert}}]{khudorozhkov2021hilbert}%
  \BibitemOpen
  \bibfield  {author} {\bibinfo {author} {\bibfnamefont {A.}~\bibnamefont
  {Khudorozhkov}}, \bibinfo {author} {\bibfnamefont {A.}~\bibnamefont
  {Tiwari}}, \bibinfo {author} {\bibfnamefont {C.}~\bibnamefont {Chamon}},\
  and\ \bibinfo {author} {\bibfnamefont {T.}~\bibnamefont {Neupert}},\
  }\href@noop {} {\bibinfo {title} {Hilbert space fragmentation in a 2d quantum
  spin system with subsystem symmetries}} (\bibinfo {year} {2021}),\ \Eprint
  {https://arxiv.org/abs/2107.09690} {arXiv:2107.09690 [cond-mat.str-el]}
  \BibitemShut {NoStop}%
\bibitem [{\citenamefont {Richter}\ and\ \citenamefont
  {Pal}(2021)}]{richter2021anomalous}%
  \BibitemOpen
  \bibfield  {author} {\bibinfo {author} {\bibfnamefont {J.}~\bibnamefont
  {Richter}}\ and\ \bibinfo {author} {\bibfnamefont {A.}~\bibnamefont {Pal}},\
  }\href@noop {} {\bibinfo {title} {Anomalous hydrodynamics in a class of
  scarred frustration-free {H}amiltonians}} (\bibinfo {year} {2021}),\ \Eprint
  {https://arxiv.org/abs/2107.13612} {arXiv:2107.13612 [cond-mat.stat-mech]}
  \BibitemShut {NoStop}%
\bibitem [{\citenamefont {Morningstar}\ \emph {et~al.}(2020)\citenamefont
  {Morningstar}, \citenamefont {Khemani},\ and\ \citenamefont
  {Huse}}]{morningstar2020kineticallyconstrained}%
  \BibitemOpen
  \bibfield  {author} {\bibinfo {author} {\bibfnamefont {A.}~\bibnamefont
  {Morningstar}}, \bibinfo {author} {\bibfnamefont {V.}~\bibnamefont
  {Khemani}},\ and\ \bibinfo {author} {\bibfnamefont {D.~A.}\ \bibnamefont
  {Huse}},\ }\bibfield  {title} {\bibinfo {title} {Kinetically constrained
  freezing transition in a dipole-conserving system},\ }\href
  {https://doi.org/10.1103/PhysRevB.101.214205} {\bibfield  {journal} {\bibinfo
   {journal} {Phys. Rev. B}\ }\textbf {\bibinfo {volume} {101}},\ \bibinfo
  {pages} {214205} (\bibinfo {year} {2020})}\BibitemShut {NoStop}%
\bibitem [{\citenamefont {Yuzbashyan}\ and\ \citenamefont
  {Shastry}(2013)}]{yuzbashyan2013integrability}%
  \BibitemOpen
  \bibfield  {author} {\bibinfo {author} {\bibfnamefont {E.~A.}\ \bibnamefont
  {Yuzbashyan}}\ and\ \bibinfo {author} {\bibfnamefont {B.~S.}\ \bibnamefont
  {Shastry}},\ }\bibfield  {title} {\bibinfo {title} {Quantum integrability in
  systems with finite number of levels},\ }\href
  {https://doi.org/10.1007/s10955-013-0689-9} {\bibfield  {journal} {\bibinfo
  {journal} {Journal of Statistical Physics}\ }\textbf {\bibinfo {volume}
  {150}},\ \bibinfo {pages} {704} (\bibinfo {year} {2013})}\BibitemShut
  {NoStop}%
\bibitem [{\citenamefont {Scaramazza}\ \emph {et~al.}(2016)\citenamefont
  {Scaramazza}, \citenamefont {Shastry},\ and\ \citenamefont
  {Yuzbashyan}}]{scaramazza2016integrable}%
  \BibitemOpen
  \bibfield  {author} {\bibinfo {author} {\bibfnamefont {J.~A.}\ \bibnamefont
  {Scaramazza}}, \bibinfo {author} {\bibfnamefont {B.~S.}\ \bibnamefont
  {Shastry}},\ and\ \bibinfo {author} {\bibfnamefont {E.~A.}\ \bibnamefont
  {Yuzbashyan}},\ }\bibfield  {title} {\bibinfo {title} {Integrable matrix
  theory: Level statistics},\ }\href
  {https://doi.org/10.1103/PhysRevE.94.032106} {\bibfield  {journal} {\bibinfo
  {journal} {Phys. Rev. E}\ }\textbf {\bibinfo {volume} {94}},\ \bibinfo
  {pages} {032106} (\bibinfo {year} {2016})}\BibitemShut {NoStop}%
\bibitem [{\citenamefont {Landsman}(1998)}]{landsman1998lecture}%
  \BibitemOpen
  \bibfield  {author} {\bibinfo {author} {\bibfnamefont {N.~P.}\ \bibnamefont
  {Landsman}},\ }\bibfield  {title} {\bibinfo {title} {Lecture notes on
  {C}*-algebras, {H}ilbert {C}*-modules, and quantum mechanics},\ }\href@noop
  {} {\bibfield  {journal} {\bibinfo  {journal} {arXiv preprint
  math-ph/9807030}\ } (\bibinfo {year} {1998})}\BibitemShut {NoStop}%
\bibitem [{\citenamefont {Nussinov}\ and\ \citenamefont
  {Ortiz}(2009)}]{nussinov2009bond}%
  \BibitemOpen
  \bibfield  {author} {\bibinfo {author} {\bibfnamefont {Z.}~\bibnamefont
  {Nussinov}}\ and\ \bibinfo {author} {\bibfnamefont {G.}~\bibnamefont
  {Ortiz}},\ }\bibfield  {title} {\bibinfo {title} {Bond algebras and exact
  solvability of {H}amiltonians: {S}pin ${S}=\frac{1}{2}$ multilayer systems},\
  }\href {https://doi.org/10.1103/PhysRevB.79.214440} {\bibfield  {journal}
  {\bibinfo  {journal} {Phys. Rev. B}\ }\textbf {\bibinfo {volume} {79}},\
  \bibinfo {pages} {214440} (\bibinfo {year} {2009})}\BibitemShut {NoStop}%
\bibitem [{\citenamefont {Cobanera}\ \emph {et~al.}(2010)\citenamefont
  {Cobanera}, \citenamefont {Ortiz},\ and\ \citenamefont
  {Nussinov}}]{cobanera2010unified}%
  \BibitemOpen
  \bibfield  {author} {\bibinfo {author} {\bibfnamefont {E.}~\bibnamefont
  {Cobanera}}, \bibinfo {author} {\bibfnamefont {G.}~\bibnamefont {Ortiz}},\
  and\ \bibinfo {author} {\bibfnamefont {Z.}~\bibnamefont {Nussinov}},\
  }\bibfield  {title} {\bibinfo {title} {Unified approach to quantum and
  classical dualities},\ }\href
  {https://doi.org/10.1103/PhysRevLett.104.020402} {\bibfield  {journal}
  {\bibinfo  {journal} {Phys. Rev. Lett.}\ }\textbf {\bibinfo {volume} {104}},\
  \bibinfo {pages} {020402} (\bibinfo {year} {2010})}\BibitemShut {NoStop}%
\bibitem [{\citenamefont {Cobanera}\ \emph {et~al.}(2011)\citenamefont
  {Cobanera}, \citenamefont {Ortiz},\ and\ \citenamefont
  {Nussinov}}]{cobenera2011bond}%
  \BibitemOpen
  \bibfield  {author} {\bibinfo {author} {\bibfnamefont {E.}~\bibnamefont
  {Cobanera}}, \bibinfo {author} {\bibfnamefont {G.}~\bibnamefont {Ortiz}},\
  and\ \bibinfo {author} {\bibfnamefont {Z.}~\bibnamefont {Nussinov}},\
  }\bibfield  {title} {\bibinfo {title} {The bond-algebraic approach to
  dualities},\ }\href {https://doi.org/10.1080/00018732.2011.619814} {\bibfield
   {journal} {\bibinfo  {journal} {Advances in Physics}\ }\textbf {\bibinfo
  {volume} {60}},\ \bibinfo {pages} {679} (\bibinfo {year} {2011})}\BibitemShut
  {NoStop}%
\bibitem [{\citenamefont {Harlow}(2017)}]{harlow2017}%
  \BibitemOpen
  \bibfield  {author} {\bibinfo {author} {\bibfnamefont {D.}~\bibnamefont
  {Harlow}},\ }\bibfield  {title} {\bibinfo {title} {The {R}yu--{T}akayanagi
  formula from quantum error correction},\ }\href
  {https://doi.org/10.1007/s00220-017-2904-z} {\bibfield  {journal} {\bibinfo
  {journal} {Communications in Mathematical Physics}\ }\textbf {\bibinfo
  {volume} {354}},\ \bibinfo {pages} {865} (\bibinfo {year}
  {2017})}\BibitemShut {NoStop}%
\bibitem [{\citenamefont {Zanardi}(2001)}]{zanardi2001virtual}%
  \BibitemOpen
  \bibfield  {author} {\bibinfo {author} {\bibfnamefont {P.}~\bibnamefont
  {Zanardi}},\ }\bibfield  {title} {\bibinfo {title} {Virtual quantum
  subsystems},\ }\href {https://doi.org/10.1103/PhysRevLett.87.077901}
  {\bibfield  {journal} {\bibinfo  {journal} {Phys. Rev. Lett.}\ }\textbf
  {\bibinfo {volume} {87}},\ \bibinfo {pages} {077901} (\bibinfo {year}
  {2001})}\BibitemShut {NoStop}%
\bibitem [{\citenamefont {Bartlett}\ \emph {et~al.}(2007)\citenamefont
  {Bartlett}, \citenamefont {Rudolph},\ and\ \citenamefont
  {Spekkens}}]{bartlett2007reference}%
  \BibitemOpen
  \bibfield  {author} {\bibinfo {author} {\bibfnamefont {S.~D.}\ \bibnamefont
  {Bartlett}}, \bibinfo {author} {\bibfnamefont {T.}~\bibnamefont {Rudolph}},\
  and\ \bibinfo {author} {\bibfnamefont {R.~W.}\ \bibnamefont {Spekkens}},\
  }\bibfield  {title} {\bibinfo {title} {Reference frames, superselection
  rules, and quantum information},\ }\href
  {https://doi.org/10.1103/RevModPhys.79.555} {\bibfield  {journal} {\bibinfo
  {journal} {Rev. Mod. Phys.}\ }\textbf {\bibinfo {volume} {79}},\ \bibinfo
  {pages} {555} (\bibinfo {year} {2007})}\BibitemShut {NoStop}%
\bibitem [{\citenamefont {Lidar}(2014)}]{lidar2014dfs}%
  \BibitemOpen
  \bibfield  {author} {\bibinfo {author} {\bibfnamefont {D.~A.}\ \bibnamefont
  {Lidar}},\ }\bibinfo {title} {Review of decoherence-free subspaces, noiseless
  subsystems, and dynamical decoupling},\ in\ \href
  {https://doi.org/https://doi.org/10.1002/9781118742631.ch11} {\emph {\bibinfo
  {booktitle} {Quantum Information and Computation for Chemistry}}}\ (\bibinfo
  {publisher} {John Wiley \& Sons, Ltd},\ \bibinfo {year} {2014})\ pp.\
  \bibinfo {pages} {295--354}\BibitemShut {NoStop}%
\bibitem [{\citenamefont {Fulton}\ and\ \citenamefont
  {Harris}(2013)}]{fulton2013representation}%
  \BibitemOpen
  \bibfield  {author} {\bibinfo {author} {\bibfnamefont {W.}~\bibnamefont
  {Fulton}}\ and\ \bibinfo {author} {\bibfnamefont {J.}~\bibnamefont
  {Harris}},\ }\href@noop {} {\emph {\bibinfo {title} {Representation theory: a
  first course}}},\ Vol.\ \bibinfo {volume} {129}\ (\bibinfo  {publisher}
  {Springer Science \& Business Media},\ \bibinfo {year} {2013})\BibitemShut
  {NoStop}%
\bibitem [{\citenamefont {Read}\ and\ \citenamefont
  {Saleur}(2007)}]{readsaleur2007}%
  \BibitemOpen
  \bibfield  {author} {\bibinfo {author} {\bibfnamefont {N.}~\bibnamefont
  {Read}}\ and\ \bibinfo {author} {\bibfnamefont {H.}~\bibnamefont {Saleur}},\
  }\bibfield  {title} {\bibinfo {title} {Enlarged symmetry algebras of spin
  chains, loop models, and s-matrices},\ }\href
  {https://doi.org/https://doi.org/10.1016/j.nuclphysb.2007.03.007} {\bibfield
  {journal} {\bibinfo  {journal} {Nuclear Physics B}\ }\textbf {\bibinfo
  {volume} {777}},\ \bibinfo {pages} {263 } (\bibinfo {year}
  {2007})}\BibitemShut {NoStop}%
\bibitem [{\citenamefont {Kitaev}(2003)}]{kitaev2003fault}%
  \BibitemOpen
  \bibfield  {author} {\bibinfo {author} {\bibfnamefont {A.}~\bibnamefont
  {Kitaev}},\ }\bibfield  {title} {\bibinfo {title} {Fault-tolerant quantum
  computation by anyons},\ }\href
  {https://doi.org/https://doi.org/10.1016/S0003-4916(02)00018-0} {\bibfield
  {journal} {\bibinfo  {journal} {Annals of Physics}\ }\textbf {\bibinfo
  {volume} {303}},\ \bibinfo {pages} {2 } (\bibinfo {year} {2003})}\BibitemShut
  {NoStop}%
\bibitem [{\citenamefont {Kitaev}\ and\ \citenamefont
  {Laumann}(2010)}]{kitaev2010topological}%
  \BibitemOpen
  \bibfield  {author} {\bibinfo {author} {\bibfnamefont {A.}~\bibnamefont
  {Kitaev}}\ and\ \bibinfo {author} {\bibfnamefont {C.}~\bibnamefont
  {Laumann}},\ }\bibfield  {title} {\bibinfo {title} {Topological phases and
  quantum computation},\ }\href@noop {} {\bibfield  {journal} {\bibinfo
  {journal} {Exact Methods in Low-dimensional Statistical Physics and Quantum
  Computing: Lecture Notes of the Les Houches Summer School: Volume 89, July
  2008}\ ,\ \bibinfo {pages} {101}} (\bibinfo {year} {2010})}\BibitemShut
  {NoStop}%
\bibitem [{\citenamefont {Nandkishore}\ and\ \citenamefont
  {Hermele}(2019)}]{fractonreview}%
  \BibitemOpen
  \bibfield  {author} {\bibinfo {author} {\bibfnamefont {R.~M.}\ \bibnamefont
  {Nandkishore}}\ and\ \bibinfo {author} {\bibfnamefont {M.}~\bibnamefont
  {Hermele}},\ }\bibfield  {title} {\bibinfo {title} {Fractons},\ }\href
  {https://doi.org/10.1146/annurev-conmatphys-031218-013604} {\bibfield
  {journal} {\bibinfo  {journal} {Annual Review of Condensed Matter Physics}\
  }\textbf {\bibinfo {volume} {10}},\ \bibinfo {pages} {295} (\bibinfo {year}
  {2019})}\BibitemShut {NoStop}%
\bibitem [{\citenamefont {Pretko}\ \emph {et~al.}(2020)\citenamefont {Pretko},
  \citenamefont {Chen},\ and\ \citenamefont {You}}]{pretko2020fracton}%
  \BibitemOpen
  \bibfield  {author} {\bibinfo {author} {\bibfnamefont {M.}~\bibnamefont
  {Pretko}}, \bibinfo {author} {\bibfnamefont {X.}~\bibnamefont {Chen}},\ and\
  \bibinfo {author} {\bibfnamefont {Y.}~\bibnamefont {You}},\ }\bibfield
  {title} {\bibinfo {title} {Fracton phases of matter},\ }\href
  {https://doi.org/10.1142/S0217751X20300033} {\bibfield  {journal} {\bibinfo
  {journal} {International Journal of Modern Physics A}\ }\textbf {\bibinfo
  {volume} {35}},\ \bibinfo {pages} {2030003} (\bibinfo {year}
  {2020})}\BibitemShut {NoStop}%
\bibitem [{\citenamefont {Chen}\ and\ \citenamefont
  {Iadecola}(2021)}]{chen2021emergent}%
  \BibitemOpen
  \bibfield  {author} {\bibinfo {author} {\bibfnamefont {I.-C.}\ \bibnamefont
  {Chen}}\ and\ \bibinfo {author} {\bibfnamefont {T.}~\bibnamefont
  {Iadecola}},\ }\bibfield  {title} {\bibinfo {title} {Emergent symmetries and
  slow quantum dynamics in a {R}ydberg-atom chain with confinement},\ }\href
  {https://doi.org/10.1103/PhysRevB.103.214304} {\bibfield  {journal} {\bibinfo
   {journal} {Phys. Rev. B}\ }\textbf {\bibinfo {volume} {103}},\ \bibinfo
  {pages} {214304} (\bibinfo {year} {2021})}\BibitemShut {NoStop}%
\bibitem [{\citenamefont {Zhang}\ \emph {et~al.}(1997)\citenamefont {Zhang},
  \citenamefont {Karbach}, \citenamefont {M\"uller},\ and\ \citenamefont
  {Stolze}}]{zhang1997tJz}%
  \BibitemOpen
  \bibfield  {author} {\bibinfo {author} {\bibfnamefont {S.}~\bibnamefont
  {Zhang}}, \bibinfo {author} {\bibfnamefont {M.}~\bibnamefont {Karbach}},
  \bibinfo {author} {\bibfnamefont {G.}~\bibnamefont {M\"uller}},\ and\
  \bibinfo {author} {\bibfnamefont {J.}~\bibnamefont {Stolze}},\ }\bibfield
  {title} {\bibinfo {title} {Charge and spin dynamics in the one-dimensional
  t-{J}$_{z}$ and t-{J} models},\ }\href
  {https://doi.org/10.1103/PhysRevB.55.6491} {\bibfield  {journal} {\bibinfo
  {journal} {Phys. Rev. B}\ }\textbf {\bibinfo {volume} {55}},\ \bibinfo
  {pages} {6491} (\bibinfo {year} {1997})}\BibitemShut {NoStop}%
\bibitem [{\citenamefont {Batista}\ and\ \citenamefont
  {Ortiz}(2000)}]{batista2000tJz}%
  \BibitemOpen
  \bibfield  {author} {\bibinfo {author} {\bibfnamefont {C.~D.}\ \bibnamefont
  {Batista}}\ and\ \bibinfo {author} {\bibfnamefont {G.}~\bibnamefont
  {Ortiz}},\ }\bibfield  {title} {\bibinfo {title} {Quantum phase diagram of
  the t-{J}$_{z}$ chain model},\ }\href
  {https://doi.org/10.1103/PhysRevLett.85.4755} {\bibfield  {journal} {\bibinfo
   {journal} {Phys. Rev. Lett.}\ }\textbf {\bibinfo {volume} {85}},\ \bibinfo
  {pages} {4755} (\bibinfo {year} {2000})}\BibitemShut {NoStop}%
\bibitem [{\citenamefont {Batista}\ and\ \citenamefont
  {Ortiz}(2001)}]{batista2001generalizedJW}%
  \BibitemOpen
  \bibfield  {author} {\bibinfo {author} {\bibfnamefont {C.~D.}\ \bibnamefont
  {Batista}}\ and\ \bibinfo {author} {\bibfnamefont {G.}~\bibnamefont
  {Ortiz}},\ }\bibfield  {title} {\bibinfo {title} {Generalized
  {J}ordan-{W}igner transformations},\ }\href
  {https://doi.org/10.1103/PhysRevLett.86.1082} {\bibfield  {journal} {\bibinfo
   {journal} {Phys. Rev. Lett.}\ }\textbf {\bibinfo {volume} {86}},\ \bibinfo
  {pages} {1082} (\bibinfo {year} {2001})}\BibitemShut {NoStop}%
\bibitem [{\citenamefont {Vafek}\ \emph {et~al.}(2017)\citenamefont {Vafek},
  \citenamefont {Regnault},\ and\ \citenamefont
  {Bernevig}}]{vafek2017entanglement}%
  \BibitemOpen
  \bibfield  {author} {\bibinfo {author} {\bibfnamefont {O.}~\bibnamefont
  {Vafek}}, \bibinfo {author} {\bibfnamefont {N.}~\bibnamefont {Regnault}},\
  and\ \bibinfo {author} {\bibfnamefont {B.~A.}\ \bibnamefont {Bernevig}},\
  }\bibfield  {title} {\bibinfo {title} {{Entanglement of Exact Excited
  Eigenstates of the Hubbard Model in Arbitrary Dimension}},\ }\href
  {https://doi.org/10.21468/SciPostPhys.3.6.043} {\bibfield  {journal}
  {\bibinfo  {journal} {SciPost Phys.}\ }\textbf {\bibinfo {volume} {3}},\
  \bibinfo {pages} {043} (\bibinfo {year} {2017})}\BibitemShut {NoStop}%
\bibitem [{\citenamefont {Caha}\ and\ \citenamefont
  {Nagaj}(2018)}]{caha2018pairflip}%
  \BibitemOpen
  \bibfield  {author} {\bibinfo {author} {\bibfnamefont {L.}~\bibnamefont
  {Caha}}\ and\ \bibinfo {author} {\bibfnamefont {D.}~\bibnamefont {Nagaj}},\
  }\href@noop {} {\bibinfo {title} {The pair-flip model: a very entangled
  translationally invariant spin chain}} (\bibinfo {year} {2018}),\ \Eprint
  {https://arxiv.org/abs/1805.07168} {arXiv:1805.07168 [quant-ph]} \BibitemShut
  {NoStop}%
\bibitem [{\citenamefont {Read}\ and\ \citenamefont
  {Tutte}(1988)}]{read1988chromatic}%
  \BibitemOpen
  \bibfield  {author} {\bibinfo {author} {\bibfnamefont {R.~C.}\ \bibnamefont
  {Read}}\ and\ \bibinfo {author} {\bibfnamefont {W.~T.}\ \bibnamefont
  {Tutte}},\ }\bibfield  {title} {\bibinfo {title} {Chromatic polynomials},\
  }\href@noop {} {\bibfield  {journal} {\bibinfo  {journal} {Selected topics in
  graph theory}\ }\textbf {\bibinfo {volume} {3}},\ \bibinfo {pages} {15}
  (\bibinfo {year} {1988})}\BibitemShut {NoStop}%
\bibitem [{\citenamefont
  {\texttt{http://oeis.org/wiki/Colorings\_of\_grid\_graphs}}()}]{oeiscolorings}%
  \BibitemOpen
  \bibfield  {author} {\bibinfo {author} {\bibnamefont
  {\texttt{http://oeis.org/wiki/Colorings\_of\_grid\_graphs}}},\ }\href@noop {}
  {}\BibitemShut {NoStop}%
\bibitem [{\citenamefont {Lieb}(1967{\natexlab{a}})}]{liebentropyice1967}%
  \BibitemOpen
  \bibfield  {author} {\bibinfo {author} {\bibfnamefont {E.~H.}\ \bibnamefont
  {Lieb}},\ }\bibfield  {title} {\bibinfo {title} {Residual entropy of square
  ice},\ }\href {https://doi.org/10.1103/PhysRev.162.162} {\bibfield  {journal}
  {\bibinfo  {journal} {Phys. Rev.}\ }\textbf {\bibinfo {volume} {162}},\
  \bibinfo {pages} {162} (\bibinfo {year} {1967}{\natexlab{a}})}\BibitemShut
  {NoStop}%
\bibitem [{\citenamefont {Lieb}(1967{\natexlab{b}})}]{liebentropyiceexact1967}%
  \BibitemOpen
  \bibfield  {author} {\bibinfo {author} {\bibfnamefont {E.~H.}\ \bibnamefont
  {Lieb}},\ }\bibfield  {title} {\bibinfo {title} {Exact solution of the
  problem of the entropy of two-dimensional ice},\ }\href
  {https://doi.org/10.1103/PhysRevLett.18.692} {\bibfield  {journal} {\bibinfo
  {journal} {Phys. Rev. Lett.}\ }\textbf {\bibinfo {volume} {18}},\ \bibinfo
  {pages} {692} (\bibinfo {year} {1967}{\natexlab{b}})}\BibitemShut {NoStop}%
\bibitem [{\citenamefont {Baxter}(1970)}]{baxter3colorings1970}%
  \BibitemOpen
  \bibfield  {author} {\bibinfo {author} {\bibfnamefont {R.~J.}\ \bibnamefont
  {Baxter}},\ }\bibfield  {title} {\bibinfo {title} {Three‐colorings of the
  square lattice: A hard squares model},\ }\href
  {https://doi.org/10.1063/1.1665102} {\bibfield  {journal} {\bibinfo
  {journal} {Journal of Mathematical Physics}\ }\textbf {\bibinfo {volume}
  {11}},\ \bibinfo {pages} {3116} (\bibinfo {year} {1970})}\BibitemShut
  {NoStop}%
\bibitem [{\citenamefont {Barber}\ and\ \citenamefont
  {Batchelor}(1989)}]{barber1989spectrum}%
  \BibitemOpen
  \bibfield  {author} {\bibinfo {author} {\bibfnamefont {M.~N.}\ \bibnamefont
  {Barber}}\ and\ \bibinfo {author} {\bibfnamefont {M.~T.}\ \bibnamefont
  {Batchelor}},\ }\bibfield  {title} {\bibinfo {title} {Spectrum of the
  biquadratic spin-1 antiferromagnetic chain},\ }\href
  {https://doi.org/10.1103/PhysRevB.40.4621} {\bibfield  {journal} {\bibinfo
  {journal} {Phys. Rev. B}\ }\textbf {\bibinfo {volume} {40}},\ \bibinfo
  {pages} {4621} (\bibinfo {year} {1989})}\BibitemShut {NoStop}%
\bibitem [{\citenamefont {Batchelor}\ and\ \citenamefont
  {Barber}(1990)}]{batchelor1990spin}%
  \BibitemOpen
  \bibfield  {author} {\bibinfo {author} {\bibfnamefont {M.~T.}\ \bibnamefont
  {Batchelor}}\ and\ \bibinfo {author} {\bibfnamefont {M.~N.}\ \bibnamefont
  {Barber}},\ }\bibfield  {title} {\bibinfo {title} {Spin-s quantum chains and
  {T}emperley-{L}ieb algebras},\ }\href
  {https://doi.org/10.1088/0305-4470/23/1/004} {\bibfield  {journal} {\bibinfo
  {journal} {Journal of Physics A: Mathematical and General}\ }\textbf
  {\bibinfo {volume} {23}},\ \bibinfo {pages} {L15} (\bibinfo {year}
  {1990})}\BibitemShut {NoStop}%
\bibitem [{\citenamefont {Aufgebauer}\ and\ \citenamefont
  {Klümper}(2010)}]{aufgebauer2010quantum}%
  \BibitemOpen
  \bibfield  {author} {\bibinfo {author} {\bibfnamefont {B.}~\bibnamefont
  {Aufgebauer}}\ and\ \bibinfo {author} {\bibfnamefont {A.}~\bibnamefont
  {Klümper}},\ }\bibfield  {title} {\bibinfo {title} {Quantum spin chains of
  {T}emperley-{L}ieb type: periodic boundary conditions, spectral
  multiplicities and finite temperature},\ }\href
  {https://doi.org/10.1088/1742-5468/2010/05/p05018} {\bibfield  {journal}
  {\bibinfo  {journal} {Journal of Statistical Mechanics: Theory and
  Experiment}\ }\textbf {\bibinfo {volume} {2010}},\ \bibinfo {pages} {P05018}
  (\bibinfo {year} {2010})}\BibitemShut {NoStop}%
\bibitem [{\citenamefont {Parkinson}(1987)}]{parkinson1987integrability}%
  \BibitemOpen
  \bibfield  {author} {\bibinfo {author} {\bibfnamefont {J.~B.}\ \bibnamefont
  {Parkinson}},\ }\bibfield  {title} {\bibinfo {title} {On the integrability of
  the {S}=1 quantum spin chain with pure biquadratic exchange},\ }\href
  {https://doi.org/10.1088/0022-3719/20/36/011} {\bibfield  {journal} {\bibinfo
   {journal} {Journal of Physics C: Solid State Physics}\ }\textbf {\bibinfo
  {volume} {20}},\ \bibinfo {pages} {L1029} (\bibinfo {year}
  {1987})}\BibitemShut {NoStop}%
\bibitem [{\citenamefont {Parkinson}(1988)}]{parkinson1988biquadratic}%
  \BibitemOpen
  \bibfield  {author} {\bibinfo {author} {\bibfnamefont {J.~B.}\ \bibnamefont
  {Parkinson}},\ }\bibfield  {title} {\bibinfo {title} {The {S}=1 quantum spin
  chain with pure biquadratic exchange},\ }\href
  {https://doi.org/10.1088/0022-3719/21/20/014} {\bibfield  {journal} {\bibinfo
   {journal} {Journal of Physics C: Solid State Physics}\ }\textbf {\bibinfo
  {volume} {21}},\ \bibinfo {pages} {3793} (\bibinfo {year}
  {1988})}\BibitemShut {NoStop}%
\bibitem [{\citenamefont {Ercolessi}\ \emph {et~al.}(2014)\citenamefont
  {Ercolessi}, \citenamefont {Vodola},\ and\ \citenamefont
  {Silvia}}]{Ercolessi2014Analysis}%
  \BibitemOpen
  \bibfield  {author} {\bibinfo {author} {\bibfnamefont {E.}~\bibnamefont
  {Ercolessi}}, \bibinfo {author} {\bibfnamefont {D.~D.}\ \bibnamefont
  {Vodola}},\ and\ \bibinfo {author} {\bibfnamefont {F.}~\bibnamefont
  {Silvia}},\ }\href {https://amslaurea.unibo.it/8323/1/ferri_silvia_tesi.pdf}
  {\bibfield  {journal} {\bibinfo  {journal} {``Analysis of the spectrum of the
  spin-1 biquadratic antiferromagnetic chain" (unpublished)}\ } (\bibinfo
  {year} {2014})}\BibitemShut {NoStop}%
\bibitem [{\citenamefont {Saito}(1990)}]{saito1990noncrossing}%
  \BibitemOpen
  \bibfield  {author} {\bibinfo {author} {\bibfnamefont {R.}~\bibnamefont
  {Saito}},\ }\bibfield  {title} {\bibinfo {title} {A proof of the completeness
  of the non crossed diagrams in {S}pin 1/2 {H}eisenberg {M}odel},\ }\href
  {https://doi.org/10.1143/JPSJ.59.482} {\bibfield  {journal} {\bibinfo
  {journal} {Journal of the Physical Society of Japan}\ }\textbf {\bibinfo
  {volume} {59}},\ \bibinfo {pages} {482} (\bibinfo {year} {1990})}\BibitemShut
  {NoStop}%
\bibitem [{\citenamefont {Oguchi}\ and\ \citenamefont
  {Kitatani}(1989)}]{oguchi1989rvb}%
  \BibitemOpen
  \bibfield  {author} {\bibinfo {author} {\bibfnamefont {T.}~\bibnamefont
  {Oguchi}}\ and\ \bibinfo {author} {\bibfnamefont {H.}~\bibnamefont
  {Kitatani}},\ }\bibfield  {title} {\bibinfo {title} {Theory of the resonating
  valence bond in quantum spin system},\ }\href
  {https://doi.org/10.1143/JPSJ.58.1403} {\bibfield  {journal} {\bibinfo
  {journal} {Journal of the Physical Society of Japan}\ }\textbf {\bibinfo
  {volume} {58}},\ \bibinfo {pages} {1403} (\bibinfo {year}
  {1989})}\BibitemShut {NoStop}%
\bibitem [{\citenamefont {Veness}\ \emph {et~al.}(2017)\citenamefont {Veness},
  \citenamefont {Essler},\ and\ \citenamefont {Fisher}}]{veness2017quantum}%
  \BibitemOpen
  \bibfield  {author} {\bibinfo {author} {\bibfnamefont {T.}~\bibnamefont
  {Veness}}, \bibinfo {author} {\bibfnamefont {F.~H.~L.}\ \bibnamefont
  {Essler}},\ and\ \bibinfo {author} {\bibfnamefont {M.~P.~A.}\ \bibnamefont
  {Fisher}},\ }\bibfield  {title} {\bibinfo {title} {Quantum disentangled
  liquid in the half-filled {H}ubbard model},\ }\href
  {https://doi.org/10.1103/PhysRevB.96.195153} {\bibfield  {journal} {\bibinfo
  {journal} {Phys. Rev. B}\ }\textbf {\bibinfo {volume} {96}},\ \bibinfo
  {pages} {195153} (\bibinfo {year} {2017})}\BibitemShut {NoStop}%
\bibitem [{\citenamefont {Protopopov}\ \emph {et~al.}(2020)\citenamefont
  {Protopopov}, \citenamefont {Panda}, \citenamefont {Parolini}, \citenamefont
  {Scardicchio}, \citenamefont {Demler},\ and\ \citenamefont
  {Abanin}}]{protopopov2020nonabelian}%
  \BibitemOpen
  \bibfield  {author} {\bibinfo {author} {\bibfnamefont {I.~V.}\ \bibnamefont
  {Protopopov}}, \bibinfo {author} {\bibfnamefont {R.~K.}\ \bibnamefont
  {Panda}}, \bibinfo {author} {\bibfnamefont {T.}~\bibnamefont {Parolini}},
  \bibinfo {author} {\bibfnamefont {A.}~\bibnamefont {Scardicchio}}, \bibinfo
  {author} {\bibfnamefont {E.}~\bibnamefont {Demler}},\ and\ \bibinfo {author}
  {\bibfnamefont {D.~A.}\ \bibnamefont {Abanin}},\ }\bibfield  {title}
  {\bibinfo {title} {Non-abelian symmetries and disorder: A broad nonergodic
  regime and anomalous thermalization},\ }\href
  {https://doi.org/10.1103/PhysRevX.10.011025} {\bibfield  {journal} {\bibinfo
  {journal} {Phys. Rev. X}\ }\textbf {\bibinfo {volume} {10}},\ \bibinfo
  {pages} {011025} (\bibinfo {year} {2020})}\BibitemShut {NoStop}%
\bibitem [{\citenamefont {{Fremling}}\ \emph {et~al.}(2018)\citenamefont
  {{Fremling}}, \citenamefont {{Repellin}}, \citenamefont {{St{\'e}phan}},
  \citenamefont {{Moran}}, \citenamefont {{Slingerland}},\ and\ \citenamefont
  {{Haque}}}]{fremling2018dynamics}%
  \BibitemOpen
  \bibfield  {author} {\bibinfo {author} {\bibfnamefont {M.}~\bibnamefont
  {{Fremling}}}, \bibinfo {author} {\bibfnamefont {C.}~\bibnamefont
  {{Repellin}}}, \bibinfo {author} {\bibfnamefont {J.-M.}\ \bibnamefont
  {{St{\'e}phan}}}, \bibinfo {author} {\bibfnamefont {N.}~\bibnamefont
  {{Moran}}}, \bibinfo {author} {\bibfnamefont {J.~K.}\ \bibnamefont
  {{Slingerland}}},\ and\ \bibinfo {author} {\bibfnamefont {M.}~\bibnamefont
  {{Haque}}},\ }\bibfield  {title} {\bibinfo {title} {{Dynamics and level
  statistics of interacting fermions in the lowest {L}andau level}},\ }\href
  {https://doi.org/10.1088/1367-2630/aae73f} {\bibfield  {journal} {\bibinfo
  {journal} {New Journal of Physics}\ }\textbf {\bibinfo {volume} {20}},\
  \bibinfo {eid} {103036} (\bibinfo {year} {2018})}\BibitemShut {NoStop}%
\bibitem [{\citenamefont {Mazur}(1969)}]{mazurbound1969}%
  \BibitemOpen
  \bibfield  {author} {\bibinfo {author} {\bibfnamefont {P.}~\bibnamefont
  {Mazur}},\ }\bibfield  {title} {\bibinfo {title} {Non-ergodicity of phase
  functions in certain systems},\ }\href
  {https://doi.org/https://doi.org/10.1016/0031-8914(69)90185-2} {\bibfield
  {journal} {\bibinfo  {journal} {Physica}\ }\textbf {\bibinfo {volume} {43}},\
  \bibinfo {pages} {533 } (\bibinfo {year} {1969})}\BibitemShut {NoStop}%
\bibitem [{\citenamefont {Dhar}\ \emph {et~al.}(2020)\citenamefont {Dhar},
  \citenamefont {Kundu},\ and\ \citenamefont {Saito}}]{dhar2020revisiting}%
  \BibitemOpen
  \bibfield  {author} {\bibinfo {author} {\bibfnamefont {A.}~\bibnamefont
  {Dhar}}, \bibinfo {author} {\bibfnamefont {A.}~\bibnamefont {Kundu}},\ and\
  \bibinfo {author} {\bibfnamefont {K.}~\bibnamefont {Saito}},\ }\href@noop {}
  {\bibinfo {title} {{Revisiting the Mazur bound and the Suzuki equality}}}
  (\bibinfo {year} {2020}),\ \Eprint {https://arxiv.org/abs/2007.04562}
  {arXiv:2007.04562 [cond-mat.stat-mech]} \BibitemShut {NoStop}%
\bibitem [{\citenamefont {Suzuki}(1971)}]{suzukiequality1971}%
  \BibitemOpen
  \bibfield  {author} {\bibinfo {author} {\bibfnamefont {M.}~\bibnamefont
  {Suzuki}},\ }\bibfield  {title} {\bibinfo {title} {Ergodicity, constants of
  motion, and bounds for susceptibilities},\ }\href
  {https://doi.org/https://doi.org/10.1016/0031-8914(71)90226-6} {\bibfield
  {journal} {\bibinfo  {journal} {Physica}\ }\textbf {\bibinfo {volume} {51}},\
  \bibinfo {pages} {277 } (\bibinfo {year} {1971})}\BibitemShut {NoStop}%
\bibitem [{\citenamefont {Prosen}\ and\ \citenamefont
  {Ilievski}(2013)}]{prosen2013quasilocalfamilies}%
  \BibitemOpen
  \bibfield  {author} {\bibinfo {author} {\bibfnamefont {T.}~\bibnamefont
  {Prosen}}\ and\ \bibinfo {author} {\bibfnamefont {E.}~\bibnamefont
  {Ilievski}},\ }\bibfield  {title} {\bibinfo {title} {Families of quasilocal
  conservation laws and quantum spin transport},\ }\href
  {https://doi.org/10.1103/PhysRevLett.111.057203} {\bibfield  {journal}
  {\bibinfo  {journal} {Phys. Rev. Lett.}\ }\textbf {\bibinfo {volume} {111}},\
  \bibinfo {pages} {057203} (\bibinfo {year} {2013})}\BibitemShut {NoStop}%
\bibitem [{\citenamefont {Prosen}(2014)}]{prosen2014quasilocal}%
  \BibitemOpen
  \bibfield  {author} {\bibinfo {author} {\bibfnamefont {T.}~\bibnamefont
  {Prosen}},\ }\bibfield  {title} {\bibinfo {title} {Quasilocal conservation
  laws in {XXZ} spin-1/2 chains: {O}pen, periodic and twisted boundary
  conditions},\ }\href
  {https://doi.org/https://doi.org/10.1016/j.nuclphysb.2014.07.024} {\bibfield
  {journal} {\bibinfo  {journal} {Nuclear Physics B}\ }\textbf {\bibinfo
  {volume} {886}},\ \bibinfo {pages} {1177 } (\bibinfo {year}
  {2014})}\BibitemShut {NoStop}%
\bibitem [{\citenamefont {Ilievski}\ \emph {et~al.}(2016)\citenamefont
  {Ilievski}, \citenamefont {Medenjak}, \citenamefont {Prosen},\ and\
  \citenamefont {Zadnik}}]{ilievski2016quasilocal}%
  \BibitemOpen
  \bibfield  {author} {\bibinfo {author} {\bibfnamefont {E.}~\bibnamefont
  {Ilievski}}, \bibinfo {author} {\bibfnamefont {M.}~\bibnamefont {Medenjak}},
  \bibinfo {author} {\bibfnamefont {T.}~\bibnamefont {Prosen}},\ and\ \bibinfo
  {author} {\bibfnamefont {L.}~\bibnamefont {Zadnik}},\ }\bibfield  {title}
  {\bibinfo {title} {Quasilocal charges in integrable lattice systems},\ }\href
  {https://doi.org/10.1088/1742-5468/2016/06/064008} {\bibfield  {journal}
  {\bibinfo  {journal} {Journal of Statistical Mechanics: Theory and
  Experiment}\ }\textbf {\bibinfo {volume} {2016}},\ \bibinfo {pages} {064008}
  (\bibinfo {year} {2016})}\BibitemShut {NoStop}%
\bibitem [{\citenamefont {Zadnik}\ \emph {et~al.}(2016)\citenamefont {Zadnik},
  \citenamefont {Medenjak},\ and\ \citenamefont
  {Prosen}}]{zadnik2016quasilocal}%
  \BibitemOpen
  \bibfield  {author} {\bibinfo {author} {\bibfnamefont {L.}~\bibnamefont
  {Zadnik}}, \bibinfo {author} {\bibfnamefont {M.}~\bibnamefont {Medenjak}},\
  and\ \bibinfo {author} {\bibfnamefont {T.}~\bibnamefont {Prosen}},\
  }\bibfield  {title} {\bibinfo {title} {{Quasilocal conservation laws from
  semicyclic irreducible representations of U$_q$($\mathfrak{sl}_2$) in XXZ
  spin-1/2 chains}},\ }\href
  {https://doi.org/https://doi.org/10.1016/j.nuclphysb.2015.11.023} {\bibfield
  {journal} {\bibinfo  {journal} {Nuclear Physics B}\ }\textbf {\bibinfo
  {volume} {902}},\ \bibinfo {pages} {339 } (\bibinfo {year}
  {2016})}\BibitemShut {NoStop}%
\bibitem [{\citenamefont {Doyon}(2017)}]{doyon2017pseudolocal}%
  \BibitemOpen
  \bibfield  {author} {\bibinfo {author} {\bibfnamefont {B.}~\bibnamefont
  {Doyon}},\ }\bibfield  {title} {\bibinfo {title} {Thermalization and
  pseudolocality in extended quantum systems},\ }\href
  {https://doi.org/10.1007/s00220-017-2836-7} {\bibfield  {journal} {\bibinfo
  {journal} {Communications in Mathematical Physics}\ }\textbf {\bibinfo
  {volume} {351}},\ \bibinfo {pages} {155} (\bibinfo {year}
  {2017})}\BibitemShut {NoStop}%
\bibitem [{\citenamefont {Feng}\ and\ \citenamefont
  {Skinner}(2021)}]{feng2021hilbert}%
  \BibitemOpen
  \bibfield  {author} {\bibinfo {author} {\bibfnamefont {X.}~\bibnamefont
  {Feng}}\ and\ \bibinfo {author} {\bibfnamefont {B.}~\bibnamefont {Skinner}},\
  }\href@noop {} {\bibinfo {title} {{H}ilbert space fragmentation produces a
  ``fracton {C}asimir effect"}} (\bibinfo {year} {2021}),\ \Eprint
  {https://arxiv.org/abs/2105.11465} {arXiv:2105.11465 [quant-ph]} \BibitemShut
  {NoStop}%
\bibitem [{\citenamefont {Kim}\ \emph {et~al.}(2015)\citenamefont {Kim},
  \citenamefont {Ba\~nuls}, \citenamefont {Cirac}, \citenamefont {Hastings},\
  and\ \citenamefont {Huse}}]{kim2015slowest}%
  \BibitemOpen
  \bibfield  {author} {\bibinfo {author} {\bibfnamefont {H.}~\bibnamefont
  {Kim}}, \bibinfo {author} {\bibfnamefont {M.~C.}\ \bibnamefont {Ba\~nuls}},
  \bibinfo {author} {\bibfnamefont {J.~I.}\ \bibnamefont {Cirac}}, \bibinfo
  {author} {\bibfnamefont {M.~B.}\ \bibnamefont {Hastings}},\ and\ \bibinfo
  {author} {\bibfnamefont {D.~A.}\ \bibnamefont {Huse}},\ }\bibfield  {title}
  {\bibinfo {title} {Slowest local operators in quantum spin chains},\ }\href
  {https://doi.org/10.1103/PhysRevE.92.012128} {\bibfield  {journal} {\bibinfo
  {journal} {Phys. Rev. E}\ }\textbf {\bibinfo {volume} {92}},\ \bibinfo
  {pages} {012128} (\bibinfo {year} {2015})}\BibitemShut {NoStop}%
\bibitem [{\citenamefont {Lin}\ and\ \citenamefont
  {Motrunich}(2017)}]{lin2017explicit}%
  \BibitemOpen
  \bibfield  {author} {\bibinfo {author} {\bibfnamefont {C.-J.}\ \bibnamefont
  {Lin}}\ and\ \bibinfo {author} {\bibfnamefont {O.~I.}\ \bibnamefont
  {Motrunich}},\ }\bibfield  {title} {\bibinfo {title} {Explicit construction
  of quasiconserved local operator of translationally invariant nonintegrable
  quantum spin chain in prethermalization},\ }\href
  {https://doi.org/10.1103/PhysRevB.96.214301} {\bibfield  {journal} {\bibinfo
  {journal} {Phys. Rev. B}\ }\textbf {\bibinfo {volume} {96}},\ \bibinfo
  {pages} {214301} (\bibinfo {year} {2017})}\BibitemShut {NoStop}%
\bibitem [{\citenamefont {Chertkov}\ \emph {et~al.}(2020)\citenamefont
  {Chertkov}, \citenamefont {Villalonga},\ and\ \citenamefont
  {Clark}}]{chertkov2020engineering}%
  \BibitemOpen
  \bibfield  {author} {\bibinfo {author} {\bibfnamefont {E.}~\bibnamefont
  {Chertkov}}, \bibinfo {author} {\bibfnamefont {B.}~\bibnamefont
  {Villalonga}},\ and\ \bibinfo {author} {\bibfnamefont {B.~K.}\ \bibnamefont
  {Clark}},\ }\bibfield  {title} {\bibinfo {title} {Engineering topological
  models with a general-purpose symmetry-to-hamiltonian approach},\ }\href
  {https://doi.org/10.1103/PhysRevResearch.2.023348} {\bibfield  {journal}
  {\bibinfo  {journal} {Phys. Rev. Research}\ }\textbf {\bibinfo {volume}
  {2}},\ \bibinfo {pages} {023348} (\bibinfo {year} {2020})}\BibitemShut
  {NoStop}%
\bibitem [{\citenamefont {Moudgalya}\ and\ \citenamefont
  {Motrunich}()}]{scarsinprep}%
  \BibitemOpen
  \bibfield  {author} {\bibinfo {author} {\bibfnamefont {S.}~\bibnamefont
  {Moudgalya}}\ and\ \bibinfo {author} {\bibfnamefont {O.~I.}\ \bibnamefont
  {Motrunich}},\ }\href@noop {} {\bibinfo {title} {(in
  preparation)}}\BibitemShut {NoStop}%
\bibitem [{\citenamefont {Albert}\ and\ \citenamefont
  {Jiang}(2014)}]{albert2014symmetries}%
  \BibitemOpen
  \bibfield  {author} {\bibinfo {author} {\bibfnamefont {V.~V.}\ \bibnamefont
  {Albert}}\ and\ \bibinfo {author} {\bibfnamefont {L.}~\bibnamefont {Jiang}},\
  }\bibfield  {title} {\bibinfo {title} {{Symmetries and conserved quantities
  in Lindblad master equations}},\ }\href
  {https://doi.org/10.1103/PhysRevA.89.022118} {\bibfield  {journal} {\bibinfo
  {journal} {Phys. Rev. A}\ }\textbf {\bibinfo {volume} {89}},\ \bibinfo
  {pages} {022118} (\bibinfo {year} {2014})}\BibitemShut {NoStop}%
\bibitem [{\citenamefont {Bu{\v{c}}a}\ and\ \citenamefont
  {Prosen}(2012)}]{buca2012note}%
  \BibitemOpen
  \bibfield  {author} {\bibinfo {author} {\bibfnamefont {B.}~\bibnamefont
  {Bu{\v{c}}a}}\ and\ \bibinfo {author} {\bibfnamefont {T.}~\bibnamefont
  {Prosen}},\ }\bibfield  {title} {\bibinfo {title} {{A note on symmetry
  reductions of the Lindblad equation: transport in constrained open spin
  chains}},\ }\href {https://doi.org/10.1088/1367-2630/14/7/073007} {\bibfield
  {journal} {\bibinfo  {journal} {New Journal of Physics}\ }\textbf {\bibinfo
  {volume} {14}},\ \bibinfo {pages} {073007} (\bibinfo {year}
  {2012})}\BibitemShut {NoStop}%
\bibitem [{\citenamefont {Essler}\ and\ \citenamefont
  {Piroli}(2020)}]{essler2020integrability}%
  \BibitemOpen
  \bibfield  {author} {\bibinfo {author} {\bibfnamefont {F.~H.~L.}\
  \bibnamefont {Essler}}\ and\ \bibinfo {author} {\bibfnamefont
  {L.}~\bibnamefont {Piroli}},\ }\bibfield  {title} {\bibinfo {title}
  {{Integrability of one-dimensional Lindbladians from operator-space
  fragmentation}},\ }\href {https://doi.org/10.1103/PhysRevE.102.062210}
  {\bibfield  {journal} {\bibinfo  {journal} {Phys. Rev. E}\ }\textbf {\bibinfo
  {volume} {102}},\ \bibinfo {pages} {062210} (\bibinfo {year}
  {2020})}\BibitemShut {NoStop}%
\bibitem [{\citenamefont {Robertson}\ and\ \citenamefont
  {Essler}(2021)}]{robertson2021exact}%
  \BibitemOpen
  \bibfield  {author} {\bibinfo {author} {\bibfnamefont {J.}~\bibnamefont
  {Robertson}}\ and\ \bibinfo {author} {\bibfnamefont {F.}~\bibnamefont
  {Essler}},\ }\href@noop {} {\bibinfo {title} {Exact solution of a quantum
  asymmetric exclusion process with particle creation and annihilation}}
  (\bibinfo {year} {2021}),\ \Eprint {https://arxiv.org/abs/2105.08828}
  {arXiv:2105.08828 [cond-mat.stat-mech]} \BibitemShut {NoStop}%
\bibitem [{\citenamefont {B{\"a}cker}(2003)}]{backer2003numerical}%
  \BibitemOpen
  \bibfield  {author} {\bibinfo {author} {\bibfnamefont {A.}~\bibnamefont
  {B{\"a}cker}},\ }\bibfield  {title} {\bibinfo {title} {Numerical aspects of
  eigenvalue and eigenfunction computations for chaotic quantum systems},\ }in\
  \href@noop {} {\emph {\bibinfo {booktitle} {The Mathematical Aspects of
  Quantum Maps}}},\ \bibinfo {editor} {edited by\ \bibinfo {editor}
  {\bibfnamefont {M.~D.}\ \bibnamefont {Esposti}}\ and\ \bibinfo {editor}
  {\bibfnamefont {S.}~\bibnamefont {Graffi}}}\ (\bibinfo  {publisher} {Springer
  Berlin Heidelberg},\ \bibinfo {address} {Berlin, Heidelberg},\ \bibinfo
  {year} {2003})\ pp.\ \bibinfo {pages} {91--144}\BibitemShut {NoStop}%
\bibitem [{\citenamefont {Esposti}\ and\ \citenamefont
  {Winn}(2005)}]{esposti2005quantum}%
  \BibitemOpen
  \bibfield  {author} {\bibinfo {author} {\bibfnamefont {M.~D.}\ \bibnamefont
  {Esposti}}\ and\ \bibinfo {author} {\bibfnamefont {B.}~\bibnamefont {Winn}},\
  }\bibfield  {title} {\bibinfo {title} {The quantum perturbed cat map and
  symmetry},\ }\href {https://doi.org/10.1088/0305-4470/38/26/005} {\bibfield
  {journal} {\bibinfo  {journal} {Journal of Physics A: Mathematical and
  General}\ }\textbf {\bibinfo {volume} {38}},\ \bibinfo {pages} {5895}
  (\bibinfo {year} {2005})}\BibitemShut {NoStop}%
\bibitem [{\citenamefont {Moudgalya}\ \emph {et~al.}(2019)\citenamefont
  {Moudgalya}, \citenamefont {Devakul}, \citenamefont {von Keyserlingk},\ and\
  \citenamefont {Sondhi}}]{moudgalya2019operator}%
  \BibitemOpen
  \bibfield  {author} {\bibinfo {author} {\bibfnamefont {S.}~\bibnamefont
  {Moudgalya}}, \bibinfo {author} {\bibfnamefont {T.}~\bibnamefont {Devakul}},
  \bibinfo {author} {\bibfnamefont {C.~W.}\ \bibnamefont {von Keyserlingk}},\
  and\ \bibinfo {author} {\bibfnamefont {S.~L.}\ \bibnamefont {Sondhi}},\
  }\bibfield  {title} {\bibinfo {title} {Operator spreading in quantum maps},\
  }\href {https://doi.org/10.1103/PhysRevB.99.094312} {\bibfield  {journal}
  {\bibinfo  {journal} {Phys. Rev. B}\ }\textbf {\bibinfo {volume} {99}},\
  \bibinfo {pages} {094312} (\bibinfo {year} {2019})}\BibitemShut {NoStop}%
\bibitem [{\citenamefont {Mondaini}\ \emph {et~al.}(2018)\citenamefont
  {Mondaini}, \citenamefont {Mallayya}, \citenamefont {Santos},\ and\
  \citenamefont {Rigol}}]{mondainicomment2018}%
  \BibitemOpen
  \bibfield  {author} {\bibinfo {author} {\bibfnamefont {R.}~\bibnamefont
  {Mondaini}}, \bibinfo {author} {\bibfnamefont {K.}~\bibnamefont {Mallayya}},
  \bibinfo {author} {\bibfnamefont {L.~F.}\ \bibnamefont {Santos}},\ and\
  \bibinfo {author} {\bibfnamefont {M.}~\bibnamefont {Rigol}},\ }\bibfield
  {title} {\bibinfo {title} {{Comment on ``Systematic Construction of
  Counterexamples to the Eigenstate Thermalization Hypothesis''}},\ }\href
  {https://doi.org/10.1103/PhysRevLett.121.038901} {\bibfield  {journal}
  {\bibinfo  {journal} {Phys. Rev. Lett.}\ }\textbf {\bibinfo {volume} {121}},\
  \bibinfo {pages} {038901} (\bibinfo {year} {2018})}\BibitemShut {NoStop}%
\bibitem [{\citenamefont {Benjamin}\ \emph {et~al.}(2008)\citenamefont
  {Benjamin}, \citenamefont {Plott},\ and\ \citenamefont
  {Sellers}}]{benjamin2008tiling}%
  \BibitemOpen
  \bibfield  {author} {\bibinfo {author} {\bibfnamefont {A.~T.}\ \bibnamefont
  {Benjamin}}, \bibinfo {author} {\bibfnamefont {S.~S.}\ \bibnamefont
  {Plott}},\ and\ \bibinfo {author} {\bibfnamefont {J.~A.}\ \bibnamefont
  {Sellers}},\ }\bibfield  {title} {\bibinfo {title} {Tiling proofs of recent
  sum identities involving pell numbers},\ }\href@noop {} {\bibfield  {journal}
  {\bibinfo  {journal} {Annals of Combinatorics}\ }\textbf {\bibinfo {volume}
  {12}},\ \bibinfo {pages} {271} (\bibinfo {year} {2008})}\BibitemShut
  {NoStop}%
\bibitem [{\citenamefont {Sloane}(2007)}]{oeis}%
  \BibitemOpen
  \bibfield  {author} {\bibinfo {author} {\bibfnamefont {N.~J.~A.}\
  \bibnamefont {Sloane}},\ }\bibfield  {title} {\bibinfo {title} {The on-line
  encyclopedia of integer sequences},\ }in\ \href@noop {} {\emph {\bibinfo
  {booktitle} {Towards Mechanized Mathematical Assistants}}},\ \bibinfo
  {editor} {edited by\ \bibinfo {editor} {\bibfnamefont {M.}~\bibnamefont
  {Kauers}}, \bibinfo {editor} {\bibfnamefont {M.}~\bibnamefont {Kerber}},
  \bibinfo {editor} {\bibfnamefont {R.}~\bibnamefont {Miner}},\ and\ \bibinfo
  {editor} {\bibfnamefont {W.}~\bibnamefont {Windsteiger}}}\ (\bibinfo
  {publisher} {Springer Berlin Heidelberg},\ \bibinfo {address} {Berlin,
  Heidelberg},\ \bibinfo {year} {2007})\ pp.\ \bibinfo {pages}
  {130--130}\BibitemShut {NoStop}%
\bibitem [{\citenamefont {Crosswhite}\ and\ \citenamefont
  {Bacon}(2008)}]{crosswhite2008fsa}%
  \BibitemOpen
  \bibfield  {author} {\bibinfo {author} {\bibfnamefont {G.~M.}\ \bibnamefont
  {Crosswhite}}\ and\ \bibinfo {author} {\bibfnamefont {D.}~\bibnamefont
  {Bacon}},\ }\bibfield  {title} {\bibinfo {title} {Finite automata for caching
  in matrix product algorithms},\ }\href
  {https://doi.org/10.1103/PhysRevA.78.012356} {\bibfield  {journal} {\bibinfo
  {journal} {Phys. Rev. A}\ }\textbf {\bibinfo {volume} {78}},\ \bibinfo
  {pages} {012356} (\bibinfo {year} {2008})}\BibitemShut {NoStop}%
\bibitem [{\citenamefont {Motruk}\ \emph {et~al.}(2016)\citenamefont {Motruk},
  \citenamefont {Zaletel}, \citenamefont {Mong},\ and\ \citenamefont
  {Pollmann}}]{motruk2016density}%
  \BibitemOpen
  \bibfield  {author} {\bibinfo {author} {\bibfnamefont {J.}~\bibnamefont
  {Motruk}}, \bibinfo {author} {\bibfnamefont {M.~P.}\ \bibnamefont {Zaletel}},
  \bibinfo {author} {\bibfnamefont {R.~S.~K.}\ \bibnamefont {Mong}},\ and\
  \bibinfo {author} {\bibfnamefont {F.}~\bibnamefont {Pollmann}},\ }\bibfield
  {title} {\bibinfo {title} {Density matrix renormalization group on a cylinder
  in mixed real and momentum space},\ }\href
  {https://doi.org/10.1103/PhysRevB.93.155139} {\bibfield  {journal} {\bibinfo
  {journal} {Phys. Rev. B}\ }\textbf {\bibinfo {volume} {93}},\ \bibinfo
  {pages} {155139} (\bibinfo {year} {2016})}\BibitemShut {NoStop}%
\end{thebibliography}%
\appendix 
\onecolumngrid
\section{A Sufficient Condition for Classical Fragmentation}\label{app:diagonalcommutants}
In this appendix, we prove a sufficient condition for classical fragmentation, i.e., for all the operators in the commutant algebra to be diagonal in the product state basis. 
For a spin system with an $m$-state local Hilbert space (i.e., spin-$(m-1)/2$), we label the states on site $j$ as $\{\ket{\alpha}_j\}$ for $1 \leq \alpha \leq m$, and we define on-site operators $N^\alpha_j \defn (\ket{\alpha}\bra{\alpha})_j$.
The sufficient condition is as follows.
If the operators $\{N^\alpha_j\}$ for all $\alpha$ and $j$ are all part of the bond algebra $\mA$, then any operator $\hO$ that is part of the corresponding commutant algebra $\mC$ is diagonal in the product state basis, i.e., it can be expressed as a polynomial in terms of the operators $\{N^\alpha_j\}$. 
To show this, we first perform a Schmidt decomposition of the operator $\hO$ about a bipartition of the full Hilbert space as $\mH = \mH_j \otimes \mH_{\textrm{rest}}$, where $\mH_j$ and $\mH_{\textrm{rest}}$ are the Hilbert spaces on site $j$ and the rest of the system respectively.  
The decomposition reads $\hO = \sum_\beta{\lambda_\beta \hO^\beta_{j} \otimes \hO^\beta_{\textrm{rest}}}$, where $\{\hO^\beta_j\}$ and $\{\hO_{\textrm{rest}}\}$ are sets of orthogonal operators that have supports on site $j$ and the rest of the system respectively, and $\lambda_\beta \geq 0$ are the Schmidt values.  
Since $\hO$ is a part of the commutant, we have $[\hO, N^\alpha_j] = 0$ for all $\alpha, j$, hence also $\sum_\beta{\lambda_\beta [\hO^\beta_j, N^\alpha_j] \otimes \hO^\beta_{\textrm{rest}}} = 0$. 
However, since the operators $\{\hO^\beta_{\textrm{rest}}\}$ are linearly-independent (since they are orthogonal), we necessarily have $[\hO^\beta_j, N^\alpha_j] = 0$ for all $\alpha$ and $\beta$, which is only possible if $\hO^\beta_j$ is a polynomial function of $N^\alpha_j$ (here simply any diagonal matrix in the employed basis). 
Since the argument applies to all sites $j$, we obtain that $\hO$ should be a polynomial function of $\{N^\alpha_j\}$, and hence is diagonal in the product state basis. 
Since the identity operator $\mathds{1}$ is always a part of the bond algebra, for a spin-1 system to exhibit classical fragmentation, it suffices if the operators $\{S^z_j\}$ and $\{(S^z_j)^2\}$ are part of the bond algebra. 
\section{\texorpdfstring{$t-J_z$}{} Commutant Algebra}\label{app:tJzcommutantorth}
In this appendix, we discuss properties of the commutant algebra corresponding to the family of $t-J_z$ models of Eq.~(\ref{eq:tJzhamil}).  
\subsection{Orthogonal Basis}
We first set up an orthogonal basis for the commutant algebra $\Cobc$ of the $t-J_z$ model discussed in Sec.~\ref{sec:tJz}.
To begin, we define operators $\hP_j$, $\hZ_j$, $\hO_j$ on site $j$, and a string operator $\hS_{i,j}$ between sites $i$ and $j$ as follows
\begin{gather}
    \hP_j \defn N^\uparrow_j + N^\downarrow_j,\;\;\;\hZ_j \defn N^\uparrow_j - N^\downarrow_j,\;\;\;\hO_j \defn \mathds{1} - \hP_j,\;\;\;\hS_{i,j} \defn \prodal{k = i}{j}{\hO_k},\;\;\; 1 \leq i \leq j \leq L, \nn \\
    \hI_{i,j} \defn \prodal{k = i}{j}{(\hP_k + \hO_k)} = \mathds{1},\;\;\; 1 \leq i \leq j \leq L,
\label{eq:localops}
\end{gather}
where $N^\uparrow_j$ and $N^\downarrow_j$ are defined in Eq.~(\ref{eq:numberdefn}), and $\mathds{1}$ denotes the identity operator. 
Note that although $\hI_{i,j}$ is simply the identity operator $\mathds{1}$ for any $i$ and $j$, we will use it to denote ``insertions" of the identity in terms of operators $\{\hP_j\}$ and $\{\hO_j\}$. 
These operators satisfy the following properties:
\begin{gather}
    \hZ_j^2 = \hP_j,\;\;\; \hP_j^2 = \hP_j,\;\;\; \hO_j^2 = \hO_j,\;\;\;\hP_j \hZ_j = \hZ_j = \hZ_j \hP_j,\;\;\; \hO_j \hP_j = \hP_j \hO_j = 0 = \hZ_j \hO_j = \hO_j \hZ_j, \nn \\
    \textrm{Tr}(\hP_j) = 2,\;\;\textrm{Tr}(\hZ_j) = 0,\;\;\textrm{Tr}(\hO_j) = 1.
\label{eq:properties}
\end{gather}
In order to express the IoMs of Eq.~(\ref{eq:tJzconserved}) in terms of operators $\{\hP_j, \hZ_j, \hO_j\}$, we need to introduce additional shorthand notation.
We define ``words" of length $l$ as follows
\begin{equation}
    A^{(1)} \cdots A^{(l)} \defn \sumal{j_1 < \cdots < j_l}{}{\hS_{1,j_1-1} \left(\prodal{k = 1}{l}{A^{(k)}_{j_k} \hS_{j_k+1,j_{k+1}-1}}\right) A^{(l)}_{j_l} \hS_{j_l + 1, L}},\;\;\; A^{(j)} \in \{P, Z\},
\label{eq:worddefn}
\end{equation}
where $0 \leq l \leq L$, and we have used the same shorthand notation for the sum as in Eq.~(\ref{eq:tJzconserved}), and the only word of length $0$ is defined to be $\hS_{1,L}$, which we will denote by $P^0$. 
To avoid ambiguities, we hereby denote the usual operator product operation when acting on the set of words by $\ast$.
Using the definition of Eq.~(\ref{eq:worddefn}) and the properties of Eq.~(\ref{eq:properties}), it is easy to verify that the product of two words is given by
\begin{equation}
    (A^{(1)} \cdots A^{(l)}) \ast (B^{(1)} \cdots B^{(k)}) = \twopartdef{((A^{(1)} \ast B^{(1)}) \cdots (A^{(k)} \ast B^{(k)}))}{k = l}{0}{k \neq l}, 
\label{eq:wordalgebra}
\end{equation}
where, as a consequence of Eq.~(\ref{eq:properties}), we have defined the product for individual ``letters'' as 
\begin{equation}
    P \ast Z \defn Z \ast P \defn Z,\;\; Z \ast Z \defn P,\;\;\; P \ast P \defn P.
\label{eq:wordproduct}
\end{equation}
The set of words of length $l$ is hence closed under products, and the set of words of length $l$ is an Abelian algebra $\mC^{(l)}$ of dimension $\textrm{dim}(\mC^{(l)}) = 2^l$.
In what follows, whenever one sees a sequence of $P$'s and $Z$'s without hats and without any $*$'s, it denotes a word, i.e., operator of the form in Eq.~(\ref{eq:worddefn}).
We will also frequently write, say, $P^j$ as a shorthand for a sequence of $j$ letters $P$ (which may be part of a larger sequence).
Each word is an equal-weight superposition of specific operator strings where each string has the form $\prod_j (\hP_j \text{~or~} \hZ_j \text{~or~} \hO_j)$ and the number and pattern of $\hP$'s and $\hZ$'s---read from left to right and omitting any intervening $\hO$'s---is fixed by the word.
The properties Eq.~(\ref{eq:properties}) make the on-site operators $\hP_j$, $\hZ_j$, and $\hO_j$ orthogonal in the Hilbert-Schmidt inner product, so distinct string operators of the above form are orthogonal, and hence distinct words are orthogonal -- this is the reason for working in terms of these on-site operators.
Furthermore, the equal-weight superposition structure in the definition of the words as well as specific expressions for $\hP_j$, $\hZ_j$, and $\hO_j$ in terms of $N_j^\uparrow$, $N_j^\downarrow$, and $\mathds{1}$ allow one to show that each word of length $l$ can be written as a linear superposition of the IoMs $N^{\sigma_1\dots\sigma_k}$, $0 \leq k \leq L$ defined in Eq.~(\ref{eq:tJzconserved}).

Similarly, we can also express the operators $\{N^{\sigma_1 \cdots \sigma_k}\}$ of Eq.~(\ref{eq:tJzconserved}) in terms of words defined in Eq.~(\ref{eq:worddefn}).
Starting with $k = 1$, we obtain
\begin{equation}
    N^{\sigma_1} = \sumal{i = 1}{L}{\left(\prodal{j = 1}{i-1}{(\hP_j + \hO_j)}\right)\frac{(\hP_i + \sigma_1 \hZ_i)}{2}\left(\prodal{j = i+1}{L}{(\hP_j + \hO_j)}\right)} = \frac{1}{2}\sumal{j, k}{}{P^j P P^k} + \frac{\sigma_1}{2}\sumal{j, k}{}{P^j Z P^k},
\end{equation}
where $P^j P P^k$ and $P^j Z P^k$ are words of length $j+k+1$ and the sums run over all values $j,k \geq 0$ and $j+k \leq L-1$, and we have abused notation and defined $\sigma_j = \uparrow/\downarrow = +1/-1$.
Note that $P^j P P^k$ here is simply a word with all $j+k+1$ letters being $P$, and the first sum can be rewritten in terms of distinct such words, $\sumal{j, k}{}{P^j P P^k} = \sum_{a=1}^L a P^a$, but exhibiting such rewriting is not necessary for our purposes here and below.
Similarly, we can express  $N^{\sigma_1 \sigma_2}$ as
\begin{align}
    &N^{\sigma_1 \sigma_2} = \sumal{i_1 < i_2}{}{\hI_{1, i_1-1}\left(\frac{(\hP_{i_1} + {\sigma_1} \hZ_{i_1})}{2} \hI_{i_1+1, i_{2}-1}\right) \frac{(\hP_{i_2} + {\sigma_2} \hZ_{i_2})}{2}}\hI_{i_2 + 1, L} \nn \\
    &= \frac{1}{4}\sumal{j, k, l}{}{P^j P P^k P P^l} +  \frac{\sigma_1}{4}\sumal{j, k, l}{}{P^j P P^k Z P^l} + \frac{\sigma_2}{4}\sumal{j, k, l}{}{P^j Z P^k P P^l} +  \frac{{\sigma_1 \sigma_2}}{4}\sumal{j, k, l}{}{P^j Z P^k Z P^l},
\end{align}
where the sums run over all values of $j, k, l$ such that $j, k, l \geq 0$ and $j + k + l \leq L -2$.
Similarly, we can also express $N^{\sigma_1 \cdots \sigma_k}$ as 
\begin{equation}
    N^{\sigma_1\cdots \sigma_k} = \sumal{i_1 < \cdots < i_k}{}{\hI_{1, i_1-1}\prodal{l = 1}{k-1}{\left(\frac{(P_{i_l} + {\sigma_l} Z_{i_l})}{2} \hI_{i_l+1, i_{l+1}-1}\right)} \frac{(P_{i_k} + {\sigma_k} Z_{i_k})}{2}} \hI_{i_k +1, L},
\label{eq:Nkexpression}
\end{equation}
which can clearly be expanded in terms of words of Eq.~(\ref{eq:worddefn}), in particular, the ones that contain at most $k$ $Z$'s, although we do not attempt to write down the exact expression.
We have thus shown that the linear span of the IoMs $N^{\sigma_1 \cdots \sigma_k}$ is the same as the linear span of words of length $l$, $0 \leq l \leq L$.
The basis of words of length $l$, $0 \leq l \leq L$ also form an orthogonal basis for the commutant algebra $\Cobc$, as discussed earlier.
Hence the full commutant algebra $\Cobc$ is then simply given by
\begin{equation}
    \Cobc = \bigoplus_{l = 0}^L{\mC^{(l)}},\;\;\;\textrm{dim}\left(\Cobc\right) = 2^{L + 1} - 1.
\label{eq:commutantalgebra}
\end{equation}
This is consistent with the fact that there are $2^{L+1} - 1$ operators $N^{\sigma_1 \cdots \sigma_k}$, which is same as $\textrm{dim}(\Cobc)$. 
The words of Eq.~(\ref{eq:worddefn}) can be generalized to PBC, where we define PBC words as
\begin{equation}
    [A^{(1)} \cdots A^{(l)}] \defn \sumal{m = 0}{l-1}{\sumal{j_1 < \cdots < j_l}{}{\hS_{1,j_1-1} \left(\prodal{k = 1}{l-1}{A^{(m + k)}_{j_k} \hS_{j_k+1,j_{k+1}-1}}\right) A^{(m+l)}_{j_l} \hS_{j_l+1, L}}},\;\;\; A^{(j)} \in \{P, Z\},
\label{eq:worddefnPBC}
\end{equation}
for $0 \leq l \leq L-1$, and we assume $A^{(l + j)} \defn A^{(j)}$ for $1 \leq j \leq l$.
Similar to the OBC case, the only word of length $0$ is $\hS_{1,L}$, which we will denote by $P^0$. 
As a consequence of Eq.~(\ref{eq:worddefnPBC}), all cyclic permutations of the words are identical, i.e., $[A^{(1)} \cdots A^{(l)}] = [A^{(2)} \cdots A^{(l)} A^{(1)}] = \cdots =[A^{(l)} A^{(1)} \cdots A^{(l-1)}]$.
Similar to Eq.~(\ref{eq:wordalgebra}), it is easy to verify that the products of two PBC words of length $l \leq L -1$ is given by
\begin{equation}
    [A^{(1)} \cdots A^{(l)}] \ast [B^{(1)} \cdots B^{(k)}] = \twopartdef{\sumal{m = 0}{l-1}{[(A^{(m + 1)} \ast B^{(1)}) \cdots (A^{(m+k)} \ast B^{(k)})]}}{k = l}{0}{k \neq l}, 
\label{eq:wordalgebraPBC}
\end{equation}
where we have again assumed $A^{(l + j)} \defn A^{(j)}$ for $1 \leq j \leq l$.
For $l = L$, however, we define PBC words as 
\begin{equation}
    [A^{(1)} \cdots A^{(L)}] \defn \prodal{j = 1}{L}{A^{(j)}_j},
\label{eq:lengthLwords}
\end{equation}
since such products by themselves (i.e., without any sums) are distinct elements in the commutant algebra $\Cpbc$. 
Hence, by our definition, the cyclic permutations of PBC words of length $L$ are not identical.
These words of length $L$ are orthogonal to the words of length $l \leq L - 1$ of Eq.~(\ref{eq:worddefnPBC}), and their products are given by Eq.~(\ref{eq:wordalgebra}).  
These words of Eqs.~(\ref{eq:worddefnPBC}) and (\ref{eq:lengthLwords}) form a complete basis for the PBC commutant algebra $\Cpbc$, and similar to the OBC case, the IoMs of Eq.~(\ref{eq:tJzconservedPBC}) can be expressed as linear combinations of the PBC words.
Similar to Eq.~(\ref{eq:commutantalgebra}), the commutant $\Cpbc$ is a direct sum of the algebra of words of length $l$, $0 \leq l \leq L$, and it is clear that $\text{dim}(\Cpbc)$ grows exponentially with $L$. 
\subsection{Algebra Generated by the SLIOMs}\label{app:SLIOMalgebra}
We now construct the algebra generated by the (left) SLIOMs $\{\hq^{(\ell)}_l\}$ discussed in Sec.~\ref{subsec:SLIOMconnection}. 
Using the definition of the SLIOMs $\hq^{(\ell)}_l$ in Eq.~(\ref{eq:tJzSLIOMs}), and the operators defined in Eq.~(\ref{eq:localops}), we can express them in terms of these words of Eq.~(\ref{eq:worddefn}) as follows
\begin{gather}
    \hq^{(\ell)}_1 = \sumal{j}{}{\hS_{1,j-1} \hZ_j \hI_{j+1, L}},\;\hq^{(\ell)}_2 = \sumal{j_1 < j_2}{}{\hS_{1,j_1-1} \hP_{j_1} \hS_{j_1+1,j_2-1} \hZ_{j_2}\hI_{j_2 + 1, L}},\;\cdots,\;\hq^{(\ell)}_l = \sumal{j_1 < \cdots < j_l}{}{\hS_{1,j_1-1} \left(\prodal{k = 1}{l-1}{\hP_{i_k} \hS_{j_k+1,j_{k+1}-1}}\right) \hZ_{j_l} \hI_{j_l + 1, L}},\nn \\
    \implies \hq^{(\ell)}_l = P^{l-1}Z + P^{l-1}Z P + P^{l-1}Z P^2 + \cdots + P^{l-1}Z P^{L-l} = \sumal{\alpha = 0}{L-l}{P^{l-1} Z P^\alpha}.
\label{eq:SLIOMwords}
\end{gather}
Using Eq.~(\ref{eq:SLIOMwords}), we now show that the sums and products of the left SLIOMs along with the $\mathds{1}$ operator generate the entire algebra $\Cobc$.
Note that a simple application of Eq.~(\ref{eq:wordalgebra}) gives
\begin{gather}
    (\hq^{(\ell)}_{l})^2 - (\hq^{(\ell)}_{l+1})^2 = P^l,\;\;1 \leq l \leq L- 1,\;\;\; (\hq^{(\ell)}_L)^2 = P^L,\nn \\
    \hq^{(\ell)}_l ((\hq^{(\ell)}_{l + m})^2 - (\hq^{(\ell)}_{l + m + 1})^2) = P^{l-1} Z P^{m},\;\;1 \leq l \leq L-1,\;\; 0 \leq m \leq L-l-1; \quad \hq^{(\ell)}_l (\hq^{(\ell)}_{L})^2 = P^{l-1} Z P^{L-l}.
\label{eq:qlproperties}
\end{gather}
Using the properties of Eq.~(\ref{eq:qlproperties}), and the fact that $\mathds{1} = \sum_{l = 0}^L{P^l}$, we obtain that $P^{l}$ and $P^{l} Z P^{m}$ can be generated from the left SLIOMs for any $l, m$. 
%, 
Now, using Eq.~(\ref{eq:wordalgebra}) it is straightforward to show that all words of the form of Eq.~(\ref{eq:worddefn}) can be generated by products of these words.
Hence, the left SLIOMs along with $\mathds{1}$ generate the full algebra of words of length $l$, $0 \leq l \leq L$, which is the commutant algebra $\Cobc$.
Similarly, the (right) SLIOMs $\{\hq^{(r)}_l\}$ defined in Eq.~(\ref{eq:tJzrightSLIOMs}) generates the entire commutant algebra $\Cobc$. This can be shown in a similar way, starting from the expressions of $\hq^{(r)}_l$ in terms of words of Eq.~(\ref{eq:worddefn})
\begin{gather}
    \hq^{(r)}_1 = \sumal{j}{}{\hI_{1, j-1} \hZ_j \hS_{j+1,L}},\;\hq^{(r)}_2 = \sumal{j_1 < j_2}{}{\hI_{1, j_1-1} \hZ_{j_1} \hS_{j_1+1,j_2-1} \hP_{j_2} \hS_{j_2+1,L} },\;\cdots,\;\hq^{(r)}_l = \sumal{j_1 < \cdots < j_l}{}{\hI_{1, j_l - 1}\hZ_{j_1}\left(\prodal{k = 1}{l-1}{\hS_{j_k+1,j_{k+1}-1} \hP_{j_{k+1}}}\right)\hS_{j_{l} + 1, L}}\nn \\
    \implies \hq^{(r)}_l = Z P^{l-1} + P Z P^{l-1} + \cdots + P^{L-l} Z P^{l-1} = \sumal{\alpha = 0}{L-l}{P^\alpha Z P^{l-1}}.
\label{eq:rightSLIOMwords}
\end{gather}
Eq.~(\ref{eq:rightSLIOMwords}) shows that these operators are distinct from the left SLIOMs of Eq.~(\ref{eq:SLIOMwords}).
Nevertheless, we can similarly show that $\{\hq^{(r)}_l\}$ along with $\mathds{1}$ generate the algebra $\Cobc$, and hence the left and right SLIOMs are different sets of generators of the same algebra.  
\section{Number and Dimensions of Krylov Subpaces in the Pair-Flip Model}\label{app:PFcounting}
In this appendix, we review exact results on the counting of the number and dimensions of Krylov subspaces in the PF model of Eq.~(\ref{eq:PFHamil}), which prove useful in performing quick analytical calculations and consistency checks in numerical calculations.
We first introduce an alternative interpretation of the Hilbert space and Krylov subspaces that enabled Ref.~\cite{caha2018pairflip} to exactly count their number and dimensions.
Any product state in the Hilbert space of the spin-$(m-1)/2$ (i.e., $m$-level) chain of length $L$ can be interpreted as a ``walk" of length $L$ starting from a fixed vertex (which we refer to as the origin) on a Bethe lattice of coordination number $m$.
We can think of the $m$ edges at any vertex of the Bethe lattice as labelling $m$ different values of the local spin such that no two edges that share a common vertex have the same label. 
The product state $\ket{\alpha_1 \cdots \alpha_{L}}$ is then a walk starting from the origin on the Bethe lattice where the first step is along the edge labeled by $\alpha_1$, second step is along the edge labeled by $\alpha_2$, and so on until the $L$-th step is along the edge labeled by $\alpha_{L}$. 
As a consequence of the labelling of the Bethe lattice, note that any walk (product state) with a repetition such as $\ket{\cdots \alpha \alpha \cdots}$ indicates that the walk retraces a step back. 
In the language of walks, the action of $\hF^{\alpha,\beta}_{i,j}$ in Eq.~(\ref{eq:PFHamil}) only changes the retraced edges of the walk (i.e., results in transitions $\ket{\alpha \alpha} \leftrightarrow \ket{\beta\beta}$), and hence does not change the end point of the walk. 
Furthermore, since $H^{(m)}_{PF}$ allows for transitions between any such retraced edges, it is easy to see that all walks with the same end point are connected by the transitions allowed by $H^{(m)}_{PF}$.
The Krylov subspaces of $H^{(m)}_{PF}$ are thus uniquely labeled by end points of the walks, or, equivalently, by the unique path on the Bethe lattice from the origin to the end point without any retracing. 
We refer to these shortest paths as ``representative walks," and the spins along these paths correspond to the ``dots" discussed in Sec.~\ref{subsec:PFfragmentation1d}.
The Krylov subspaces with $j$ dots correspond to walks that end at distance $j$ from the origin for $0 \leq j \leq L$, which involve $(L - j)/2$ retracings.  
\subsection{Counting dimensions}\label{sec:NIScounting}
In the following, we denote the number of Krylov subspaces with $j$ dots by $n_{j}$, and the dimension of each such subspace for a system size of $L$ by $D_{L, j}$. 
Since all the Krylov subspaces span the full Hilbert space, we should have
\begin{equation}
    \sumal{j}{}{n_{j} D_{L, j}} = m^{L}. 
\label{eq:consistency}
\end{equation}
While Eq.~(\ref{eq:consistency}) might be reminiscent of Eq.~(\ref{eq:dimensions}), these are different, and $d_\lambda = 1$ for all $\lambda$ in the PF model (i.e., the most general PF model has no degeneracies among different Krylov subspaces with the same number of dots.)
Following the discussion in Sec.~\ref{subsec:PFfragmentation1d}, we directly obtain the number of distinct Krylov subspaces with $j$ dots with OBC by imposing the condition that the colors of adjacent dots to be unequal 
\begin{equation}
    n_{j} = \twopartdef{1}{j = 0}{m (m -1)^{j - 1}}{j \geq 1}.
\label{eq:Djexpression}
\end{equation}
The dimension counting of such subspaces $D_{L, j}$ is not so straightforward, which involves counting the number of walks of length $L$ with a fixed end point on a Bethe lattice, and this was studied in great detail in Ref.~\cite{caha2018pairflip} using connections to the counting of colored Dyck paths.
Defining a generating function for the Krylov subspace dimension with $j$ dots as
\begin{equation}
    G_{j}(z) \defn \sumal{\ell = 0}{\infty}{D_{\ell, j} z^{\ell}},\;\;\; D_{0, 0} \defn 1,\;\; D_{\ell, j} = 0\;\;\textrm{if}\;\;j > \ell,
\label{eq:generatingfunctiondefn}
\end{equation}
they obtain
\begin{equation}
    G_{j}(z) = \frac{2(m-1)}{m-2 + m \sqrt{1 - 4(m-1) z^2}}\left(\frac{1 - \sqrt{1 - 4(m-1) z^2}}{2(m-1) z}\right)^{j},
\label{eq:generatingfunction}
\end{equation}
where we have changed the conventions and notations of Ref.~\cite{caha2018pairflip} to express the final result in the language used in this work.
The generating function in Eq.~(\ref{eq:generatingfunction}) also encodes the following obvious constraints
\begin{itemize}
    \item $D_{\ell, j} = 0$ if $j > \ell$, i.e., there cannot be more than $\ell$ dots in a system of size $\ell$. 
    \item $D_{\ell, j} = 0$ if $j$ and $\ell$ have the opposite parity, i.e., a system of even (resp.\ odd) size cannot have Krylov subspaces with odd (resp.\ even) number of dots. 
\end{itemize}
Using Eq.~(\ref{eq:Djexpression}), we can verify that $\sumal{j = 0}{\infty}{G_{j}(z) n_{j}} = (1 - m z)^{-1} = \sumal{\ell = 0}{\infty}{m^{\ell} z^{\ell}}$, which, due to Eq.~(\ref{eq:generatingfunctiondefn}), is a verification of Eq.~(\ref{eq:consistency}).
\section{Independence of Conserved Quantities in the Pair-Flip Model}\label{app:PFdependence}
In this appendix, we show that not all the operators in Eq.~(\ref{eq:stringops}) are linearly independent. %
As discussed in Sec.~\ref{subsec:defn}, the operators $\{N^\alpha_j\}$ are not linearly independent from $\mathds{1}$, and they satisfy $\sum_{\alpha = 1}^{m}{N_j^\alpha} = \mathds{1}$ for all $j$.
As a consequence, for any even system size $L$, the ``one-index" conserved quantities $\{N^{\alpha_1}\}$ can be expressed in terms of ``two-index" conserved quantities $\{N^{\gamma_1 \gamma_2}\}$ as follows
\begin{gather}
    N^{\alpha_1} = \sumal{j_1 = 1}{L}{(-1)^{j_1} N^{\alpha_1}_{j_1}} = \sumal{j_1 = 1}{L}{(-1)^{j_1} N^{\alpha_1}_{j_1} \left[\frac{(-1)^{j_1+1} + (-1)^L}{2} - \frac{(-1)^{j_1-1} + (-1)}{2}\right]} \nn \\
    =  \sumal{j_1 < p = 1}{L}{(-1)^{j_1+p} N^{\alpha_1}_{j_1}} - \sumal{p < j_1 = 1}{L}{(-1)^{j_1+p} N^{\alpha_1}_{j_1}} =  \sumal{\beta = 1}{m}{\left[\sumal{j_1 < p = 1}{L}{(-1)^{j_1+p} N^{\alpha_1}_{j_1} N^{\beta}_{p}} - \sumal{p < j_1 = 1}{L}{(-1)^{j_1+p} N^{\beta}_{p} N^{\alpha_1}_{j_1}}\right]} \nn \\
    =\sumal{\beta}{}{\left(N^{\alpha_1 \beta} - N^{\beta \alpha_1}\right)}
    = \sumal{\beta \neq \alpha_1}{}{\left(N^{\alpha_1 \beta} - N^{\beta \alpha_1}\right)},
\end{gather}
where we have used the fact that $(-1)^{j+1} = (-1)^{j-1}$, $(-1)^L = 1$, and the identity
\begin{equation}
    \sumal{p = q}{r}{(-1)^p} = \frac{(-1)^q + (-1)^r}{2}.
\label{eq:sumidentity}
\end{equation}
Using the same manipulations, any ``odd-index" conserved quantity $\{N^{\alpha_1 \cdots \alpha_{2k-1}}\}$ can be expressed in terms of ``even-index" ones $\{N^{\gamma_1 \cdots \gamma_{2k}}\}$ as follows
\begin{gather}
    N^{\alpha_1 \cdots \alpha_{2k-1}} = \sumal{j_1 < \cdots < j_{2k-1}}{}{(-1)^{j_1 + \cdots + j_{2k-1}} N^{\alpha_1}_{j_1} \cdots N^{\alpha_{2k-1}}_{j_{2k-1}}} = \sumal{j_1 < \cdots < j_{2k-1}}{}{(-1)^{j_1 + \cdots + j_{2k-1}} N^{\alpha_1}_{j_1} \cdots N^{\alpha_{2k-1}}_{j_{2k-1}}}\times\\
    \left[\frac{(-1)^L + (-1)^{j_{2k-1} + 1}}{2} - \frac{(-1)^{j_{2k-1}-1} + (-1)^{j_{2k-2}+1}}{2} + \frac{(-1)^{j_{2k-2}-1} + (-1)^{j_{2k-3}+1}}{2} - \cdots - \frac{(-1)^{j_{1}-1} + (-1)}{2}\right] \nn \\
    = \sumal{\beta}{}{(N^{\alpha_1 \cdots \alpha_{2k-2} \alpha_{2k-1} \beta} - N^{\alpha_1 \cdots \alpha_{2k-2} \beta \alpha_{2k-1}} + N^{\alpha_1 \cdots \beta \alpha_{2k-2} \alpha_{2k-1} } - \cdots - N^{\beta \alpha_1 \cdots \alpha_{2k-1}})}.
\label{eq:oddeven}
\end{gather}
Note that terms of the form $N^{\gamma_1 \cdots \gamma_{2k}}$ with any $\gamma_l = \gamma_{l+1}$ for some $1 \leq l \leq 2k-1$ are automatically excluded in the final sum in Eq.~(\ref{eq:oddeven}), either due to constraints on the indices in $N^{\alpha_1\dots\alpha_{2k-1}}$ on the LHS or due to cancellations in the sum over $\beta$.
Similarly, for an odd system size $L$, the even index conserved quantities $\{N^{\alpha_1 \cdots \alpha_{2k}}\}$ can be expressed in terms of the odd index conserved quantities $\{N^{\gamma_1 \cdots \gamma_{2k+1}}\}$.
For $k = 0$, this directly follows since the ``zero-index" conserved quantity $\mathds{1}$ is simply a sum of the ``one-index" conserved quantities $\{N^{\alpha_1}\}$.
For general $k$, we obtain
\begin{gather}
    N^{\alpha_1 \cdots \alpha_{2k}} = \sumal{j_1 < \cdots < j_{2k}}{}{(-1)^{j_1 + \cdots + j_{2k}} N^{\alpha_1}_{j_1} \cdots N^{\alpha_{2k}}_{j_{2k}}} = \sumal{j_1 < \cdots < j_{2k}}{}{(-1)^{j_1 + \cdots + j_{2k}} N^{\alpha_1}_{j_1} \cdots N^{\alpha_{2k}}_{j_{2k}}}\times \nn \\
    \left[-\frac{(-1)^L + (-1)^{j_{2k} + 1}}{2} + \frac{(-1)^{j_{2k}-1} +
    (-1)^{j_{2k-1}+1}}{2} - \frac{(-1)^{j_{2k-1}-1} + (-1)^{j_{2k-2}+1}}{2} + \cdots - \frac{(-1)^{j_{1}-1} + (-1)}{2}\right] \nn \\
    = \sumal{\beta}{}{(-N^{\alpha_1 \cdots \alpha_{2k-1} \alpha_{2k} \beta} + N^{\alpha_1 \cdots \alpha_{2k-1} \beta \alpha_{2k}} - N^{\alpha_1 \cdots \beta \alpha_{2k-1} \alpha_{2k} } + \cdots - N^{\beta \alpha_1 \cdots \alpha_{2k}})},
\label{eq:evenodd}
\end{gather}
where again any term $N^{\gamma_1 \cdots \gamma_{2k+1}}$ with any $\gamma_l = \gamma_{l+1}$ for some $1 \leq l \leq 2k$ is automatically excluded.
\section{Commutants of the Temperley-Lieb Models}\label{app:TLcommutants}
In this appendix, we sketch the structure of additional conserved quantities in the commutant of the TL models, apart from the ones in Eq.~(\ref{eq:TLJkconservation}), and we refer readers to Ref.~\cite{readsaleur2007} for a more detailed analysis. 
We begin with the commutation relation
\begin{equation}
    \left[\left(M_j\right)^{\beta_1}_{\alpha_1} \left(M_{k}\right)^{\beta_2}_{\alpha_2}, \he_{j,k}\right]  =\twopartdef{-\delta^{\beta_2}_{\alpha_1}(\ket{\beta_1 \alpha_2}\bra{\psising})_{j,k} + \delta^{\beta_1}_{\alpha_2}(\ket{\psising}\bra{\alpha_1 \beta_2})_{j,k}}{\textrm{$j$ odd and $k$ even}}{-\delta^{\beta_1}_{\alpha_2}(\ket{\alpha_1 \beta_2}\bra{\psising})_{j,k} + \delta^{\beta_2}_{\alpha_1}(\ket{\psising}\bra{\beta_1 \alpha_2})_{j,k}}{\textrm{$j$ even and $k$ odd}},
\label{eq:ncommSUm}
\end{equation}
which reduces to Eq.~(\ref{eq:nncommTLuneq}) when $\alpha_1 \neq \beta_2$ and $\beta_1 \neq \alpha_2$. 
To construct additional IoMs we begin by contracting various indices in Eq.~(\ref{eq:ncommSUm}) to obtain
\begin{align}
    &\left[\sumal{\gamma = 1}{m}{\left(M_j\right)^{\gamma}_{\alpha_1} \left(M_{k}\right)^{\beta_2}_{\gamma}}, \he_{j,k}\right]  =\twopartdef{-\delta^{\beta_2}_{\alpha_1}\he_{j,k} + m(\ket{\psising}\bra{\alpha_1 \beta_2})_{j,k}}{\textrm{$j$ odd and $k$ even}}{-m(\ket{\alpha_1 \beta_2}\bra{\psising})_{j,k} + \delta^{\beta_2}_{\alpha_1} \he_{j,k}}{\textrm{$j$ even and $k$ odd}}, \nn \\
    &\left[\sumal{\gamma = 1}{m}{\left(M_j\right)^{\beta_1}_{\gamma} \left(M_k\right)^{\gamma}_{\alpha_2}}, \he_{j,k}\right]  =\twopartdef{-m(\ket{\beta_1 \alpha_2}\bra{\psising})_{j,k} + \delta^{\beta_1}_{\alpha_2}\he_{j,k}}{\textrm{$j$ odd and $k$ even}}{-\delta^{\beta_1}_{\alpha_2}\he_{j,k} + m(\ket{\psising}\bra{\beta_1 \alpha_2})_{j,k}}{\textrm{$j$ even and $k$ odd}},\nn \\
    &\left[\sumal{\epsilon,\gamma = 1}{m}{\left(M_i\right)^{\gamma}_{\epsilon} \left(M_j\right)^{\epsilon}_{\gamma}}, \he_{j,k}\right] = 0.
\label{eq:ncommSUmcontract}
\end{align}
Using Eqs.~(\ref{eq:ncommSUm}) and (\ref{eq:ncommSUmcontract}), we can construct quadratic IoMs in a ``traceless" form~\cite{readsaleur2007} as
\begin{equation}
    M^{\beta_1, \beta_2}_{\alpha_1, \alpha_2} = \sumal{j_1 < j_2}{}{\left[\left(M_{j_1}\right)^{\beta_1}_{\alpha_1}\left(M_{j_2}\right)^{\beta_2}_{\alpha_2} - \frac{1}{m}\sumal{\gamma = 1}{m}{\left(\left(M_{j_1}\right)^{\beta_1}_{\gamma}\left(M_{j_2}\right)^{\gamma}_{\alpha_2}\delta^{\beta_2}_{\alpha_1} + \left(M_{j_1}\right)^{\gamma}_{\alpha_1}\left(M_{j_2}\right)^{\beta_2}_{\gamma}\delta^{\beta_1}_{\alpha_2}\right)} + \frac{1}{m^2} \delta^{\beta_1}_{\alpha_2}\delta^{\beta_2}_{\alpha_1}\sumal{\epsilon,\gamma = 1}{m}{\left(M_{j_1}\right)^{\gamma}_\epsilon \left(M_{j_2}\right)^{\epsilon}_\gamma}\right] }
\label{eq:J2conservedfull}
\end{equation}
When $\alpha_1 \neq \beta_2$ and $\beta_1 \neq \alpha_2$ this reduces to Eq.~(\ref{eq:TLJ2conservation}), while when one or both of these inequalities are not satisfied  we obtain new IoMs that are independent of the ones in Eq.~(\ref{eq:TLJ2conservation}) (up to relations such as $\sum_{\gamma=1}^m M^{\beta_1, \gamma}_{\gamma, \alpha_2} = 0$, etc.).
Similar new IoMs with a larger number of indices can also be constructed, and although their derivations are more involved, they have simple expressions in terms of Jones-Wenzl projectors, see Ref.~\cite{readsaleur2007} for details.

\section{Canonical Configurations in the Spin-1 Dipole-Conserving Model}\label{app:canonicalconfigs}
To understand the canonical configurations that generate the Krylov subspaces in the spin-1 dipole-conserving model of Eq.~(\ref{eq:spin1dip}), we introduce a diagrammatic representation for product states, and the mapping proceeds in two steps.
First, we represent sites using ``dots" -- we denote $\ket{0}$ by an unfilled dot $\tket{\udot{0}}$, and $\ket{+}$ and $\ket{-}$ both by filled dots $\tket{\fdot{0}}$.
Second, we then connect any alternating pattern of $\ket{+}$ and $\ket{-}$ by ``links," and we refer to any set of sites connected by links as a ``cluster" 
For example, we have the mapping
\begin{equation}
    \ket{0 + 0 + - 0 + 0 - 0 -} = \tket{\udot{0}\fdot{0.5}\udot{1}\dimer{1.5}{2}\udot{2.5}\arcdimer{2}{3}\udot{3.5}\arcdimer{3}{4}\udot{4.5}\fdot{5}} = \ket{0 - 0 - + 0 - 0 + 0 +}
\label{eq:dipoleexample}
\end{equation}
Note that the transitions of the dipole Hamiltonian in Eq.~(\ref{eq:spin1dipoletransitions}) are invariant under a global spin flip $\ket{+} \leftrightarrow \ket{-}$.
Hence, for any Krylov subspace (except for the one-dimensional subspace $\ket{0 0 \cdots 0 0}$), there is always a conjugate Krylov subspace with the spins flipped.
Any pattern of unfilled dots, filled dots, and links thus uniquely represents a product state in the Hilbert space up to an overall spin flip.
Although we do not distinguish between the conjugate subspaces in the diagrammatic representation, we will take this multiplicity into account while counting the number of Krylov subspaces. 
We note that this mapping of configurations to ``dots" and ``clusters" is related to a mapping to ``defects" discussed in Ref.~\cite{sliom2020}.
In particular, the leftmost dot in a cluster in the former mapping corresponds to a defect in the latter.\footnote{We thank Tibor Rakovszky and Pablo Sala for pointing this out.}
To understand the dynamics of the Hamiltonian $\hdip$ of Eq.~(\ref{eq:spin1dip}) in the diagrammatic representation, we note that the transitions of Eq.~(\ref{eq:spin1dipoletransitions}) allowed by the terms $\hP_{[j-1, j+1]}$ can be written as
\begin{equation}
    \tket{\ldasharc{0.5}\udot{0}\dimer{0.5}{1}\rdasharc{1}} \leftrightarrow \tket{\ldasharc{0}\rdasharc{0.5}\dimer{0}{0.5}\udot{1}},\;\;\;\;\tket{\ldasharc{0.5}\rdasharc{0.5}\udot{0}\fdot{0.5}\udot{1}} \leftrightarrow \tket{\ldasharc{0}\dimer{0}{0.5}\dimer{0.5}{1}\rdasharc{1}},
\label{eq:hdiptransitionsdiag}
\end{equation}
where the dotted lines indicate possible links to the left and right of the filled dots. 
Using Eq.~(\ref{eq:hdiptransitionsdiag}), we observe the following:
\begin{enumerate}
    \item Any product state is composed of a number of disconnected clusters, where each cluster is a connected set of (filled) dots.
    For example, the state of Eq.~(\ref{eq:dipoleexample}) is composed of three clusters of filled dots. 
    \item The actions of the terms $\{\hP_{[j-1,j+1]}\}$ do not change (i) the number of clusters, (ii) the charge and the closely related parity of the number of filled dots in any cluster (even-parity clusters have charge zero while odd-parity clusters have charge $\pm 1$), and (iii) the dipole moment of any cluster. Furthermore, the charge of the left-most filled dot in any cluster remains unchanged, and so does the right-most charge.
    \item Using the transitions of Eq.~(\ref{eq:hdiptransitionsdiag}), any cluster can be ``reduced" to a cluster that is either a filled dot or a link, depending on whether the number of filled dots in a cluster is odd or even. 
    For example, we can reduce the middle cluster in Eq.~(\ref{eq:dipoleexample}) as follows
    \begin{equation}
        \tket{\dimer{1.5}{2}\udot{2.5}\arcdimer{2}{3}\udot{3.5}\arcdimer{3}{4}} \leftrightarrow \tket{\udot{1.5}\dimer{2}{2.5}\dimer{2.5}{3}\udot{3.5}\arcdimer{3}{4}} \leftrightarrow \tket{\udot{1.5}\udot{2}\udot{3}\udot{3.5}\arcdimer{2.5}{4}}
    \label{eq:reductionexample}
    \end{equation}
    The dipole moment of the original cluster completely fixes the location of the final dot for odd parity or the length of the link for even pairty (but not its location).
    \item After the reduction of clusters, a canonical configuration for the state can be obtained by moving any cluster with a single link adjacent to the cluster on its right (or the right end of the chain) using the transitions of Eq.~(\ref{eq:hdiptransitionsdiag}).
    It is easy to see that this is always possible. 
    For example, after the reduction of Eq.~(\ref{eq:reductionexample}), the configuration Eq.~(\ref{eq:dipoleexample}) can be brought to a canonical form as follows
    \begin{gather}
        \tket{\udot{0}\fdot{0.5}\udot{1}\udot{1.5}\udot{2}\udot{3}\udot{3.5}\arcdimer{2.5}{4}\udot{4.5}\fdot{5}} \leftrightarrow \tket{\udot{0}\fdot{0.5}\udot{1}\udot{1.5}\udot{2}\udot{3}\arcdimer{2.5}{3.5}\dimer{3.5}{4}\dimer{4}{4.5}\fdot{5}} \leftrightarrow \tket{\udot{0}\fdot{0.5}\udot{1}\udot{1.5}\udot{2}\udot{3}\dimer{2.5}{3}\dimer{3}{3.5}\udot{4}\arcdimer{3.5}{4.5}\fdot{5}}\nn \\
        \leftrightarrow \tket{\udot{0}\fdot{0.5}\udot{1}\udot{1.5}\udot{2}\udot{2.5}\udot{3.5}\udot{4}\arcdimer{3}{4.5}\fdot{5}}
    \label{eq:reductionexample2}
    \end{gather}
\end{enumerate}
These observations allow us to bring any product state into a canonical configuration that consists of unfilled and filled dots, as well as clusters with a single link, which we refer to as ``dimers."
Furthermore, all the links are located immediately to the left of a filled dot or the right end of the chain. 
Each of these canonical configurations generates a different Krylov subspace, and the number of Krylov subspace is hence simply the number of such canonical configurations.
To count the number of canonical configurations, we uniquely map each of them to a tiling pattern of $L$ sites by three objects: $\btp\udot{0}\etp$, $\btp \fdot{0}\etp$, $\btp\dimer{0}{0.5}\etp$. 
We do this by retaining any filled and unfilled dots in the configuration, and by ``shortening" any dimer to a nearest neighboring dimer while keeping its left end fixed. 
For example, the invariant string of Eq.~(\ref{eq:reductionexample2}) maps to a tiling pattern as follows
\begin{equation}
    \tket{\udot{0}\fdot{0.5}\udot{1}\udot{1.5}\udot{2}\udot{2.5}\udot{3.5}\udot{4}\arcdimer{3}{4.5}\fdot{5}} : \fbox{\begin{tabular}{c}\btp \udot{0}\fdot{0.5}\udot{1}\udot{1.5}\udot{2}\udot{2.5}\udot{4.5}\udot{4}\dimer{3}{3.5}\fdot{5}\etp\end{tabular}}.
\label{eq:tilingmap}
\end{equation}
Note that with the convention we have used that any dimer in the physical configuration always has either the end of the chain or a filled dot to its right, the mapping to the tiling pattern is one to one. 
The number of canonical configurations is then simply the number of tilings of an $L \times 1$ ``grid" with two types of $1 \times 1$ ``squares" ($\btp\udot{0}\etp$ and $\btp\fdot{0}\etp$) and one type of $2 \times 1$ ``dominoes" ($\btp\dimer{0}{0.5}\etp$).
This is known to be the $(L+1)^{th}$ Pell number $P_{L+1}$~\cite{benjamin2008tiling}, where $P_0 = 0$, $P_1 = 1$, and  easy to understand recursion relation $P_{L + 1} =  2P_L + P_{L-1}$. 
Accounting for the fact that each configuration has conjugate (except for the one with unfilled dots on all sites), the number of Krylov subspaces $D_L$ (i.e., the dimension of the commutant algebra) of an $L$-site system with OBC is given by
\begin{equation}
    D_L = 2 P_{L+1} - 1 = \frac{(1+\sqrt{2})^{L+1} - (1-\sqrt{2})^{L+1}}{\sqrt{2}} - 1 
    \sim \frac{1}{\sqrt{2}} (\sqrt{2} + 1)^{L +1} \;\;\textrm{for large $L$}.
\label{eq:commutantdimdipole}
\end{equation}
Note that the mapping to dots, links and clusters can also be done for PBC, although care has to be taken to count the ones that cannot be connected to each other using the rules of Eq.~(\ref{eq:hdiptransitionsdiag}).

\section{Mazur bounds in the \texorpdfstring{$t-J_z$}{} model}\label{app:MazurtJz}
In this appendix, we compute three Mazur bounds for the OBC $t-J_z$ model, obtained by considering the left SLIOMs $\{\hq^{(\ell)}_l\}$,  right SLIOMs $\{\hq^{(r)}_l\}$, and the full commutant algebra $\Cobc$ respectively.
It is convenient to express the SLIOMs and operators in $\Cobc$ in terms of words of Eq.~(\ref{eq:worddefn}), and use the properties of Eqs.~(\ref{eq:properties}) and (\ref{eq:wordalgebra}) to compute norms and overlaps required for the Mazur bounds. 
\subsection{SLIOM Mazur Bounds}\label{subsec:SLIOMMazur}
To begin, we compute the norms and overlaps of the left and right SLIOMs with the $Z_j$ operators. 
Using Eqs.~(\ref{eq:SLIOMwords}) and (\ref{eq:rightSLIOMwords}), with simple combinatorics we obtain
\begin{gather}
    \langle \hq^{(x)}_l \hq^{(x)}_m \rangle = \delta_{l,m} \times \frac{1}{3^L}\sumal{\alpha = 0}{L-l}{\textrm{Tr}\left(P^{l + \alpha}\right)} = \delta_{l,m} \times \sumal{\alpha = 0}{L- l}{\frac{2^{l+\alpha}}{3^{L}} \binom{L}{l + \alpha}} = \delta_{l,m} \times \sumal{\alpha = l}{L}{\frac{2^l}{3^\alpha}\binom{\alpha - 1}{l-1}} \leq 1,\;\;\;x \in \{\ell, r\}\nn \\
    \langle Z_j \hq^{(\ell)}_l\rangle = \frac{1}{3^L}\sumal{\alpha = 0}{L-l}{\textrm{Tr}\left(Z_j P^{l-1}Z P^{\alpha}\right)} = \sumal{\alpha = 0}{L-l}{\frac{2^{l + \alpha}}{3^L}\binom{j - 1}{l-1}\binom{L-j}{\alpha}} = \frac{2^l}{3^j} \binom{j-1}{l-1} \nn \\
    \langle Z_j \hq^{(r)}_l \rangle = \frac{1}{3^L}\sumal{\alpha = 0}{L-l}{\textrm{Tr}\left(Z_j P^{\alpha}Z P^{l-1}\right)} = \sumal{\alpha = 0}{L-l}{\frac{2^{l + \alpha}}{3^L}\binom{j - 1}{\alpha}\binom{L-j}{l-1}} = \frac{2^l}{3^{L-j+1}} \binom{L - j}{l-1},
\label{eq:SLIOMnormoverlap}
\end{gather}
where $P^{l + \alpha}$, $P^{l-1} Z P^\alpha$, and $P^\alpha Z P^{l-1}$ denote ``words" defined in Eq.~(\ref{eq:worddefn}). 
Using these expressions, the corresponding Mazur bounds $M^{(\ell)}_{Z_j}$ and $M^{(r)}_{Z_j}$ defined in Eq.~(\ref{eq:mazurqtys}) read
\begin{equation}
    M^{(\ell)}_{Z_j} =  \sumal{l = 1}{L}{\frac{\langle Z_j \hq^{(\ell)}_l\rangle^2}{\langle \hq^{(\ell)}_l \hq^{(\ell)}_l\rangle}} = \frac{1}{3^{2j}}\sumal{l = 1}{j}{\frac{2^l \binom{j -1}{l-1}^2}{\sumal{\alpha = l}{L}{\frac{1}{3^\alpha}\binom{\alpha - 1}{l-1}}}}, \;\;M^{(r)}_{Z_j} = \sumal{l = 1}{L}{\frac{\langle Z_j \hq^{(r)}_l\rangle^2}{\langle \hq^{(r)}_l \hq^{(r)}_l\rangle}} = \frac{1}{3^{2(L-j+1)}}\sumal{l = 1}{L-j+1}{\frac{2^l \binom{L - j}{l-1}^2}{\sumal{\alpha = l}{L}{\frac{1}{3^\alpha}\binom{\alpha - 1}{l-1}}}}. 
\label{eq:mazursliom}
\end{equation}
%
%
%Furthermore, these do not seem to exactly saturate the late-time values of the autocorrelation functions in the bulk of the chain, as can be deduced using the exact diagonalization results shown in Ref.~\cite{sliom2020}. 
%
%
\subsection{OBC Commutant Mazur Bounds}\label{subsec:commutantMazur}
We now compute the bound $M^{(\textrm{obc})}_{Z_j}$ defined in Eq.~(\ref{eq:mazurqtys}), where $\{Q_\alpha\}$ is any orthogonal basis for the commutant $\Cobc$.
Here we choose $\{Q_\alpha\}$ to be the words defined in Eq.~(\ref{eq:worddefn}). 
Since we are interested in obtaining a bound on the autocorrelation function $Z_j$, the only words that have a non-zero overlap with $Z_j$ are the ones with a single $Z$ in the word, we refer to them as $W_{\alpha,\beta} \defn P^\alpha Z P^\beta$. 
Note that the left and right SLIOMs defined in Eqs.~(\ref{eq:SLIOMwords}) and (\ref{eq:rightSLIOMwords}) are subsets of superpositions of $\{W_{\alpha,\beta}\}$ and related by $\hq^{(\ell)}_l = \sumal{\alpha = 0}{L-l}{W_{l-1, \alpha}}$.
Using the properties of Eqs.~(\ref{eq:worddefn})-(\ref{eq:wordproduct}), we obtain the following relations
\begin{gather}
    \langle W_{\alpha,\beta} W_{\gamma,\delta} \rangle = \delta_{\alpha,\gamma} \delta_{\beta,\delta} \times \frac{1}{3^L}\textrm{Tr}\left(P^{\alpha + 1 + \beta}\right) = \delta_{\alpha,\gamma}\delta_{\beta,\delta} \times \frac{2^{\alpha + \beta + 1}}{3^L}\binom{L}{\alpha + \beta + 1}\nn \\
    \langle W_{\alpha,\beta} Z_j \rangle = \frac{2^{\alpha + \beta + 1}}{3^L}\binom{j-1}{\alpha}\binom{L-j}{\beta}.
\label{eq:Qalpbetdefn}
\end{gather}
The Mazur bound $M^{(\textrm{obc})}_{Z_j}$ then reads
\begin{equation}
    M^{(\textrm{obc})}_{Z_j} =  \sumal{\alpha = 0}{L-1}{\ \sumal{\beta = 0}{L-1-\alpha}{\frac{\langle Z_j W_{\alpha,\beta}\rangle^2}{\langle W_{\alpha,\beta} W_{\alpha,\beta}\rangle}}} = \sumal{\alpha = 0}{j-1}{\ \sumal{\beta = 0}{L-j}{\frac{2^{\alpha + \beta + 1}\binom{j-1}{\alpha}^2 \binom{L-j}{\beta}^2}{3^L \binom{L}{\alpha + \beta + 1}}}}.
\label{eq:mazurboundfull}
\end{equation}
The expression of Eq.~(\ref{eq:mazurboundfull}) also allows for an asymptotic scaling analysis of $M^{(\textrm{obc})}_{Z_j}$.
Defining $x \defn j/L$, $p \defn \alpha/L$, $q \defn \beta/L$, for large $L$, we can write (after tedious simplifications)
\begin{gather}
    M^{(\textrm{obc})}(x) = \sqrt{L}\int_0^x\int_0^{1-x}{dp\ dq\ C(x,p,q)\exp\left(L F(x,p,q)\right)} \nn \\
    F(x,p,q) \defn p \log\left(\frac{2(p+q)(x-p)^2}{(1-p-q)p^2}\right) +  q \log\left(\frac{2(p+q)(1-x-q)^2}{(1-p-q) q^2}\right)+ 2x\log\left(\frac{x(1-x-q)}{(x-p)(1-x)}\right) + \log\left(\frac{(1-p-q)(1-x)^2}{3(1-x-q)^2}\right)\nn \\
    C(x,p,q) \defn \frac{(x-p)(1-x)(p+q)^{3/2}}{(2\pi)^{3/2}x p q(1-x-q)(1-p-q)^{1/2}},
\label{eq:mazurfullintegral}
\end{gather}
where $M^{(\textrm{obc})}(x)$ is the continuum approximation for $M^{(\textrm{obc})}_{Z_j}$ for large $L$, and we have used Stirling's approximation to substitute
\begin{equation}
    \binom{y L}{z L} = \frac{1}{\sqrt{2\pi L}}\sqrt{\frac{y}{z(y-z)}}\exp\left(y\log y - z\log z - (y-z)\log(y-z)\right)\;\;\textrm{as}\;\;L \rightarrow \infty,\;\;\;0 < y \leq 1,\;\; 0 < z \leq y.
\end{equation}
Performing the standard saddle point approximation on the integral in Eq.~(\ref{eq:mazurfullintegral}) in the large $L$ limit, we obtain
\begin{equation}
     M^{(\textrm{obc})}(x) = \frac{2 \pi\ C(x, p_0, q_0)}{\sqrt{L\ \det H(x, p_0, q_0)}}\exp\left(L F(x,p_0,q_0)\right) = \frac{1}{3}\sqrt{\frac{2}{\pi L x(1-x)}}. 
\label{eq:mazursaddle}
\end{equation}
where $(p, q) = (p_0, q_0)$ is the saddle points of $F(x, p, q)$, $\det H(x, p, q)$ is the determinant of the Hessian matrix of $F(x, p, q)$, and we have used
\begin{gather}
    \left.\left(\frac{\partial F}{\partial p}, \frac{\partial F}{\partial q}\right)\right|_{(p,q) = (p_0, q_0)} = (0, 0) \;\;\implies (p_0, q_0) = \left(\frac{2x}{3}, \frac{2(1-x)}{3}\right),\nn \\
    \det H(x, p_0, q_0) = \frac{81}{2 x(1-x)},\;\; C(x, p_0, q_0) = \frac{3}{\pi^{\frac{3}{2}} 2 x(1-x)} ,\;\; F(x, p_0, q_0) = 0. 
\label{eq:p0q0soln}
\end{gather}
Eq.~(\ref{eq:mazursaddle}) rigorously proves the numerical observations in Ref.~\cite{sliom2020}. 
\subsection{PBC Commutant Mazur Bounds}\label{subsec:PBCcommutantMazur}
We now compute the Mazur bound $M^{(\textrm{pbc})}_{Z_j}$ defined in Eq.~(\ref{eq:PBCmazur}), where $\{Q_\alpha\}$ is an orthogonal basis for $\Cpbc$. 
Here, we choose $\{Q_\alpha\}$ to be the words of Eqs.~(\ref{eq:worddefnPBC}) and (\ref{eq:lengthLwords}).
Similar to the OBC case, we only focus on words that have a non-zero overlap with $Z_j$, which are the ones with a single $Z$ in the word.
Due to the different definitions of words of length $l \leq L - 1$ and words of length $L$, words with a single $Z$ can either have the form $W_{\alpha} \defn [P^{\alpha} Z]$ for $0 \leq \alpha \leq L - 2$, or $W_{L,\alpha} \defn [P^{\alpha} Z P^{L-\alpha-1}]$ for $0 \leq \alpha \leq L-1$. 
These words have the following properties
\begin{align}
    &\langle W_{\alpha} W_{\beta} \rangle = \delta_{\alpha,\beta} \times \frac{2^{\alpha+1}}{3^L}\binom{L}{\alpha +1}(\alpha + 1),\;\;\; \langle Z_j W_\alpha\rangle =\frac{2^{\alpha + 1}}{3^L}\binom{L-1}{\alpha}\;\; \textrm{for}\;\;0 \leq \alpha, \beta \leq L - 2\nn \\
    &\langle W_{L, \alpha} W_{L,\beta}\rangle = \delta_{\alpha, \beta} \times \left(\frac{2}{3}\right)^L,\;\;\;\langle Z_j W_{L, \alpha}\rangle = \delta_{j, \alpha+1} \times \left(\frac{2}{3}\right)^L\;\;\;\textrm{for}\;\; 0 \leq \alpha,\beta \leq L-1. 
\label{eq:PBCconservedproperties}
\end{align}
Using Eq.~(\ref{eq:PBCconservedproperties}) and (\ref{eq:PBCmazur}), the PBC Mazur bound reads
\begin{equation}
    M^{(\textrm{pbc})}_{Z_j} =  \sumal{\alpha = 0}{L-2}{\frac{\langle Z_j W_{\alpha}\rangle^2}{\langle W_{\alpha} W_{\alpha}\rangle}} + \sumal{\alpha = 0}{L-1}{\frac{\langle Z_j W_{L, \alpha}\rangle^2}{\langle W_{L, \alpha} W_{L, \alpha}\rangle}} = \sumal{\alpha = 0}{L-2}{\frac{2^{\alpha + 1}\binom{L-1}{\alpha}^2 }{3^L (\alpha + 1)\binom{L}{\alpha + 1}}} + \left(\frac{2}{3}\right)^L = \frac{2}{3L} + \left(1 - \frac{1}{L}\right)\left(\frac{2}{3}\right)^L.
\label{eq:mazurboundPBCfull}
\end{equation}
To explore if this is the case with other autocorrelation functions, we consider the Mazur bound for the operator $Z_j Z_{j+1}$.
Similar to the previous case, we choose $\{Q_\alpha\}$ to be the words of Eqs.~(\ref{eq:worddefnPBC}) and (\ref{eq:lengthLwords}), and only focus on words that have a non-zero overlap with $Z_j Z_{j+1}$, which are the ones with a two $Z$'s in the word.
Such words can either have the form $\tW_{\alpha} \defn [P^{\alpha} Z^2]$ for $0 \leq \alpha \leq L - 3$, or $\tW_{L,\alpha} \defn [P^{\alpha} Z^2 P^{L-\alpha-2}]$ for $0 \leq \alpha \leq L-2$. 
These words have the following properties
\begin{align}
    &\langle \tW_{\alpha} \tW_{\beta} \rangle = \delta_{\alpha,\beta} \times \frac{2^{\alpha+2}}{3^L}\binom{L}{\alpha +2}(\alpha + 2),\;\;\; \langle Z_j Z_{j+1} \tW_\alpha\rangle =\twopartdef{\frac{2^{\alpha + 2}}{3^L}\binom{L-2}{\alpha}}{\alpha \geq 1}{\frac{4}{3^L}\times 2}{\alpha = 0}\;\; \textrm{for}\;\;0 \leq \alpha, \beta \leq L - 3\nn \\
    &\langle \tW_{L, \alpha} \tW_{L,\beta}\rangle = \delta_{\alpha, \beta} \times \left(\frac{2}{3}\right)^L,\;\;\;\langle Z_j Z_{j+1} W_{L, \alpha}\rangle = \delta_{j, \alpha+1} \times \left(\frac{2}{3}\right)^L\;\;\;\textrm{for}\;\; 0 \leq \alpha,\beta \leq L-2. 
\label{eq:PBCconservedproperties2}
\end{align}
The Mazur bound $M^{(\textrm{pbc})}_{Z_j Z_{j+1}}$ is then given as follows
\begin{gather}
    M^{(\textrm{pbc})}_{Z_j Z_{j+1}} = \sumal{\alpha = 0}{L-3}{\frac{\langle Z_j Z_{j+1} \tW_{\alpha}\rangle^2}{\langle \tW_{\alpha} \tW_{\alpha}\rangle}} + \sumal{\alpha = 0}{L-1}{\frac{\langle Z_j Z_{j+1} \tW_{L, \alpha}\rangle^2}{\langle \tW_{L, \alpha} \tW_{L, \alpha}\rangle}} = \frac{16}{3^L \times L (L-1)} +  \sumal{\alpha = 1}{L-3}{\frac{\binom{L-2}{\alpha}^2 2^{\alpha+2}}{3^L (\alpha+2) \binom{L}{\alpha+2}}} + \left(\frac{2}{3}\right)^L\nn \\
    = \frac{8}{27 (L-1)} - \frac{4}{27 L (L-1)} + \left(\frac{2}{3}\right)^L\left(1-\frac{1}{L}\right) + \frac{12}{3^L \times L (L-1)} \sim \frac{1}{L}.
\label{eq:ZZmazurpbc}
\end{gather}
\section{Mazur Bounds in the Spin-1 Dipole-Conserving Model}\label{app:dipolemazur}
In this appendix, we compute the contribution of a class of blockaded Krylov subspaces to the Mazur bound of the local spin operator $Z_j$ in the spin-1 dipole-conserving model.
As discussed in Sec.~\ref{subsubsec:dipolemazur}, the Mazur bound contribution $M^{(\textrm{block})}_{Z_j}$ of all the subspaces in which site $j$ frozen is given by Eq.~(\ref{eq:dipblockmazur}).
Hence it is sufficient to compute $\sum_{\alpha}{D_{\mK_{j,\alpha}}}$, the total dimension of the Krylov subspaces $\mK_{j,\alpha}$, where site $j$ is frozen and $Z_j = \pm 1$. 
In the following, we do this by studying the blockaded subspaces in the dots and links language that is convenient to describe the dynamics in the spin-1 dipole-conserving model (see App.~\ref{app:canonicalconfigs}). 
Recall that $\btp\fdot{0}\etp$ denotes spins $+$ or $-$, $\btp\udot{0}\etp$ denotes spin $0$, and links connect adjacent filled dots if they have opposite $Z_j$'s.
We first note that in the language of dots and links, we can always represent states in a blockaded Krylov subspace in following form
\begin{equation}
    \tket{\raisebox{-2mm}{\tbox{-1.3}{1.7}\betdot{-1.1}\betdot{-0.8}\betdot{-0.5}\ldasharc{-0.2}\fdot{-0.2}\udot{0.3}\betdot{0.6}\betdot{0.9}\betdot{1.2}\udot{1.5}
    \tbox{1.8}{4.7}\fdot{2}\fdot{2.5}\rdasharc{2.5}\betdot{2.8}\betdot{3.1}\betdot{3.4}\betdot{3.7}\ldasharc{4}\fdot{4}\fdot{4.5}
    \tbox{4.8}{7.8}\udot{5}\betdot{5.3}\betdot{5.6}\betdot{5.9}\udot{6.2}\fdot{6.7}\rdasharc{6.7}\betdot{7}\betdot{7.3}\betdot{7.6}}},
\label{eq:blockcanonical}
\end{equation}
where the boxes denote the configurations on the left region ($\ket{L^{(\ell)}}$), the blockade ($\ket{B^{(b)}}$), and the right region ($\ket{R^{(r)}}$) with lengths $\ell$, $b$, and $r = L - \ell - b$ respectively (see Sec.~\ref{subsec:invariantstrings}), with more details about these regions given below, and the dotted lines indicate possible links to the left or right of the filled dots. 
The form in Eq.~(\ref{eq:blockcanonical}) has two main implications.
First, the leftmost two spins of $\ket{B^{(b)}}$ are always $+ +$ or $- -$, and the same is true about the rightmost two spins.
Second, the rightmost (resp. leftmost) non-zero spin (i.e., that are $+$ or $-$) of $\ket{L^{(\ell)}}$ (resp. $\ket{R^{(r)}}$) is the same as the leftmost (resp. rightmost) spin of $\ket{B^{(b)}}$. Note that it is also possible to have all $0$'s in the left and right regions. 
As a consequence, there are no links that connect the left or right region to the blockaded region. 
It is easy to verify that as long as the entire middle region is frozen, the rules of Eq.~(\ref{eq:hdiptransitionsdiag}) preserve the form of Eq.~(\ref{eq:blockcanonical}).
Given a frozen site $j$, we can always choose the blockade configuration $\ket{B^{(b)}}$ of the form of Eq.~(\ref{eq:blockcanonical}) that includes site $j$.
Further, without loss of generality, we can require that the only $\btp\fdot{0}\etp$'s (that are not connected by a link) in the blockade $\ket{B^{(b)}}$ are on its leftmost and rightmost ends, and we refer to such blockades  as ``irreducible blockades" $\ket{I^{(b)}}$.
It is easy to verify that any irreducible blockade configurations should have the following form
\begin{equation}
    \ket{I^{(b)}} = \tket{\fdot{0}
    \arcdimer{0.5}{3}\udot{1}\betdot{1.3}\betdot{1.6}\betdot{1.9}\betdot{2.2}\udot{2.5}
    \arcdimer{3.5}{6}\udot{4}\betdot{4.3}\betdot{4.6}\betdot{4.9}\betdot{5.2}\udot{5.5}
    \betdot{6.3}\betdot{6.6}\betdot{6.9}\betdot{7.2}
    \arcdimer{7.5}{10}\udot{8}\betdot{8.3}\betdot{8.6}\betdot{8.9}\betdot{9.2}\udot{9.5}
    \fdot{10.5}},
\label{eq:irreducibleblockade}
\end{equation}
where there can be any number of dimers of varying lengths, and the only constraint is that they should be ``maximally packed" with no $\btp\udot{0}\etp$'s between two dimers.
Note that we also include the case with zero dimers, i.e., $\tket{\fdot{0}\fdot{0.5}}$ is an irreducible blockade.
However, given a frozen site $j$, the choice of the irreducible blockade is not always unique, and it could be a part of two such irreducible blockades $\ket{I^{(b)}}$ and $\ket{I^{(b')}}$.
In such a case, we refer to the full blockade region as $\ket{I^{(b)} \cup I^{(b')}}$, which could have a configuration such as 
\begin{equation}
    \ket{I^{(b)} \cup I^{(b')}} = \tket{\fdot{0}
    \arcdimer{0.5}{3}\udot{1}\betdot{1.3}\betdot{1.6}\betdot{1.9}\betdot{2.2}\udot{2.5}
    \betdot{3.3}\betdot{3.6}\betdot{3.9}\betdot{4.2}
    \arcdimer{4.5}{7}\udot{5}\betdot{5.3}\betdot{5.6}\betdot{5.9}\betdot{6.2}\udot{6.5}
    \fdot{7.5}
    \arcdimer{8}{10.5}\udot{8.5}\betdot{8.8}\betdot{9.1}\betdot{9.4}\betdot{9.7}\udot{10}
    \betdot{10.8}\betdot{11.1}\betdot{11.4}\betdot{11.7}
    \arcdimer{12}{14.5}\udot{12.5}\betdot{12.8}\betdot{13.1}\betdot{13.4}\betdot{13.7}\udot{14}
    \fdot{15}},
\label{eq:irreducibleovercount}
\end{equation}
where $j$ is the $\btp\fdot{0}\etp$ in the middle, and the blockades to the left and the right of $j$ (including site $j$) are $\ket{I^{(b)}}$ and $\ket{I^{(b')}}$ respectively.
As we discuss below, such configurations will be important to correct for the overcounting in our counting procedure.
The main strategy to compute the total dimension of the described blockaded Krylov subspaces is as follows.
We first fix $\ket{B^{(b)}}$ in the middle box in Eq.~(\ref{eq:blockcanonical}) to be an irreducible blockade $\ket{I^{(b)}}$ containing site $j$. 
The total dimension of all Krylov subspaces with that blockade configuration reduces to a simple counting of the number of allowed configurations of the left and right regions that are compatible with the blockade region (i.e., they have the form of Eq.~(\ref{eq:blockcanonical})). 
The total dimension of all blockaded Krylov subspaces is then obtained by summing over all possible irreducible blockade configurations, and correcting for overcounting due to subspaces in which $j$ can be a part of multiple irreducible blockades.  
In the following, we continue to denote the number of spins in the left, blockade, and right regions of the chain by $\ell$, $b$, and $r$ respectively, such that $\ell + b + r = L$. 
We label all Krylov subspaces with $\ell$ sites in the left region and blockade configuration $\ket{B^{(b)}}$ by $\mK(B^{(b)}, \ell)$, where we have in mind configurations of the form of Eq.~(\ref{eq:blockcanonical}) (note that we are using one symbol to denote many subspaces), and we denote the dimension of all such subspaces by $D_{\mK(B^{(b)}, \ell)}$.
Using the form of Eq.~(\ref{eq:blockcanonical}), this is simply given by 
\begin{equation}
    D_{\mK(B^{(b)}, \ell)} = D_{\ell} \times D_{r},\;\;\;D_{x} = \frac{3^x + 1}{2},\;\;\;x \in \{\ell, r\},\;\;\;r = L - \ell - b.
\label{eq:blockdimensions}
\end{equation}
where $D_\ell$ (resp. $D_r$) is the number of allowed configurations in the left (resp. right) regions with the rightmost (resp. leftmost) non-zero spin being the same as the leftmost (resp. rightmost) of $\ket{B^{(b)}}$. 
The expressions for $D_{\ell}$ and $D_r$ in Eq.~(\ref{eq:blockdimensions}) can be verified straightforwardly, and are independent of the particular configuration $\ket{B^{(b)}}$.  
Further, if $j$ is deep in the bulk of the chain, we have $1 \ll \ell \ll L$ (i.e., $\ell = yL$ for $0 < y < 1$), and $D_{\mK(B^{(b)}, \ell)}$ for a blockade of range $b$ can be estimated as follows (measured relative to the total Hilbert space dimension)
\begin{equation}
    \frac{1}{3^L} D_{\mK(B^{(b)}, \ell)} =
    \frac{1}{3^L} D_{\ell} D_{L - \ell - b} = \frac{1}{4 \times 3^L}\left(3^{\ell} + 1\right)\left(3^{L - \ell - b} + 1\right) = \frac{1}{4\times 3^b} + \mathcal{O}(e^{-L}),
\label{eq:eachterm}
\end{equation}
where $b$ is the size of the blockade. Note that this calculation is for fixed (but thermodynamically large) $\ell$, which however drops out in the thermodynamic limit. The result depends on only on the size of the assumed blockade region but is independent of the internal structure of this region, which in particular can be reducible.
Turning to Mazur bound calculations, heuristically, the Mazur bound $M^{(\textrm{block})}_{Z_j}$ of Eq.~(\ref{eq:dipblockmazur}) is expressed as follows
\begin{equation}
    M^{(\textrm{block})}_{Z_j} = \frac{1}{3^L}\sumal{\alpha}{}{D_{\mK_{j,\alpha}}} = \frac{1}{3^L}\sumal{(I^{(b)}, \ell)}{}{D_{\mK(I^{(b)}, \ell)} \delta_{Z_j, \pm}} - \frac{1}{3^L}\sumal{I^{(b)}, I^{(b')}}{}{D_{\mK(I^{(b)} \cup I^{(b')}, j - b)}}.
\label{eq:blockmazur}
\end{equation}
In the first sum, $\{(I^{(b)}, \ell)\}$ runs over irreducible blocks and their positions covering $j$, and $\delta_{Z_j, \pm}$ ensures that the frozen site $j$ is either $+$ or $-$ (so that it has a non-vanishing contribution to the Mazur bound, see discussion in Sec.~\ref{subsubsec:dipolemazur}).
Since we are interested only in the Krylov subspaces where site $j$ belongs to the blockaded region, we have $j - b \leq \ell \leq j - 1$.
In the second sum in Eq.~(\ref{eq:blockmazur}), $I^{(b)}$ and $I^{(b')}$ run over irreducible blocks; this term accounts for blockade configurations of the form of Eq.~(\ref{eq:irreducibleovercount}), which have been double-counted in the first sum.
In other words, the restriction to irreducible blocks in the first sum takes care of most possibilities of overcounting, and the only overcounting that appears from naive summation over $\ell$ is cancelled by the second sum.
Before proceeding, we note that according to Eq.~(\ref{eq:eachterm}), the contributions of the various blockades in Eq.~(\ref{eq:blockmazur}) depend only on the blockade size, hence it is useful to arrange the sums there according to blockade size, i.e., 
\begin{equation}
    M^{(\textrm{block})}_{Z_j} = \sumal{b}{}{M^{(b)}_{Z_j}} - \sumal{b,b'}{}{O^{(b,b')}_{Z_j}},
\label{eq:blocksizewise}
\end{equation}
where the sum is over blockade sizes, and $M^{(b)}_{Z_j}$ is the first sum in Eq.~(\ref{eq:blockmazur}) with $I^{(b)}$'s restricted to size $b$, and $O^{(b,b')}_{Z_j}$ is the second sum with $I^{(b)}$'s and $I^{(b')}$'s restricted to sizes $b$ and $b'$.
Since their contributions decay exponentially with blockade size, we expect smaller values of $b$ to dominate the sums in Eq.~(\ref{eq:blocksizewise}).
We start with computing the contribution of two-site blockades.
According to Eq.~(\ref{eq:irreducibleblockade}), the only two-site irreducible blockade is $\ket{I^{(2)}} = \tket{\fdot{0}\fdot{0.5}}$.
In order for the site $j$ to be within the blockade subspace and $Z_j = \pm$, we have two choices for $\ell$, i.e., $\ell = j - 2$ or $\ell = j - 1$. 
However, by including the contribution of both these choices of $\ell$, we have double-counted the contribution from all the Krylov subspaces with the three-site blockade $\ket{I^{(2)} \cup I^{(2)}} = \tket{\fdot{0}\fdot{0.5}\fdot{1}}$ with $j$ on the middle site (hence $\ell = j-2$). 
Using Eqs.~(\ref{eq:blocksizewise}), (\ref{eq:blockmazur}), and (\ref{eq:eachterm}), we obtain the total contribution due to two-site blockades to be
\begin{gather}
    M^{(2)}_{Z_j} = 2 \times \frac{2}{4 \times 3^2},\;\;\;O^{(2,2)}_{Z_j} = 2 \times \frac{1}{4 \times 3^3},\;\;\;\left.M^{(\textrm{block})}_{Z_j}\right|_{\textrm{two-site}} \defn M^{(2)}_{Z_j} - O^{(2,2)}_{Z_j} = \frac{5}{54} \approx 0.0926,
\label{eq:2siteblockade}
\end{gather}
where the overall factors of $2$ accounts for the two possibilities $\btp\fdot{0}\etp = \pm$.
Moving on to three-site blockades, we find that there are no irreducible blockade configurations, hence $M^{(3)}_{Z_j} = O^{(3, b)}_{Z_j} = O^{(b, 3)}_{Z_j} = 0$. 
Next, we consider four-site blockades, where the unique irreducible blockade is $\ket{I^{(4)}} = \tket{\fdot{0}\dimer{0.5}{1}\fdot{1.5}}$.
In this case, we have four possible values of $\ell$ ($j-4 \leq \ell \leq j-1$) such that $Z_j = \pm 1$.
However, by including all their contributions, we are double counting the contributions of the following blockades:
\begin{equation}
    \ket{I^{(4)}\cup I^{(4)}} = \underset{j}{\tket{\fdot{0}\dimer{0.5}{1}\fdot{1.5}\dimer{2}{2.5}\fdot{3}}},\;\;\;\ket{I^{(4)} \cup I^{(2)}} = \underset{\hspace{10mm}j}{\tket{\fdot{0}\dimer{0.5}{1}\fdot{1.5}\fdot{2}}},\;\;\ket{I^{(2)} \cup I^{(4)}} = \underset{j\hspace{10mm}}{\tket{\fdot{0}\fdot{0.5}\dimer{1}{1.5}\fdot{2}}}.
\end{equation}
Accounting for the overcounting and doubling to account for the two possible spin patterns, using Eqs.~(\ref{eq:blocksizewise}) and (\ref{eq:eachterm}) we obtain
\begin{gather}
    M^{(4)}_{Z_j} = 2 \times \frac{4}{4 \times 3^4},\;\;\;O^{(2,4)}_{Z_j} = O^{(4,2)}_{Z_j} = 2 \times \frac{1}{4 \times 3^5},\;\;\;O^{(4,4)}_{Z_j} = 2 \times \frac{1}{4 \times 3^7},\nn \\
    \left.M^{(\textrm{block})}_{Z_j}\right|_{\textrm{four-site}} \defn M^{(4)}_{Z_j} - O^{(4,2)}_{Z_j} - O^{(2,4)}_{Z_j} - O^{(4,4)}_{Z_j} = \frac{89}{4374} \approx 0.02035.
\label{eq:4siteblockade}
\end{gather}
Similarly, we can consider irreducible blockades $\ket{I^{(b)}}$ of the form of Eq.~(\ref{eq:irreducibleblockade}) that span $b$ sites. 
Simple combinatorics gives the number of such blockades with $n$ dimers to be
\begin{equation}
    N_{b, n} = \binom{b-n-3}{n-1},\;\;\;0 \leq n \leq \floor*{\frac{b}{2}}-1. \;\;\;
\label{eq:numblockades}
\end{equation}
Note that although the combinatorics is not directly valid for $n = 0$ or $b \leq 3$, we nevertheless find that $N_{b,n}$ of Eq.~(\ref{eq:numblockades}) is the correct number of blockaded configurations even in those cases.  
For each blockade with $n$ dimers, there are $(2n + 2)$ available choices of $\ell$ such that $Z_j = \pm$. 
However, similar to the $b = 2$ and $b = 4$ cases, by including all such positions, we are overcounting the contribution of the $(b + b' - 1)$-site blockade configurations of the from $\ket{I^{(b)} \cup I^{(b')}}$.
[The overcount only happens when using the blockade probability Eq.~(\ref{eq:eachterm}) for choices $\ell$ where $j$ is one of the endpoints of an irreducible blockade of size $b$, since part of the blockade satisfaction condition in Eq.~(\ref{eq:eachterm}) comes from immediately adjacent irreducible blockade configurations of various sizes $b'$ sharing site $j$. There is no overcount when $j$ belongs to one of the dimers in an irreducible blockade.]
Similar to Eqs.~(\ref{eq:2siteblockade}) and (\ref{eq:4siteblockade}), we then obtain
\begin{equation}
    M^{(b)}_{Z_j} = 2 \times \sumal{n = 0}{\floor*{\frac{b}{2}}-1}{N_{b,n}\left(\frac{2n+2}{4 \times 3^b}\right)} = \frac{F^{(2)}_{b-3} + F_{b-3}}{3^b},\;\;\;O^{(b,b')}_{Z_j} = 2\times \frac{1}{4 \times 3^{b + b'-1}}\left(\sumal{n = 0}{\floor*{\frac{b}{2}}-1}{N_{b,n}}\right)\left(\sumal{n' = 0}{\floor*{\frac{b'}{2}}-1}{N_{b',n'}}\right)  = \frac{F_{b-3} F_{b'-3}}{2 \times 3^{b + b'-1}},
\label{eq:mazurbblock}
\end{equation}
where $\{F_n\}$ are the Fibonacci numbers and $\{F^{(2)}_n\}$ are the second-order Fibonacci numbers (defined by the recursion relation $F^{(2)}_n = F^{(2)}_{n-1} + F^{(2)}_{n-2} + F_{n-2}$), and we have used the conventions $F_0 = F^{(2)}_0 = F^{(2)}_{-1} = 0$ and $F_{-1} = 1$.\footnote{We can also see the appearance of the Fibonacci numbers by reproducing the above counts without first organizing by the number of dimers:  The number $P_k$ of all possible close packings of dimers on $k = b-2$ sites in Eq.~(\ref{eq:irreducibleblockade}) satisfies $P_k = \sum_{m=2}^k P_{k-m}$, where we organized by the number of sites $m$ covered by the first dimer.  
This is equivalent to standard Fibonacci recursion, with initial conditions such that $P_k = F_{k-1}$.
We also need the sum over all possible close packings weighted by the number of dimers in the packing.
Denoting this by $W_k$ on $k$ sites, we have $W_k = \sum_{m=2}^k (W_{k-m} + P_{k-m})$, which with the desired initial conditions leads to identification $W_k = F_{k-1}^{(2)}$.}
In the above, we have already used $L \rightarrow \infty$ expressions for the blockade probabilities in Eq.~(\ref{eq:eachterm}).
Working in the thermodynamic limit, using Eqs.~(\ref{eq:blocksizewise}) and (\ref{eq:mazurbblock}), the Mazur bound due to all the blockaded subspaces is then given by
\begin{equation}
    M^{(\textrm{block})}_{Z_j} = \sumal{b = 2}{\infty}{\left(\frac{F^{(2)}_{b-3}}{3^b} + \frac{F_{b-3}}{3^b}\right)} - \frac{3}{2}\sumal{b,b' = 2}{\infty}{\frac{F_{b-3} F_{b'-3}}{3^{b + b'}}} = \left.x^3 G^{(2)}(x)\right|_{x = \frac{1}{3}} + \left.(x^2 + x^3 G(x))\right|_{x = \frac{1}{3}} - \frac{3}{2}\left.(x^2 + x^3 G(x))^2\right|_{x = \frac{1}{3}},
\label{eq:mazurblockseries}
\end{equation}
where $G(x)$ and $G^{(2)}(x)$ are generating functions corresponding to the Fibonacci and second-order Fibonacci numbers, which are known to be~\cite{oeis}
\begin{equation}
    G(x) \defn \sumal{n = 0}{\infty}{F_n x^n} = \frac{x}{1-x-x^2},\;\;\;G^{(2)}(x) \defn \sumal{n = 0}{\infty}{F^{(2)}_n x^n} = \frac{x\ (1-x)}{(1 - x - x^2)^2}.
\label{eq:generatingfunctions}
\end{equation}
Substituting Eq.~(\ref{eq:generatingfunctions}) into Eq.~(\ref{eq:mazurblockseries}), we obtain
\begin{equation}
    M^{(\textrm{block})}_{Z_j} = \frac{2}{15} \approx 0.1333.
\label{eq:blockadeexact}
\end{equation}
\section{Mazur Bounds in the Temperley-Lieb Model}\label{app:TLmazur}
In this appendix, we provide details on the Mazur bound computation for the edge energy operator $\he_{1,2}$ in the Temperley-Lieb models of Eq.~(\ref{eq:TLHamil}). 
We start by simplifying Eq.~(\ref{eq:Mej}) as
\begin{equation}
M^{(\textrm{TL})}_{\he_{j,j+1}} = \frac{m^2}{m^L} \sum_\alpha \frac{[N_{\mK_\alpha}^{\text{dimer@}(j,j+1)}]^2}{D_{\mK_\alpha}} = \frac{m^2}{m^L} \sum_{\lambda=0}^{L/2-1} d_\lambda \frac{[N_\lambda^{\text{dimer@}(j,j+1)}]^2}{D_\lambda} ~.
\label{eq:Mejapp}
\end{equation}
where $N_{\mK_\alpha}^{\text{dimer@}(j,j+1)}$ is the number of configurations in $\mK_\alpha$ with a dimer between sites $j$ and $j+1$, and $D_{\mK_\alpha}$ is the dimension of the Krylov subspace $\mK_\alpha$.
Further, we have taken $L$ even for concreteness and have used results from Eq.~(\ref{eq:TLdimensions}): given $2\lambda$ dots, there are $d_\lambda$ distinct Krylov subspaces, each of dimension $D_\lambda$, and with the same action of the TL algebra in terms of the dots and dimers pictures; in particular, each such Krylov subspace has the same number of configurations with a dimer between sites $j$ and $j+1$, which we have denoted by $N^{\text{dimer@}(j,j+1)}_\lambda$.
Note that the sum terminates at $\lambda = L/2 - 1$ since the $2\lambda$ dots must be on the sites other than $j,j+1$.
At the edge, $j=1$, $N^{\text{dimer@}(1,2)}_\lambda$ is the same as the number of dots and dimers pictures with $2\lambda$ dots and %dimers 
restricted to the remaining $L-2$ sites, which is given by $D_\lambda$ in Eq.~(\ref{eq:TLdimensions}) with $L$ replaced by $L-2$.
Marking the appropriate chain lengths with superscripts on $D_\lambda$, we have
\begin{equation}
M^{(\textrm{TL})}_{\he_{1,2}} = \frac{m^2}{m^L} \sum_{\lambda=0}^{L/2-1} d_\lambda \frac{[D_\lambda^{(L-2)}]^2}{D_\lambda^{(L)}} = \frac{m^2}{m^L} \sum_{\lambda=0}^{L/2-1} d_\lambda D_\lambda^{(L-2)} \frac{(L+1)^2 - (2\lambda+1)^2}{4L(L-1)} ~,
\label{eq:MTLexpand}
\end{equation}
where in the last equation we factored out one $D_\lambda^{(L-2)}$ and used a simple expression for $D_\lambda^{(L-2)}/D_\lambda^{(L)}$ derived from Eq.~(\ref{eq:TLdimensions}).
This writing enables exact evaluation of the Mazur bound by utilizing the knowledge that 
\begin{equation}
    \sum_{\lambda=0}^{K/2} d_\lambda D_\lambda^{(K)} = m^K,
\label{eq:TLfullHilbert}
\end{equation}
the full Hilbert space dimension on $K$ sites (assumed even for concreteness, see Eqs.~(\ref{eq:dimensions}) and (\ref{eq:TLdimensions})).
We then compute $\sum_{\lambda=0}^{K/2} (2\lambda+1)^2 d_\lambda D_\lambda^{(K)}$, by taking appropriate derivatives of Eq.~(\ref{eq:TLfullHilbert}) with respect to the formal quantum group parameter $q$ that is an argument of $d_\lambda$ (see Eq.~(\ref{eq:TLdimensions})) and that enters the right-hand-side via $m = q + q^{-1}$, see Eq.~(\ref{eq:qdefn}).
It is then easy to verify that 
\begin{align}
\sum_{\lambda=0}^{K/2} (2\lambda+1)^2 d_\lambda D_\lambda^{(K)} &= \frac{q}{q-q^{-1}} \frac{\mathrm{d}}{\mathrm{d}q} \left[q \frac{\mathrm{d}}{\mathrm{d}q} [m^K (q-q^{-1})] \right],\;\;\;m = q + q^{-1},\;\;\;d_\lambda = [2\lambda + 1]_q\nn \\
&= m^K [(K+1)^2 - 4K(K-1)/m^2]
\label{eq:TLderiv}
\end{align}
Using Eqs.~(\ref{eq:MTLexpand}), (\ref{eq:TLfullHilbert}), and (\ref{eq:TLderiv}), we obtain the connected Mazur bound of Eq.~(\ref{eq:Mconndefn}) to be
\begin{equation}
M^\text{conn}_{\he_{1,2}} = \frac{1}{L-1} \left[1 - \frac{4}{m^2} + \frac{6}{m^2 L} \right] ~.
\label{eq:Mconnapp}
\end{equation}

\section{Matrix Product Operators (MPO) forms of the Commutant Algebra Basis Elements}\label{app:MPOforms}
In this appendix, we show that the IoMs in the commutant algebras in the $t-J_z$, PF, and TL models (with OBC) in the main text have simple MPO expressions. 
Any operator $\hat{\mathcal{O}}$ is said to have an efficient MPO representation if they can be written as
\begin{equation}
    \hat{\mathcal{O}} = \sum\limits_{\{s_n\}, \{t_n\}}{[{b_A^l}^T A_1^{[s_1 t_1]} A_2^{[s_2 t_2]} \dots A_L^{[s_L t_L]} b_A^r]}\ket{\{s_n\}}\bra{\{t_n\}},
\label{eq:generalOBCMPO}
\end{equation}
where $A$ can be thought of as $\chi \times \chi$ matrices with elements expressed as $d \times d$ matrices acting on the physical indices.
$\chi$ is referred to as the bond dimension of the MPO and the corresponding vector space is the auxiliary space, and $b_{A}^l$ and $b_{A}^r$ are $\chi$-dimensional boundary vectors of the MPO in the auxiliary space, which are usually set to $b^l_A = (1 \;\; 0 \;\; \cdots \;\; 0)^T$ and $b^r_A = (0 \;\; \cdots \;\; 0 \;\; 1)^T$ respectively.
The MPO expressions for local operators is straightforward to construct, e.g., the $U(1)$ generators $N^{\sigma}$'s of Eq.~(\ref{eq:tJzsymmetries}) can be expressed as an MPO with bond dimension $\chi = 2$, where
\begin{equation}
    A_j = 
    \begin{pmatrix}
        \mathds{1} & N^\sigma_j \\
        0 & \mathds{1}
    \end{pmatrix},\;\;\;\sigma \in \{\uparrow, \downarrow\}.
\label{eq:1MPOexpression}
\end{equation}
For the OBC $t-J_z$ model, the MPO expressions for the conserved quantities $N^{\sigma_1 \cdots \sigma_k}$ of Eq.~(\ref{eq:tJzconserved}) can be constructed using systematic methods in the literature, see e.g., Refs.~\cite{crosswhite2008fsa,motruk2016density, moudgalya2018b}.
We find that $N^{\sigma_1 \cdots \sigma_k}$ has the form of Eq.~(\ref{eq:generalOBCMPO}), where $A_j$ has bond dimension $\chi = k+1$, and is given by
\begin{equation}
    A_j = 
    \begin{pmatrix}
        \mathds{1} & N^{\sigma_1}_j & 0 & \cdots & 0 \\
        0 & \mathds{1} & N^{\sigma_2}_j & \ddots & \vdots \\
        \vdots & \ddots & \ddots  & \ddots & 0 \\
        \vdots & \ddots & \ddots  & \mathds{1} & N^{\sigma_k}_j \\
        0 & \cdots & \cdots  & 0 & \mathds{1}
    \end{pmatrix}.
\label{eq:kMPOexpression}
\end{equation}
The MPO representation of Eq.~(\ref{eq:kMPOexpression}) also helps to directly show that $N^{\sigma_1 \cdots \sigma_k}$ is a conserved quantity of the $t-J_z$ model.
This is evident from the two-site MPO which reads
\begin{equation}
    A_j A_{j+1} = 
    \begin{pmatrix}
        \mathds{1} & N^{\sigma_1}_j + N^{\sigma_1}_{j+1} & N^{\sigma_1}_j N^{\sigma_2}_{j+1} & 0 & \cdots & 0 \\
        0 & \mathds{1} & N^{\sigma_2}_j + N^{\sigma_2}_{j+1} & N^{\sigma_2}_j N^{\sigma_3}_{j+1} & \ddots & \vdots \\
        \vdots & \ddots & \ddots  & \ddots & \ddots & 0  \\
        \vdots & \ddots & \ddots  &\mathds{1} & N^{\sigma_{k-1}}_j + N^{\sigma_{k-1}}_{j+1} & N^{\sigma_{k-1}}_j N^{\sigma_{k}}_{j+1} \\
        \vdots & \ddots & \ddots  &\ddots & \mathds{1} & N^{\sigma_k}_j + N^{\sigma_k}_{j+1} \\
        0 & \cdots & \cdots  &\cdots & 0 & \mathds{1}
    \end{pmatrix}.
\label{eq:kMPOexpression2}
\end{equation}
In Eq.~(\ref{eq:kMPOexpression2}), each element within $A_j A_{j+1}$ commutes with the terms $\hT_{j,j+1}$ and $\hV_{j,j+1}$ of $H_{t-J_z}$ as a consequence of Eqs.~(\ref{eq:localcomm}) and (\ref{eq:nncomm}), and thus the MPO generated by $A_j$ of Eq.~(\ref{eq:kMPOexpression}) commutes with all the $\{\hT_{k,k+1}\}$ and $\{\hV_{k,k+1}\}$ and is guaranteed to be a global conserved quantity of $H_{t-J_z}$. 
Similarly, the IoMs $N^{\alpha_1 \cdots \alpha_k}$ and $M^{\beta_1 \cdots \beta_k}_{\alpha_1 \cdots \alpha_k}$ of Eqs.~(\ref{eq:stringops}) and (\ref{eq:TLJkconservation}) in the OBC PF and TL models respectively also have simple MPO representations.
The MPO matrices $A_j$ in those cases have the same form as Eq.~(\ref{eq:kMPOexpression}) with the substitutions $N^{\sigma_l}_j \rightarrow (-1)^j N^{\alpha_l}_j$ and $N^{\sigma_l}_j \rightarrow (M_j)^{\beta_l}_{\alpha_l}$ respectively. 
Moreover, for the TL models, the additional quadratic IoMs $M^{\beta_1, \beta_2}_{\alpha_1, \alpha_2}$ of Eq.~(\ref{eq:J2conservedfull}) also have simple MPO representations with bond dimension $\chi = 3 + m (\delta^{\beta_1}_{\alpha_2} + \delta^{\beta_2}_{\alpha_1})$ as follows
\begin{equation}
    A_j = 
    \begin{pmatrix}
        \mathds{1} & \left(J_j\right)^{\beta_1}_{\alpha_1} & -\frac{\delta^{\beta_2}_{\alpha_1}}{\sqrt{m}}\left(J_j\right)^{\beta_1}_1 & \cdots & -\frac{\delta^{\beta_2}_{\alpha_1}}{\sqrt{m}}\left(J_j\right)^{\beta_1}_m  & -\frac{\delta^{\beta_1}_{\alpha_2}}{\sqrt{m}}\left(J_j\right)^{1}_{\alpha_1} & \cdots & -\frac{\delta^{\beta_1}_{\alpha_2}}{\sqrt{m}}\left(J_j\right)^{m}_{\alpha_1} & 0\\
        0 & \mathds{1} & 0 & \cdots & \cdots & \cdots & \cdots & 0 & \left(J_j\right)^{\beta_2}_{\alpha_2} \\
        \vdots & \ddots & \ddots & \ddots & \ddots & \ddots & \ddots & \vdots & \frac{\delta^{\beta_2}_{\alpha_1}}{\sqrt{m}}\left(J_j\right)^{1}_{\alpha_1} \\
        \vdots & \ddots & \ddots & \ddots & \ddots & \ddots & \ddots & \vdots & \vdots \\
        \vdots & \ddots & \ddots & \ddots & \ddots & \ddots & \ddots & \vdots & \frac{\delta^{\beta_2}_{\alpha_1}}{\sqrt{m}}\left(J_j\right)^{m}_{\alpha_1} \\
        \vdots & \ddots & \ddots & \ddots & \ddots & \ddots & \ddots & \vdots & \frac{\delta^{\beta_1}_{\alpha_2}}{\sqrt{m}}\left(J_j\right)^{\beta_2}_{1} \\
        \vdots & \ddots & \ddots & \ddots & \ddots & \ddots & \ddots & 0 & \vdots \\
        \vdots & \ddots & \ddots & \ddots & \ddots & \ddots & \ddots & \mathds{1} & \frac{\delta^{\beta_1}_{\alpha_2}}{\sqrt{m}}\left(J_j\right)^{\beta_2}_{m} \\
        0 & \cdots  & \cdots & \cdots & \cdots & \cdots & \cdots & 0 &\mathds{1} \\
    \end{pmatrix}.
\label{eq:MconserveMPO}
\end{equation}
\end{document}